\documentclass[useAMS,usegraphicx,usenatbib]{mn2e}

\newcommand{\Teff}{\ensuremath{T_{\rm eff}}}               
\newcommand{\ms}{\,m\,s$^{-1}$}                            
\newcommand{\mc}[1]{\multicolumn{2}{c}{#1}}
\newcommand{\er}[3]{\ensuremath{#1^{+#2}_{-#3}}}
\newcommand{\erm}[3]{\mc{\ensuremath{#1^{+#2}_{-#3}}}}
\newcommand{\Nobs}{\ensuremath{N_{\rm obs}}}
\newcommand{\Porb}{\ensuremath{P_{\rm orb}}}                
\newcommand{\reff}[1]{{#1}}                                  

\title[Homogeneous studies of transiting extrasolar planets. I.]
      {Homogeneous studies of transiting extrasolar planets. I. Light curve analyses}

\author[John Southworth]
       {John Southworth\thanks{E-mail: jkt@astro.keele.ac.uk} \\
        Department of Physics, University of Warwick, Coventry, CV4 7AL, UK}

\begin{document} \maketitle 

\begin{abstract}
I present an homogeneous analysis of the transit light curves of fourteen well-observed transiting extrasolar planets. The light curves are modelled using {\sc jktebop}, random errors are measured using Monte Carlo simulations, and the effects of correlated noise are included using a residual-permutation algorithm. The importance of stellar limb darkening on the light curve solutions and parameter uncertainties is investigated using five different limb darkening laws and including different numbers of coefficients as fitted parameters. The linear limb darkening law cannot adequately fit the HST photometry of HD\,209458, but the other four laws give very similar results to each other for all transit light curves. In most cases fixing the limb darkening coefficients at theoretically predicted values does not bias the results, but {\em does} cause the error estimates to be too small. The available theoretical limb darkening coefficients clearly disagree with empirical values measured from the HST light curves of HD\,209458; limb darkening must be included as fitted parameters when analysing high-quality light curves.\\
In most cases the results of my analysis agree with the values found by other authors, but the uncertainties I find can be significantly larger (by factors of up to three). Despite these greater uncertainty estimates, the analyses of sets of independent light curves for both HD\,189733 and HD\,209458 lead to results which do not agree with each other. This discrepancy is worst for the ratio of the radii (6.7\hspace{1pt}$\sigma$ for HD\,189733 and 3.7\hspace{1pt}$\sigma$ for HD\,209458), which depends primarily on the depth of the transit. It is therefore not due to the analysis method but is present in the light curves. These underlying systematic errors cannot be detected from the reduced data alone unless at least three independent light curves are available for an individual planetary system.\\
The surface gravities of transiting extrasolar planets are known to be correlated with their orbital periods. New surface gravity values, calculated from the light curve results and the stellar spectroscopic orbits, show that this correlation is still present. New high-precision light curves are needed for HD\,149026, OGLE-TR-10, OGLE-TR-56, OGLE-TR-132 and GJ\,436, and new radial velocity curves for the XO-1, WASP-1, WASP-2 and the OGLE planetary systems.
\end{abstract}

\begin{keywords}
stars: planetary systems --- stars: binaries: eclipsing --- stars: fundamental parameters
\end{keywords}


\section{Introduction}                                                            \label{sec:intro}

The discovery of extrasolar planets was one of the great scientific achievements of the twentieth century \citep{MayorQueloz95natur}, made possible by technological advances in stellar spectroscopy \reff{\citep[e.g.][]{Butler+96pasp}}. At this point over 200 exoplanets have been discovered through radial velocity monitoring of their parent stars, making this technique very successful \citep{UdrySantos07araa}. The limitation of the method is that it is not possible to directly measure any planetary properties: for each system only lower limits on the mass and orbital separation of the planet can be obtained.

The solution to this problem arrived five years later, with the first detection of the transit of an extrasolar planet across the disc of its parent star, HD\,209458 \citep{Charbonneau+00apj,Henry+00apj}. Since then over thirty transiting extrasolar planetary systems (TEPs) have been discovered, the majority by wide-field photometric variability surveys. Much more information is available for TEPs because the light curve shape is sensitive to the sizes of the star and planet and the inclination of the orbit.

From the study of the light curve of a TEP, the spectroscopic orbit of the star, and one additional piece of information (often a constraint on the stellar mass using theoretical models), it is possible to precisely measure the masses, radii, surface gravities and densities of both the star and the planet. This powerful procedure then allows detailed investigations into the chemical composition of the planet and the physical processes which govern its properties and evolution.

The characteristics of the known TEPs have been derived in this way by many authors, but using several analysis techniques and a wide variety of observational and theoretical constraints. Considering only the light curve analyses, various studies have used different limb darkening functions and in many cases the coefficients of these functions have been fixed at theoretically predicted values without any consideration of whether this approach might introduce systematic errors into the results. It is important to pay attention to the effects of limb darkening when modelling the light curves of TEPs because the coefficients may be correlated with the other parameters and so affect the measurement of the physical properties of the system as a whole. In light of this, it is appropriate to concentrate on studying the known TEPs to obtain their properties as accurately and as homogeneously as possible.

In this work I present light curve analyses for all fourteen TEPs for which decent light curves are available. Each system is studied using the same methods, to ensure that the results are as homogeneous and reliable as possible. A requirement of this method is that the importance of effects such as limb darkening must be quantified so the best approach can be identified and used for every system.

To enable an homogeneous study of the light curves of TEPs I investigate the following questions:
Is linear limb darkening an acceptable approximation?
Is it possible to measure accurate limb darkening coefficients from the data?
Are these coefficients consistent with theoretical predictions?
How accurately can orbital inclinations be measured from individual light curves?
How important are correlated errors in each light curve?

The results of this analysis makes it possible to answer further questions:
Do the results in this work agree with published values?
Do independent light curves of the same TEP agree with each other?
If not, what is the size of the systematic error which causes the disagreement?
How well understood are the properties of the known TEPs?

In addition to dealing with these questions, I identify which of the studied TEPs would benefit the most from further study, both photometric and spectroscopic. Finally, I measure the surface gravities of each planet directly from the light curve parameters and the velocity amplitude of the parent star, using the method introduced by \citet{Me+04mn3} and \citet{Me++07mn}. In a second paper (in preparation) the results presented here are used to determine the physical properties of the planets and stars in these systems using additional constraints from theoretical models, paying close attention to random and systematic errors \reff{and to the interagreement of independent sets of stellar evolutionary models}.



\section{Modelling the light curves of transiting extrasolar planetary systems}            \label{sec:analysis}

A TEP is simply a special case of an eclipsing binary system. The study of eclipsing binary light curves has been part of stellar astrophysics for many years, and was originally put on a solid foundation by Russell \citeyearpar{Russell12apj,Russell12apj2}. Russell's technique, sometimes referred to as ``rectification'', was successively refined (see \citealt{RussellMerrill52book,Kopal59book}) but still requires several approximations which are not appropriate for many systems.

When computers became widely available the Russell-Merrill method was superseded by models based on numerical integration, including the Wilson-Devinney code \citep{WilsonDevinney71apj,Wilson79apj,Wilson93aspc}
{\sc wink} \citep{Wood71aj,Wood73pasp}, {\sc ebop} \citep{Etzel75,PopperEtzel81aj} and others \citep{Hill79pdao,Hadrava90coska}, whose use is widespread. More recently, \citet{MandelAgol02apj} and \citet{Gimenez06aa} have produced fully analytical models which are intended for the study of TEPs. This approach allows a very fast computation, but requires approximations such as the assumption of spherical shapes.

One feature of a TEP system is that the radii of the two components are very different, which can be taxing for models based on numerical integration over the surfaces of the components. Those models which use Roche geometry (e.g.\ the Wilson-Devinney code) simulate the component surfaces using several thousands of points each. This means that they very accurately reproduce distorted surfaces, but can be crippled by numerical error when the planet is so small that it only covers a small fraction of the points on the stellar surface. However, the use of a fully analytic model, with its accompanying requirement of sphericity, can potentially lead to an unquantifiable systematic error in TEPs where the planet is distorted because it is very close to the star.

In this work I adopt the {\sc ebop} model \citep{Etzel75,PopperEtzel81aj,Etzel81conf}, which simulates the components of an eclipsing binary system using biaxial ellipsoids. It can also be restricted to spherical objects, so systematic effects arising from the assumption of a physical shape can be easily quantified. {\sc ebop} uses numerical integration in concentric annuli over the surface of each component \citep{NelsonDavis72apj}, an approach which requires only a modest amount of computation and has a much lower numerical error than other non-analytic codes \citep{Etzel75}.

I use the {\sc jktebop} code%
\footnote{{\sc jktebop} is available from \\ {\tt http://www.astro.keele.ac.uk/$\sim$jkt/codes.html}}%
, a heavily modified version of {\sc ebop} \citep{Me++04mn,Me++04mn2} which incorporates the Levenberg-Marquardt optimisation algorithm ({\sc mrqmin}; \citealt{Press+92book}, p.\,678), an improved treatment of \reff{limb darkening (LD)}, and extensive error analysis techniques. The {\sc ebop} model has been shown to work well for TEPs by a number of authors, including \citet{Me++07mn}, \citet{Wilson+06pasp}, \citet{Gimenez06aa} and \citet{Shporer+07mn}.

\subsection{What does a transit light curve tell us?}                       \label{sec:analysis:info}

An accurate orbital period, $\Porb$, and reference time of minimum light, $T_0$, is available for all systems studied in this work. Apart from $\Porb$ and $T_0$, the light curve of a TEP depends mainly on three quantities\reff{, which can be expressed in several different ways. In this work I take the quantities to be} the fractional radii of the two stars and the orbital inclination, $i$. The fractional radii are defined as%
\footnote{Throughout this work I identify stellar parameters with a subscripted `A' and planet parameters with a subscripted 'b' to conform to IAU nomenclature. I have neglected the usual subscripts `s' and `p' as these can lead to confusion (`s' for TEPs refers to the star but for eclipsing binaries has been used to designate the secondary component; 'p' likewise could mean `planet' or conversely `primary' in the different communities).}%
\begin{equation}
r_{\rm A} = \frac{R_{\rm A}}{a} \qquad \qquad r_{\rm b} = \frac{R_{\rm b}}{a}
\end{equation}
where $R_{\rm A}$ and $R_{\rm b}$ are the (absolute) stellar and planetary radii and $a$ is the orbital semimajor axis. There are three main directly measurable quantities: the transit depth, its overall duration, and duration of totality. Given a light curve with a well-defined transit shape, $r_{\rm A}$, $r_{\rm b}$ and $i$ can be measured accurately from these three main quantities.

It is common in the literature to state that ordinary light curves do not reach the accuracy required to lift the degeneracy between $i$ and $R_{\rm A}$ \citep[e.g.][]{Rowe+07xxx}. A transit light curve has three main observable properties and depends on three main physical properties, so such a statement is incorrect. The crucial advantage the study of TEPs has over that of eclipsing binaries is that the optical light produced by the planet can be assumed to be zero, thus reducing the number of light curve parameters from four to three. The results obtained in this and other works demonstrates that ground-based light curves can provide definitive results for TEPs. Note that (apart from $\Porb$) a light curve holds no information about the overall scale of the system so is not directly dependent on the mass or absolute radius of either component.

\subsection{Solution procedure}                                               \label{sec:analysis:lc}

Rather than fitting for $r_{\rm A}$ and $r_{\rm b}$ directly, {\sc jktebop} fits for the sum of the fractional radii, $(r_{\rm A}+r_{\rm b})$, and the ratio of the radii, $k$ \citep{Me+05mn}. This is because these quantities are only very weakly correlated for a wide variety of light curve shapes, which improves solution convergence. In the analyses presented in this work I have always fitted for $(r_{\rm A}+r_{\rm b})$, $k$, $i$ and $T_0$.

The period, $\Porb$, is generally known with much more accuracy than any other parameter, so I have usually fixed it at a known value. In most cases the orbital eccentricity, $e$, is set to zero (see \citealt{Laughlin+05apj} for a well-presented study of previous claims of a non-zero eccentricity for HD\,209458) and the planet is assumed to produce no light (an excellent approximation at optical wavelengths). The mass ratio, which in {\sc ebop} governs how aspherical the planet is, is fixed at an appropriate value. The mass ratio has a negligible effect on the solutions (changes by a factor of two are unimportant). I emphasise that the mass ratio values adopted here are not otherwise used in the determination of the properties of the TEPs. To obtain a high numerical accuracy I set the size of the integration annuli (parameter `{\sc intring}' in {\sc ebop}) to be 1$^\circ$. LD is also included using one of five different laws (see Section\,\ref{sec:analysis:ld}).

One potential problem in analysing TEP light curves is that any additional light contribution from a nearby star will decrease the transit depth and cause a systematic error in the measured fractional radii of the two components. There is unfortunately no way of detecting `third light' from a transit light curve alone: including it as a fitted parameter will render the solution thoroughly indeterminate. Thankfully, the possibility of a blend is one of the basic tests performed by transit search projects to ensure a candidate TEP is not a false positive \citep[e.g.][]{Weldrake+07xxx}. The spectra of all confirmed TEPs have also received intense scrutiny and none have been found to contain light contamination from a third body. Third light can therefore be ruled out at roughly the 5\% level for most systems. This amount of third light is in most cases insufficient to cause an important systematic error in the measured fractional radii.

\reff{One of the assumptions of the light curve models used here and elsewhere is that the component bodies can be represented by geometrical shapes with definite edges. This assumption does not necessarily correspond to the canonical theoretical approach of defining a stellar or planetary surface as the point at which the optical depth is $\tau = \frac{2}{3}$. This effect has been considered in detail by \citet{Burrows+07apj}, who find that for a highly irradiated low-gravity planet the difference in radius can be up to 5\%. This effect must be borne in mind during any comparisons between observed and theoretically predicted planetary radii.}

\subsection{Treatment of the stellar limb darkening}                                \label{sec:analysis:ld}

LD is an important effect in light curve analysis of TEPs as it has a small but significant effect on the entire light curve shape. This is often included using the simple linear law \citep{Russell12apj2}:
\begin{equation}
I_\mu = I_0 [1 - u(1-\mu)]
\end{equation}
where $\mu = \cos\gamma$ and $\gamma$ is the angle between the surface normal and the line of sight. $I_0$ is the intensity at the centre of the stellar disc, $I_\mu$ is the intensity at angle $\gamma$ and $u$ is the \reff{LD coefficient (LDC)}. The linear LD law is known to be a poor approximation to the predictions of model atmospheres \citep[e.g.][]{Vanhamme93aj}. There is also evidence that it performs less well than more complex laws, at least at optical blue and visual wavelengths, in eclipsing binaries \citep{Me++07aa} and in TEPs \citep{Me++07mn}.

A currently unanswered question in the study of TEPs is how important is LD, and does the use of different laws make much difference to the light curve results? I have therefore implemented a number of other LD laws into {\sc jktebop} and used them in the light curve modelling process. These include the quadratic law \citep{Kopal51}:
\begin{equation}
I_\mu = I_0 [1 - u_{\rm q}(1-\mu) - v_{\rm q}(1-\mu)^2]
\end{equation}
where the linear LDC is $u_{\rm q}$ and the nonlinear LDC is $v_{\rm q}$; the square-root law \citep{DiazGimenez92aa}:
\begin{equation}
I_\mu = I_0 [1 - u_{\rm s}(1-\mu) - v_{\rm s}(1-\sqrt{\mu})]
\end{equation}
and the logarithmic law \citep{KlinglesmithSobieski70aj}:
\begin{equation}
I_\mu = I_0 [1 - u_{\rm \ell}(1-\mu) - v_{\rm \ell}\mu\ln\mu]
\end{equation}
I have also used the cubic law (\citealt{Vantveer60book}; see also \citealt{Barban+03aa}):
\begin{equation}
I_\mu = I_0 [1 - u_{\rm c}(1-\mu) - v_{\rm c}(1-\mu)^3]
\end{equation}
but this is complicated by the unavailability of theoretically-predicted LDCs.

In \citet{Me++07aa} it was found that the square-root and quadratic LD laws gave significantly better results than the linear law for an eclipsing binary with partial eclipses. Attempts to derive both linear and nonlinear LDCs for each star were successful, but gave imprecise results because the two LDCs for one object are strongly correlated \citep[][their Fig.\,3]{Me++07aa}. A similar situation was consistently found in the current work (see Fig.\,\ref{fig:lc:tres1mc}). We can draw three conclusions from this: (1) nonlinear LD laws {\em should} be used; (2) in most cases it is sufficient to fix one LDC at a reasonable value and fit for only the second LDC; (3) using LD laws with three or more LDCs (e.g.\ the four-coefficient law introduced by \citealt{Claret00aa}) is unnecessary for data of the level of accuracy currently achievable.

When modelling the light curves in this work I have calculated solutions for all five LD laws given above. I have always tried to fit for two LDCs when using a nonlinear law. In those cases where this did not give reasonable results, I have fixed the nonlinear LDC to a reasonable (theoretically predicted) value and fitted for the linear LDC. In a few cases this has also delivered unphysical results, so both LDCs had to be fixed to theoretical values. When an LDC was fixed, its value was perturbed in the error analysis to ensure that this source of error was included in the results. \reff{In every case the applied perturbations followed a flat distribution of $\pm$0.1 around the fixed value.} Although some nonlinear LDCs were fixed to theoretically predicted values, in these cases the linear and nonlinear LDCs were so strongly correlated that this introduce only an infinitesimal dependence on theoretical model atmospheres. \reff{For a few systems I have also presented, for illustration and comparison with other studies, solutions with both LDCs fixed and also not perturbed.}

\reff{Other researchers have fitted linear combinations of LDCs, for example $2u_{\rm q}+v_{\rm q}$ and $u_{\rm q}-2v_{\rm q}$ for the quadratic law \citep[e.g.][]{Winn++07apj}. These combinations are less strongly correlated than $u_{\rm q}$ and $v_{\rm q}$ so their values can be found more precisely. Using these combinations can improve convergence when fitting observational data, but I find that they do not lead to more precise measurements for $u_{\rm q}$ and $v_{\rm q}$. I have not used these combinations for two reasons: no problems with convergence have been noticed; and $u$ and $v$ are more straightforwardly comparable to model predictions.}

Theoretically predicted LDCs were obtained by bilinear interpolation in the stellar \Teff\ and surface gravity%
\footnote{The code for obtaining theoretically predicted LDCs by bilinear interpolation is called {\sc jktld} and is available from {\tt http://www.astro.keele.ac.uk/$\sim$jkt/codes.html}},
using the tabulations provided by \citet{Vanhamme93aj}, \citet{Diaz++95aas}, \citet{Claret00aa,Claret04aa2}, \citet{ClaretHauschildt03aa} and \citet{Claret++95aas}

\subsection{Error analysis}                                                      \label{sec:analysis:errors}

The formal parameter uncertainties produced by {\sc mrqmin} are calculated from the solution covariance matrix and are unreliable in the presence of parameter correlations \citep{Popper84aj, MaceroniRucinski97pasp, MeClausen07aa}, so have been ignored in this work. To calculate the random errors on the light curve parameters I have used the Monte Carlo simulation algorithm implemented in {\sc jktebop} \citep{Me++04mn2,Me+04mn3}, which was found to give reliable results by \citet{Me+05mn}.

The Monte Carlo algorithm finds the best fit to the actual data and uses this to generate a model light curve with the same time sampling. Then Gaussian noise is added and the resulting synthetic light curve is fitted in the same way as the actual data. The latter step is performed many times, and the 1\hspace{1pt}$\sigma$ uncertainties in the measured light curve parameters are assessed from the distribution of the best fits to the simulated light curves. A vital part of this process is that substantial random perturbations are applied to the initial parameter values before fitting each simulated light curve. This allows a detailed exploration of the parameter space and of parameter correlations, and so gives similar results to the Markov Chain Monte Carlo techniques used by other researchers \citep[e.g.][]{Holman+06apj}. For each analysis I have used 1000 Monte Carlo simulations, which is a good compromise between avoiding small-number statistics and requiring a large amount of CPU time.

Whilst the random errors are straightforward to quantify, the presence of correlated `red' noise can cause systematic errors which the Monte Carlo algorithm will not properly account for. This is a widespread problem in the study of TEPs because changes in airmass, telescope pointing and focus, detector response and other effects can cause significant correlations between adjacent datapoints in a light curve. Problems can also be caused by the data reduction technique. \reff{Unfortunately, there is no general way of completely removing these effects from a light curve.} An excellent example of systematic bias is given by \citet{Gillon+07aa3}, who demonstrate that the depth of an observed transit of OGLE-TR-132 varies depending on exactly how the data reduction is performed. This bias could not be detected from the reduced light curve alone. A detailed investigation of how red noise affects the analysis of TEPs is given by \citet{Pont++06mn}

To assess the effect of systematic errors on the uncertainties of the light curve parameter measurements, I have implemented into {\sc jktebop} a residual-permutation (or ``prayer bead'') algorithm \citep{Jenkins++02apj}. In this method, the residuals around the best-fitting light curve solution are all shifted to the next datapoint. A new fit is found, then the residuals are shifted again, with those which drop off the end of the dataset being wrapped around to the start of the data. This process continues until the residuals have cycled back to where they originated, and results in a distribution of fitted values for each parameter from which its uncertainty can be estimated. The residual-permutation algorithm gives reliable and easily interpretable results, but its statistical properties are not totally clear and it is only sensitive to correlated noise on timescales different to the eclipse duration. The method has been used before on several TEPs \citep[e.g.][]{Bouchy+05aa2,Pont+05aa,Gillon+07aa1,Knutson+07xxx}.

For each light curve I have used the residual permutation algorithm to assess the importance of systematic errors. In those cases where it gives larger parameter uncertainties than the Monte Carlo algorithm, which only considers Poisson noise, I have adopted the residual permutation results, which are sensitive to both Poisson and correlated noise. The parameter error estimates presented in this work also include a contribution from the effects of using different LD laws, and in every case are 1\hspace{1pt}$\sigma$ errors. When observational errors are available I have used these to set the size of the Gaussian noise level for the Monte Carlo analysis. In cases where this resulted in a reduced $\chi^2$ ($\chi_{\rm red}^{\ 2}$) of greater than unity I have multiplied the Monte Carlo errors by $\sqrt{\chi_{\rm red}^{\ 2}}$ to compensate for the underestimated observational errors \citep{Bruntt+06aa,Me++07aa}.


\section{Results for each system}                                                          \label{sec:results}

Below I present an analysis of the available light curves of each TEP for which good or definitive data are available. For each light curve a set of solutions are tabulated for each of the five LD laws and for fitting either one or two LDCs. As these tables are quite bulky they have been shifted to an Appendix. For each system there is also a table comparing the final light curve parameters with those from previous analyses.

The TEPs are discussed in rough order of increasing complexity, starting with the TrES, XO and WASP and HAT systems. The TEPs discovered using photometry from the OGLE\footnote{\tt http://www.astrouw.edu.pl/$\sim$ogle/} project follow; these are in general a bit more difficult to analyse because they are fainter and so have less good light curves. GJ\,436 is then discussed on the basis of infrared observations from the {\it Spitzer} space telescope; this analysis is more complicated as the orbit is eccentric and the light produced by the planet is not negligible. Finally I investigate HD\,149026, HD\,189733 and HD\,209458. Whilst the very shallow transit of HD\,149026 makes it difficult to study, the brightness of the last two objects means that the available observations of them are of remarkable quality.


\subsection{TrES-1}

TrES-1 was discovered to be a TEP by the Trans-Atlantic Exoplanet Survey. \citet{Alonso+04apj} presented discovery light curves, follow-up photometry, and high-precision radial velocities which showed it to have a radius similar to that of Jupiter but a significantly lower mass. An excellent $z$-band light curve of TrES-1 was obtained by \citet{Winn++07apj} and covers three consecutive transits. The analysis of these data by \citet{Winn++07apj} used quadratic LD, with the LDCs encouraged towards theoretically predicted values by the inclusion of a mild penalty term in the figure of merit. TrES-1 has also been studied by \citet{SteffenAgol05mn}, who found no evidence of a third planet from analysis of published times of transit, and by \citet{Knutson+07pasp}, who were not able to detect a secondary eclipse from a ground-based $L$-band light curve.

\begin{table*} \caption{\label{tab:lcfit:tres1} Final parameters
of the fit to the \citet{Winn++07apj} light curve of TrES-1 from the
{\sc jktebop} analysis. These have been compared to the properties
of TrES-1 found by \citet{Winn++07apj} and \citet{Alonso+04apj}.
Quantities without quoted uncertainties were not given by those
authors but have been calculated from other parameters which were.}
\begin{tabular}{l l r@{\,$\pm$\,}l r@{\,$\pm$\,}l r@{\,$\pm$\,}l}
\hline \hline
                              & & \mc{This work} & \mc{\citet{Winn++07apj}} & \mc{\citet{Alonso+04apj}} \\
\hline
Sum of the fractional radii   & $r_{\rm A}$$+$$r_{\rm b}$   &   0.1097  & 0.0022   &    \mc{0.1088}      & \mc{ }                      \\
Ratio of the radii            & $k$                         &   0.1381  & 0.0014   &  0.13686 & 0.00082  & \erm{0.130}{0.009}{0.003}   \\
Orbital inclination (deg.)    & $i$                         &  88.67    & 0.71     &    \mc{$>$ 88.4}    & \erm{88.5}{1.5}{2.2}        \\
Fractional stellar radius     & $r_{\rm A}$                 &   0.0964  & 0.0018   &  0.0957  & 0.0014   & \mc{ }                      \\
Fractional planetary radius   & $r_{\rm b}$                 &   0.01331 & 0.00035  &  0.01308 & 0.00023  & \mc{0.01279}                \\
\hline \hline \end{tabular} \end{table*}


\begin{figure} \includegraphics[width=0.48\textwidth,angle=0]{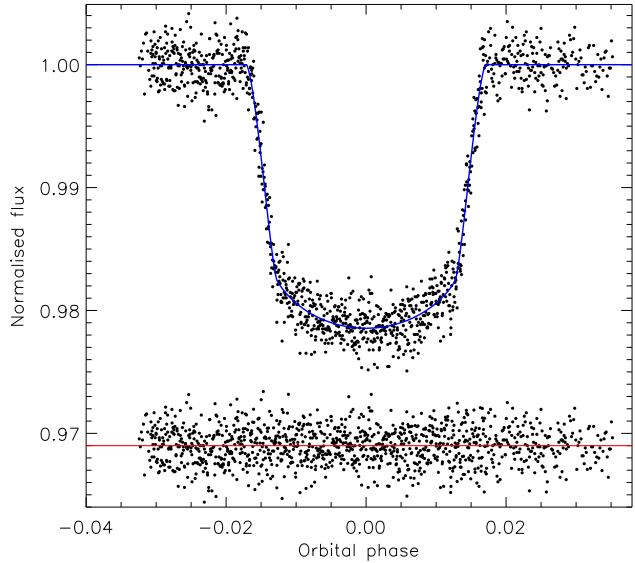}
\caption{\label{fig:lc:tres1} Phased light curve of the transits of TrES-1
from \citet{Winn++07apj} compared to the best fit found using {\sc jktebop}
and the quadratic LD law with both LDCs included as fitted parameters.
The residuals of the fit are plotted at the bottom of the figure, offset
from zero.} \end{figure}

I have modelled the \citet{Winn++07apj} light curve of TrES-1 (1149 datapoints), adopting $\Porb = 3.030065$\,d and fitting for the midpoint of the second of the three transits. A circular orbit was adopted, along with a mass ratio of $q = 0.00085$ \citep{Winn++07apj}. \reff{Solutions have been calculated for three scenarios: (1) LDCs fixed at theoretical values; (2) including the linear LDC as a fitted parameter, and fixing the nonlinear LDC at a theoretical value [and perturbing it during the Monte Carlo analysis to ensure this approach is reflected in the uncertainties]; and (3) fitting for both linear and nonlinear LDCs. The solutions are given in Table\,\ref{tab:lc:tres1}.}

\reff{TrES-1 is unusual among the systems studied in this work in that the nonlinearity of its LD is very small. The lowest residuals are, as expected, found for solutions where both LDCs are allowed to vary. Solutions with the linear LDC as a fitting parameter but the nonlinear LDC fixed are in good agreement with solutions with both LDCs fitted. However, the solutions where the LDCs are held fixed at theoretical values differ in two way: the parameters are slightly different (and in the case of $r_{\rm b}$ the discrepancy is significant) and the error bars are lower.}

\reff{The solutions in which at least one LDC is a fitted parameter are in good agreement with each other. As the values are well-defined even when both LDCs are fitted, I have used these in calculating the final solutions. The solutions with both LDCs fixed at theoretical values have been rejected, because we unfortunately cannot trust theoretical LDCs (see Section\,\ref{sec:tep:209458}). For the linear law, the best-fitting $u_{\rm A}$ does not agree with theoretical predictions. The best light curve fit for the adopted solution type and the quadratic LD law is shown in Fig.\,\ref{fig:lc:tres1} along with the observations and the residuals.}

\reff{The residual perturbation results are in good agreement with the Monte Carlo results, which indicates that systematic errors are not important for these data. The final results from this study are given in Table\,\ref{tab:lcfit:tres1} and compared to published studies. The results presented by \citet{Winn++07apj} are in good agreement, but are closer to the solutions I find when fixing both LDCs. This is expected, because \citet{Winn++07apj} included a penalty term in their figure of merit to encourage the LDCs to be close to theoretical values. Both the \citet{Winn++07apj} and \citet{Alonso+04apj} quote smaller errorbars than found in the present study. This is not surprising given the inclusion of additional error sources in the current work (e.g.\ the effect of different LD laws and the inclusion of the LDCs as fitted parameters), and confirms that the Monte Carlo analysis algorithm in {\sc jktebop} is reliable in the presence of parameter correlations.}

\begin{figure} \includegraphics[width=0.48\textwidth,angle=0]{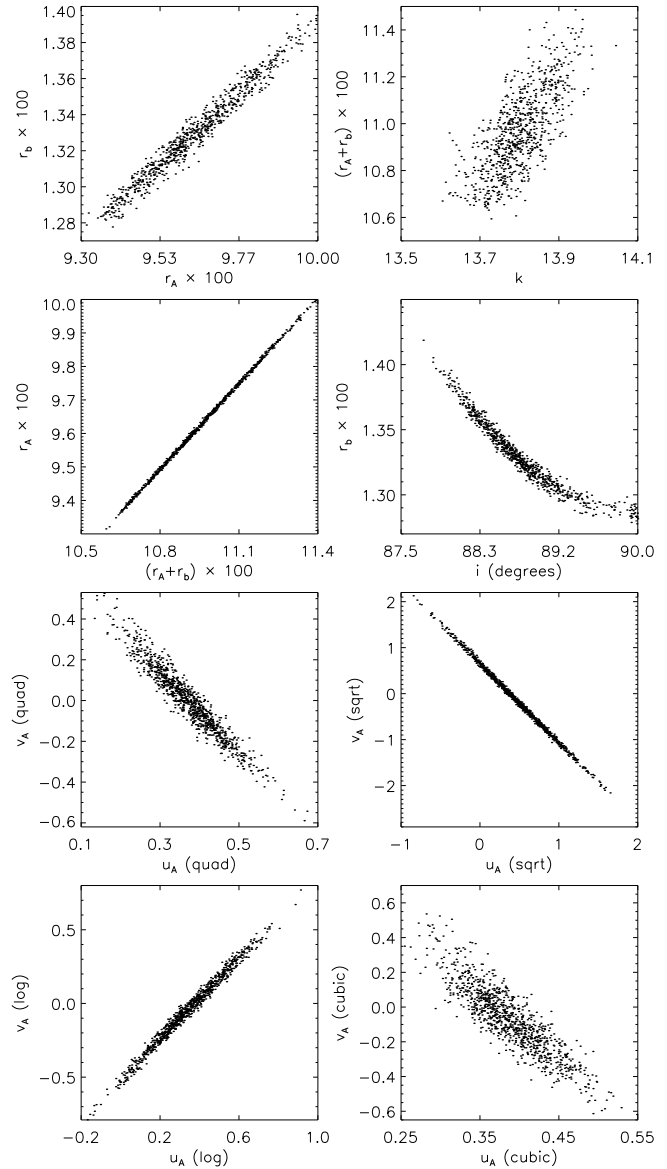}
\caption{\label{fig:lc:tres1mc} Scatter plots showing some results of the
Monte Carlo analysis used to measure the uncertainties in the light curve
parameters. The upper four panels show results obtained using linear LD:
it can be seen that $r_{\rm A}$ and ($r_{\rm A}$$+$$r_{\rm b}$) are strongly
correlated with each other in this representative case, whilst $r_{\rm b}$
and $i$ are also correlated. The lower four panels show the correlations
between the two LDCs for each of the nonlinear LD laws: the
correlation is the strongest for square-root and weakest for cubic.} \end{figure}

\begin{figure} \includegraphics[width=0.48\textwidth,angle=0]{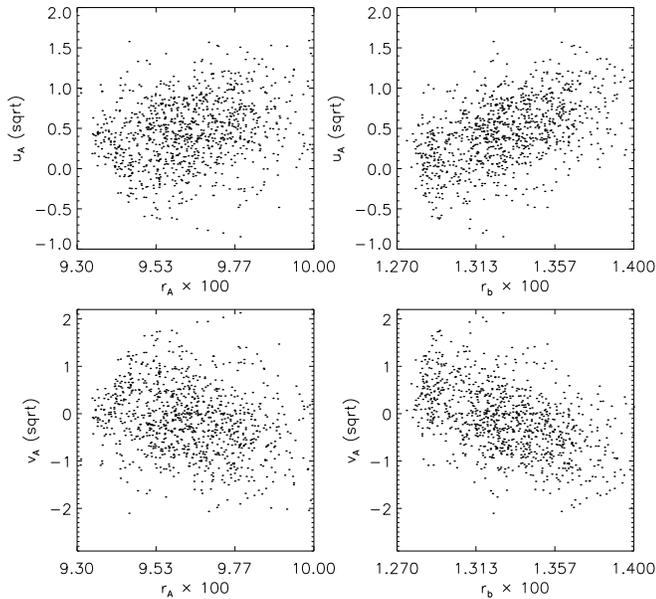}
\caption{\label{fig:lc:tres1mcb} Scatter plots similar to Fig.\,\ref{fig:lc:tres1mc}
but showing the weakness of the correlations between the LDCs and the
fractional radii of the star and planet. The square-root LD law was used for the
results in this plot as this law should show the strongest correlations.} \end{figure}

\subsubsection{Parameter correlations}

Fig.\,\ref{fig:lc:tres1mc} shows example plots of the output of the {\sc jktebop} Monte Carlo analysis algorithm for TrES-1. The first four panels show the joint distributions of parameter values for $r_{\rm A}$, $r_{\rm b}$, $k$, ($r_{\rm A}$$+$$r_{\rm b}$) and $i$. There is a clear correlation between $r_{\rm A}$ and $r_{\rm b}$ and a strong correlation between $r_{\rm A}$ and $k$. The latter correlation is expected because $k$ is very well determined (as it primarily depends on the eclipse depth). The parameters ($r_{\rm A}$$+$$r_{\rm b}$) and $k$ are only weakly correlated, vindicating their use as parameters of the fit.

The lower half of Fig.\,\ref{fig:lc:tres1mc} presents the Monte Carlo output for the LDCs for the four different nonlinear LD laws. Strong correlations are seen for all four laws, which is expected because the light curves do not contain enough information on LD to obtain well-determined values for two LDCs; this situation is very similar to that for the partially-eclipsing binary system $\beta$\,Aurigae \citep{Me++07aa}. The strongest correlation is seen for the square-root LD law and the weakest correlation for the cubic law, but these correlations have no clear effect on the other light curve parameters (Table\,\ref{tab:lc:tres1}). This means that the cubic LDCs are the best-determined, so the cubic LD law should be used if the LDCs are considered to be interesting in their own right.

Fig.\,\ref{fig:lc:tres1mcb} shows the very weak dependence of the derived $r_{\rm A}$ and $r_{\rm b}$ on the LDCs for the square-root law. Very little correlation can be seen, which confirms that the values of the LDCs have very little effect on those for the fractional radii of the components. The square-root law was used for this comparison because it is expected to have the strongest correlations (see Fig.\,\ref{fig:lc:tres1mc}).


\subsection{TrES-2}

TrES-2 was the second transiting extrasolar planetary system discovered by the Trans-Atlantic Exoplanet Survey \citep{Odonovan+06apj} and is both larger and more massive than Jupiter. Excellent $z$-band light curves of three transits were presented and analysed by the TLC project \citep{Holman+07apj}, who worked in close collaboration with \citet{Sozzetti+07apj} to obtain a highly accurate and consistent set of properties for the system. Its relatively low orbital inclination (i.e.\ high impact parameter) means that the fractional radii of the two components can be determined to a high accuracy.

I have studied the light curve obtained by \citet{Holman+07apj}, which contains 1033 datapoints and is superior to those presented in the discovery paper \citep{Odonovan+06apj}. I assumed a mass ratio of 0.0012 \citep{Holman+07apj,Sozzetti+07apj} and fitted for the orbital period as the Holman et al.\ data cover 13 orbital cycles.

\begin{table} \caption{\label{tab:lcfit:tres2} Final parameters
of the fit to the \citet{Holman+07apj} light curve of TrES-2 from
the {\sc jktebop} analysis. These have been compared to the properties
of TrES-2 found by \citet{Odonovan+06apj} and \citet{Holman+07apj}.
Quantities without quoted uncertainties were not given by those
authors but have been calculated from other parameters which were.}
\begin{tabular}{l r@{\,$\pm$\,}l r@{\,$\pm$\,}l r@{\,$\pm$\,}l}
\hline \hline
                      & \mc{This work}       & \mc{O'Donovan}      & \mc{Holman et}       \\
                      & \mc{ } & \mc{et al.\ \citeyearpar{Odonovan+06apj}} & \mc{al.\ \citeyearpar{Holman+07apj}} \\
\hline
$r_{\rm A}+r_{\rm b}$ &   0.1460  & 0.0042   &      \mc{0.1429}    &       \mc{0.1475}    \\
$k$                   &   0.1268  & 0.0032   &      \mc{0.1273}    &    0.1253 & 0.0010   \\
$i$ ($^\circ$)        &  83.71    & 0.42     &   83.90   & 0.22    &   83.57   & 0.14     \\
$r_{\rm A}$           &   0.1296  & 0.0038   &      \mc{0.1267}    &    0.1311 & 0.0021   \\
$r_{\rm b}$           &   0.01643 & 0.00046  &      \mc{0.01614}   &    0.0164 & 0.0004   \\
\hline \hline \end{tabular} \end{table}

\begin{figure} \includegraphics[width=0.48\textwidth,angle=0]{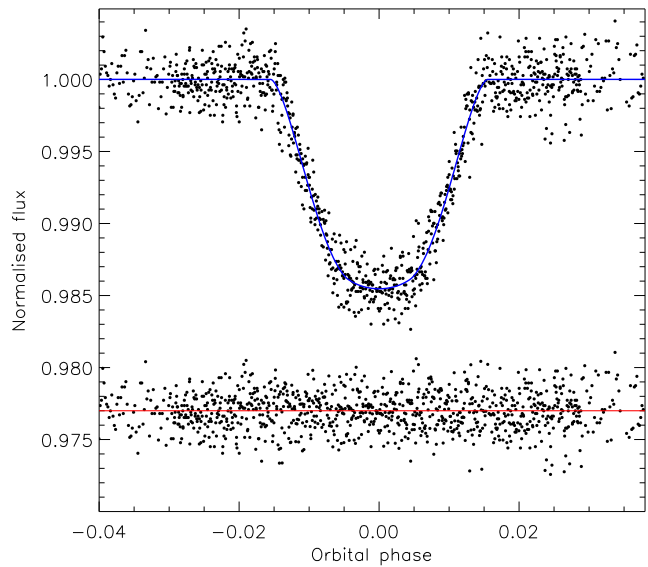}
\caption{\label{fig:lc:tres2} Phased light curve of the transits of TrES-2
from \citet{Holman+07apj} compared to the best fit found using {\sc jktebop}
and the quadratic LD law with the linear LDC included as a fitted parameter.
The residuals of the fit are plotted at the bottom of the figure, offset
from zero.} \end{figure}

\reff{The results of the {\sc jktebop} analysis are given in Table\,\ref{tab:lc:tres2}. The solutions where both LDCs are fitted return scattered results and very uncertain values for the coefficients, so these have been rejected. The solutions with one fitted LDC are well-defined and in good agreement with each other (using the different LD laws), so have been adopted when calculating the final results. Solutions with both LDCs fixed give parameter values in good agreement with the adopted results, but the errorbars are substantially smaller (as also found for TrES-1). It seems that solutions with LDCs fixed at theoretical values not only can give biased results, but give error estimates which are artificially small. These solutions should only be used when there is no alternative.}

For the uncertainties in the light curve parameters I have adopted those found using residual permutation, as they are slightly larger than those from the Monte Carlo results. My final solution is given in Table\,\ref{tab:lcfit:tres2} and compared to those of \citet{Odonovan+06apj} and \citet{Holman+07apj}. As with TrES-1 I find parameter values in good agreement with those from other studies, but my uncertainties are somewhat larger (by factors of up to 3). When studying HD\,189733 and HD\,209458 (see below) I find that my solutions of different light curves do not agree to within the uncertainties I find. This indicates that my uncertainties are not overestimated, and so that the uncertainties given in previous studies can be quite optimistic. This phenomenon is {\em not} peculiar to TrES-2 or the analyses of \citet{Odonovan+06apj} and \citet{Holman+07apj} but occurs for most of the TEPs studied in this work.


\subsection{XO-1}

This was discovered to be a TEP by the XO project \citep{Mccullough+06apj}. A light curve was also obtained by SuperWASP-North and is presented by \citet{Wilson+06pasp}. The light curves from these two small-aperture large-field instruments are not definitive. Good light curves were obtained and studied by \citet{Holman+06apj} using a variety of telescopes and passbands.

\begin{table*} \caption{\label{tab:lcfit:xo1} Final parameters
of the fit to the \citet{Holman+07apj} light curve of XO-1 from
the {\sc jktebop} analysis. These have been compared to the
properties of XO-1 found by \citet{Mccullough+06apj} and
\citet{Wilson+06pasp}, as well as those of \citet{Holman+06apj}.
Quantities without quoted uncertainties were not given by those
authors but have been calculated from other parameters which were.}
\begin{tabular}{l r@{\,$\pm$\,}l c c c}
\hline \hline
                      & \mc{This work} & \citet{Mccullough+06apj} & \citet{Wilson+06pasp} & \citet{Holman+06apj} \\
\hline
$r_{\rm A}+r_{\rm b}$ &   0.1003  & 0.0022   &    0.1080          &    0.1084           & 0.1000                 \\
$k$                   &   0.1317  & 0.0019   & 0.134 $\pm$ 0.004  & 0.138 $\pm$ 0.020   & 0.13102 $\pm$ 0.00064  \\
$i$ ($^\circ$)        &  89.06    & 0.84     &  87.7 $\pm$ 1.2    &  88.9 $\pm$ 1.0     & \er{89.31}{0.46}{0.53} \\
$r_{\rm A}$           &   0.0886  & 0.0019   &    0.0953          &   0.0953            & 0.0884                 \\
$r_{\rm b}$           &   0.01166 & 0.00035  &    0.01273         &   0.01312           & 0.01159                \\
\hline \hline \end{tabular} \end{table*}

\begin{figure} \includegraphics[width=0.48\textwidth,angle=0]{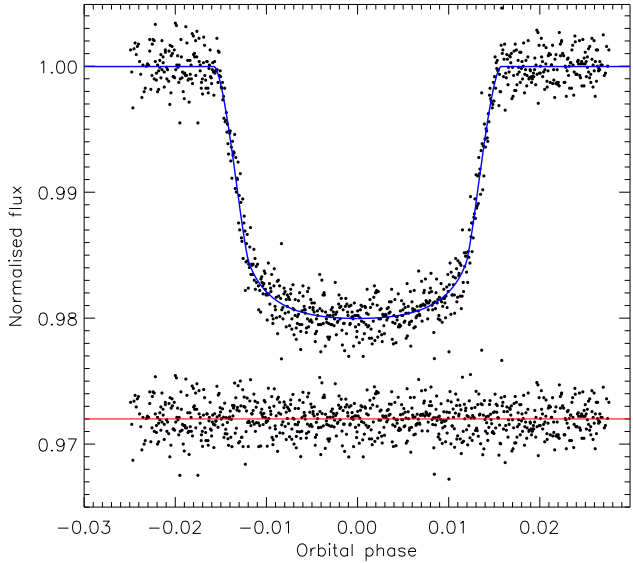}
\caption{\label{fig:lc:xo1} Phased light curve of the transits of XO-1
from \citet{Holman+06apj} compared to the best fit found using {\sc jktebop}
and the quadratic LD law with the linear LDC included as a fitted parameter.
The residuals of the fit are plotted at the bottom of the figure, offset
from zero.} \end{figure}

In this work I have analysed the two best datasets, which come from the FLWO 1.2\,m ($z$-band, two transits, 821 datapoints) and the Palomar 1.5\,m ($R$-band, 209 datapoints) telescopes. I have adopted $\Porb = 3.941534$\,d and a mass ratio of 0.001 \citep{Mccullough+06apj}. The results are given in Table\,\ref{tab:lc:xo1} for the FLWO data: the Palomar dataset gives substantially larger parameter uncertainties so has only been used as a reality check. The solutions with two fitted LDCs give the lowest scatter, and well-defined coefficients (except for the square-root law), so have been used to calculate the final results. As with TrES-1, the best-fitting $u_{\rm A}$ for the linear law does not agree with theoretical predictions. The residual-permutation uncertainties are in good agreement with the Monte Carlo results, indicating that correlated noise is not important for these data.

The quadratic-law best fit is shown in Fig.\,\ref{fig:lc:xo1} and the final light curve parameters are given in Table\,\ref{tab:lcfit:xo1}. The agreement between the parameters found here and those of \citet{Holman+06apj} is excellent, but again my uncertainties are significantly greater. The results of \citet{Mccullough+06apj} and \citet{Wilson+06pasp} also agree well, taking into account the lower quality of the light curves available to them.


\subsection{WASP-1}

WASP-1 was presented as a candidate TEP by \citet{Christian+06mn}, and its planetary nature was confirmed by \citet{Cameron+07mn}. A decent $I$-band light curve, containing 583 datapoints covering two transits, was presented by \citet{Shporer+07mn}%
\footnote{The data obtained by \citet{Shporer+07mn} are available in PDF format from the electronic version of that paper. They were converted into machine-readable format for this work using the UNIX command {\tt /usr/bin/pdftotext -layout}.}%
. A good $z$-band light curve was obtained by \citet{Charbonneau+07apj} and contains 657 datapoints.

\begin{table*} \caption{\label{tab:lcfit:wasp1} Final parameters
of the fit to the \citet{Shporer+07mn} and \citet{Charbonneau+07apj} light curves of WASP-1 from
the {\sc jktebop} analysis. These have been compared to the properties
of WASP-1 found by \citet{Shporer+07mn} and \citet{Charbonneau+07apj}.
Quantities without quoted uncertainties were not given by those
authors but have been calculated from other parameters which were.}
\begin{tabular}{l c c c c}
\hline \hline
                      & This work                   & This work (final result)    & \citet{Shporer+07mn} & \citet{Charbonneau+07apj}\\
                      & (Shporer data)              & (Charbonneau data)          &                      &                          \\
\hline
$r_{\rm A}+r_{\rm b}$ & \er{0.200}{0.015}{0.013}    & \er{0.1918}{0.0101}{0.0064} & 0.1909               &                          \\
$k$                   & \er{0.1033}{0.0016}{0.0016} & \er{0.1045}{0.0015}{0.0015} & 0.1015 $\pm$ 0.0013  & 0.1019 $\pm$ 0.0009      \\
$i$ ($^\circ$)        & 84.0 to 90.0                & 86.5 to 90.0                & 89.8 $\pm$ 1.9       & 86.1 to 90.0             \\
$r_{\rm A}$           & \er{0.182}{0.011}{0.013}    & \er{0.1737}{0.0057}{0.0089} & 0.1733 $\pm$ 0.0091  &                          \\
$r_{\rm b}$           & \er{0.0187}{0.0013}{0.0016} & \er{0.0182}{0.0007}{0.0011} & 0.0176               &                          \\
\hline \hline \end{tabular} \end{table*}

\begin{figure} \includegraphics[width=0.48\textwidth,angle=0]{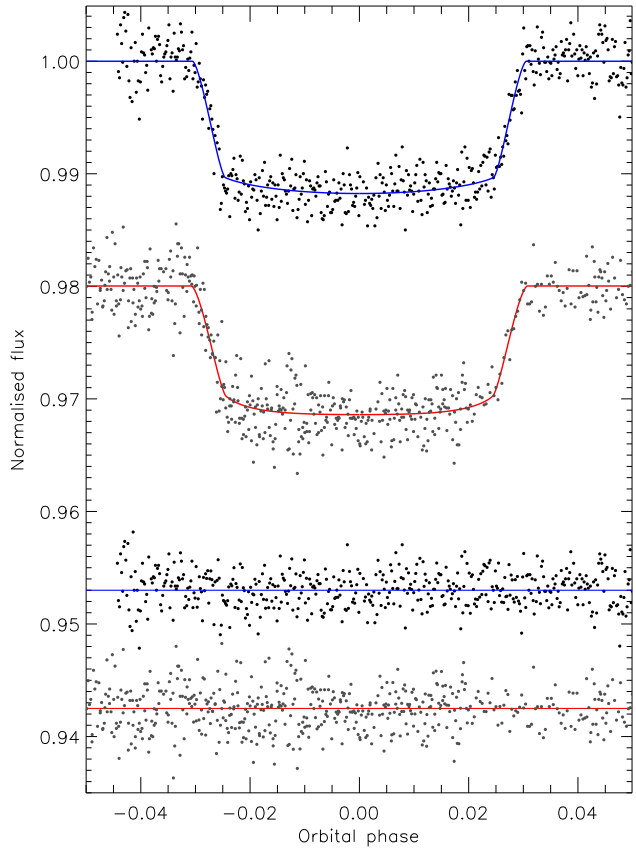}
\caption{\label{fig:lc:wasp1} Phased light curves of the transits of WASP-1
from \citet{Shporer+07mn} (grey points) and \citet{Charbonneau+07apj} (black
points) compared to the best fits found using {\sc jktebop} (red line for the
Shporer data and blue line for the Charbonneau data) and the quadratic LD law
with the linear LDC included as a fitted parameter. The residuals
of the fits are plotted at the bottom of the figures, offset from zero.}
\end{figure}

I have analysed both the Shporer and Charbonneau light curves here, adopting $\Porb = 2.519961$\,d and a mass ratio of 0.00075 \citep{Shporer+07mn}. The results for the Shporer data are given in Table\,\ref{tab:lc:wasp1:shporer}: fitting for two LDCs does not cause a significant improvement in the results so I have used the solutions for fitting one LDC. \reff{From this point on, solutions with fixed LDCs have not been calculated as they are less reliable than other approaches. This is because the imposition of fixed LDCs is an artificial constraint which can both bias the best fit and decrease the derived parameter uncertainties.}

The residual-permutation algorithm gives uncertainties which are substantially greater than the Monte Carlo uncertainties: this problem seems to arise because the WASP-1 has a high orbital inclination (low impact parameter) which is therefore poorly constrained. This difficulty also occurs when analysing the Charbonneau light curve (Table\,\ref{tab:lc:wasp1:charb}), and means that $i$ is only constrained to lie between 86$^\circ$ and 90$^\circ$.

For the final results I adopt my solution of the Charbonneau light curve with parameter uncertainties from the residual-permutation algorithm. The high-inclination nature of this system means its photometric parameters (particularly $i$) are not very well defined despite the availability of high-precision light curves. More extensive high-quality photometric observations are needed for WASP-1. My solution of the Charbonneau light curve is in reasonable agreement with the solution of the Shporer light curve and with the results given by \citet{Shporer+07mn} and \citet{Charbonneau+07apj}, although the published results have lower uncertainty estimates and slightly smaller ratios of the radii. My solutions are compared to the observations in Fig.\,\ref{fig:lc:wasp1}.


\subsection{WASP-2}

WASP-2 was presented as a TEP candidate by \citet{Street+07mn} and confirmation of its nature was presented by \citet{Cameron+07mn}. \citet{Charbonneau+07apj} obtained a good $z$-band light curve (426 datapoints) which I have analysed here. I adopted $\Porb = 2.152226$\,d from \citet{Cameron+07mn} and a mass ratio of 0.001. In contrast to WASP-1, this system has a lower inclination ($i = 84.8^\circ$) and therefore much better-defined fractional radii. This comparison makes it clear that it is much easier to measure the radii of TEPs which do not have central transits.

\begin{figure} \includegraphics[width=0.48\textwidth,angle=0]{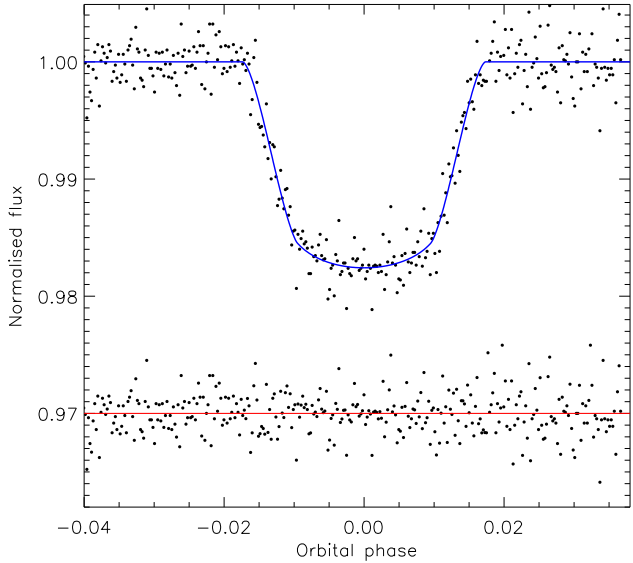}
\caption{\label{fig:lc:wasp2} Phased light curve of the transit of WASP-2
from \citet{Charbonneau+07apj} compared to the best fit found using {\sc jktebop}
and the quadratic LD law with the linear LDC included as a fitted parameter.
The residuals of the fit are plotted at the bottom of the figure, offset
from zero.} \end{figure}

Table\,\ref{tab:lc:wasp2} contains the results of the {\sc jktebop} analysis and demonstrates that allowing two LDCs to be fitted instead of one does not improve the quality of the fit at all and results in very small nonlinear LDCs. The residual-permutation method gives very similar uncertainties to the Monte Carlo analysis. For the final results I therefore adopt the solutions with one LDC fitted and with Monte Carlo uncertainties (Table\,\ref{tab:lcfit:wasp2}) --  these results are in reasonable agreement with those of \citet{Charbonneau+07apj}. The best fit is plotted in Fig.\,\ref{fig:lc:wasp2}.

\begin{table} \caption{\label{tab:lcfit:wasp2} Final parameters
of the fit to the \citet{Charbonneau+07apj} light curve of WASP-2 from
the {\sc jktebop} analysis. These have been compared to the properties
of WASP-2 found by \citet{Charbonneau+07apj}. I have not included the
parameters from \citet{Cameron+07mn} as they are very uncertain due to
the larger observational errors in the SuperWASP discovery light curve.}
\begin{tabular}{l r@{\,$\pm$\,}l r@{\,$\pm$\,}l}
\hline \hline
                      &    \mc{This work}    & \mc{\citet{Charbonneau+07apj}} \\
\hline
$r_{\rm A}+r_{\rm b}$ &   0.1409  & 0.0067   &     \mc{ }         \\
$k$                   &   0.1313  & 0.0017   &  0.1309 & 0.0015   \\
$i$ ($^\circ$)        &  84.83    & 0.53     & 84.74   & 0.39     \\
$r_{\rm A}$           &   0.1245  & 0.0058   &     \mc{ }         \\
$r_{\rm b}$           &   0.01635 & 0.00093  &     \mc{ }         \\
\hline \hline \end{tabular} \end{table}

\subsection{HAT-P-1}

This object was found to be a TEP system by \citet{Bakos+07apj} from data taken by the HAT project \citep{Bakos+02pasp,Bakos+04pasp}. Excellent light curves from the FLWO 1.2\,m ($z$-band, 1709 observations), Lick 1.0\,m ($Z$-band, 1475 datapoints) and Wise 1.0\,m telescopes have been presented and analysed by \citet{Winn+07aj2}.

\begin{table*} \caption{\label{tab:lcfit:hat1} Final parameters
of the fit to the \citet{Winn+07aj2} FLWO and Lick light curves of HAT-P-1 from the
{\sc jktebop} analysis, compared to those found by \citet{Winn+07aj2} and \citet{Bakos+07apj}.
Quantities without quoted uncertainties were not given by those
authors but have been calculated from other parameters which were.}
\begin{tabular}{l r@{\,$\pm$\,}l r@{\,$\pm$\,}l r@{\,$\pm$\,}l c r@{\,$\pm$\,}l}
\hline \hline
                      & \mc{This work (FLWO)} & \mc{This work (Lick)} & \mc{This work (final)}
                        & \citet{Bakos+07apj} & \mc{\citet{Winn+07aj2}} \\
\hline
$r_{\rm A}+r_{\rm b}$ & 0.1026 & 0.0037  & 0.1056 & 0.0056  & 0.1035 & 0.0031  & 0.1089        & \mc{0.1044}      \\
$k$                   & 0.1125 & 0.0011  & 0.1102 & 0.0047  & 0.1124 & 0.0010  & 0.122         & 0.11094& 0.00082 \\
$i$ ($^\circ$)        & 86.37  & 0.31    & 86.10  & 0.38    & 86.26  & 0.24    & 85.9 $\pm$ 0.8& 86.22  & 0.24    \\
$r_{\rm A}$           & 0.0922 & 0.0033  & 0.0951 & 0.0053  & 0.0930 & 0.0028  & 0.0971        & 0.0940 & 0.0028  \\
$r_{\rm b}$           & 0.01037& 0.00046 & 0.01048& 0.00047 & 0.01043& 0.00033 & 0.0118        & 0.0104 & 0.0004  \\
\hline \hline \end{tabular} \end{table*}

\begin{figure} \includegraphics[width=0.48\textwidth,angle=0]{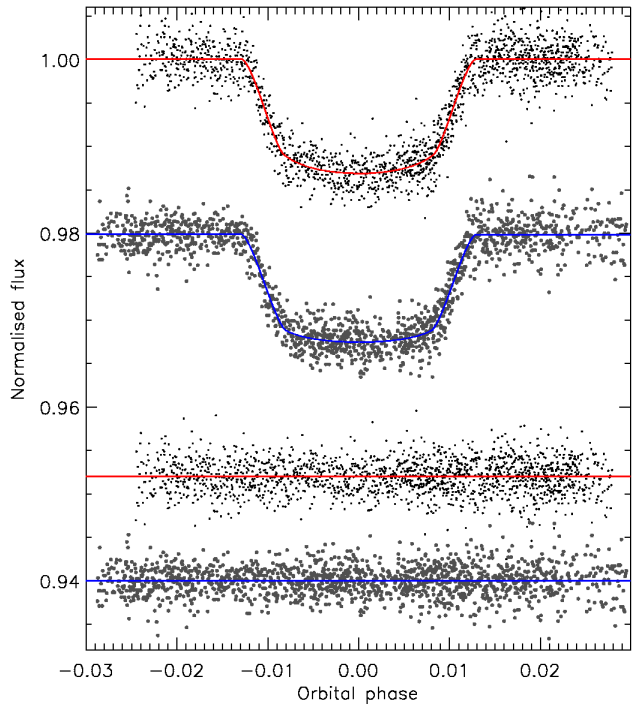}
\caption{\label{fig:lc:hat1} Phased light curves of the transit of HAT-P-1
from \citet{Winn+07aj2} compared to the best fits found using {\sc jktebop}
and the quadratic LD law. The FLWO $z$-band data are plotted at the top and
the Lick $Z$-band light curve is offset by $-0.02$. The residuals of the fit
are plotted at the bottom of the figure, offset from zero.} \end{figure}

I have studied the FLWO and Lick observations from \citet{Winn+07aj2} in this work, adopting a mass ratio of 0.00045 \citep{Bakos+07apj} and including $\Porb$ as one of the fitted parameters. For both datasets the solutions with two fitted LDCs gave no significant improvement over those with one fitted LDC (Table\,\ref{tab:lc:hat1a} and Table\,\ref{tab:lc:hat1b}), so the latter were used to calculate the light curve results. The residual-shift algorithm uncertainties were adopted as these are larger than the Monte Carlo method. The FLWO and Lick results are in excellent agreement (Table\,\ref{tab:lcfit:hat1}, Fig.\,\ref{fig:lc:hat1}) so they were combined using a weighted mean to obtain the final results. The final result is in good agreement with published analyses, and the photometric properties of HAT-P-1 are very well determined.


\subsection{OGLE-TR-10}

The transiting nature of this object was discovered by \citet{Udalski+02aca} in the course of the OGLE-III project. \citet{Konacki+03apj} and \citet{Bouchy+05aa2} subsequently found a small radial velocity motion, and its exoplanetary status was confirmed by \citet{Konacki+05apj}. It was originally thought to be abnormally large for its radius; improved photometry has removed this anomaly but it remains one of the most bloated of the known TEPs.

\reff{Dedicated photometric studies of OGLE-TR-10 have been presented by \citet{Holman+07apj2} and \citet{Pont+07aa}. The former study is based on an $I$-band light curve obtained using the Magellan telescope, which fully covers two transits separated by almost one year. The latter study includes $V$- and $R$-band data covering only part of a single transit, observed with the VLT.}

\reff{The Magellan and VLT observations do not agree on the depth of the transit. This is attributed by \citet{Pont+07aa} to systematic errors arising from the use of image subtraction by \citet{Holman+07apj2}. The OGLE data indicate a transit depth of $\sim$1.9\% ($I$-band), the Magellan data have a transit depth of 1.1\% ($I$-band), and the VLT light curve transit depth is 1.4\% ($V$ and $R$). A version of the OGLE light curve, detrended using the algorithm of \citet{KruszewskiSemeniuk03aca}, gives a transit depth in agreement with the VLT data \citep{Pont+07aa}.}

\reff{I have analysed both the Magellan \citep{Holman+07apj2} and VLT \citep{Pont+07aa} light curves for OGLE-TR-10, and separate solutions are given for both datasets. A mass ratio of 0.00056 and an initial period of 3.101278\,d was adopted \citep{Holman+07apj2}. For the Magellan data,  I have included the period as a fitted parameter. For the VLT data the orbital period is a crucial quantity: because only half of one transit was observed, \Porb\ is very highly correlated with $T_0$ and with $r_{\rm A}$ and $r_{\rm b}$. Adopting a fixed \Porb\ and $T_0$ might therefore bias both the resulting $r_{\rm A}$ and $r_{\rm b}$ {\it and} their error estimates. To avoid these problems I have taken six times of minimum light from the literature (the five times quoted by \citealt{Holman+07apj2} and the $T_0$ given by \citealt{Konacki+05apj}) and have included \Porb\ and $T_0$ as fitted parameters. {\sc jktebop} treats both the times of minimum and the light curve datapoints as observational data \citep[see][]{Me++07aa}, allowing an optimal and statistically correct solution to be found. Uncertainties in all observed quantities are included in the Monte Carlo analysis, so the error estimates also take into account the uncertainties in individual times of minimum light.}


\begin{table*} \caption{\label{tab:lcfit:ogle10} Final parameters of the
fit to the Magellan \citep{Holman+07apj2} and VLT \citep{Pont+07aa} light
curves of OGLE-TR-10 from the {\sc jktebop} analysis. These have been
compared to the properties of OGLE-TR-10 found in previous studies.
Quantities without quoted uncertainties were not given by those authors
but have been calculated from other parameters which were.}
\begin{tabular}{l r@{\,$\pm$\,}l r@{\,$\pm$\,}l r@{\,$\pm$\,}l r@{\,$\pm$\,}l c r@{\,$\pm$\,}l r@{\,$\pm$\,}l}
\hline \hline
             & \mc{Magellan} & \mc{VLT $V$} & \mc{VLT $R$} & \mc{VLT (this}
               & Konacki et & \mc{Holman et} & \mc{Pont et al.}\\
             & \mc{(this work)} & \mc{(this work)} & \mc{(this work)} & \mc{work, final)}
               & al.\ \citeyearpar{Konacki+05apj} & \mc{al.\ \citeyearpar{Holman+07apj2}} & \mc{\citeyearpar{Pont+07aa}}\\
\hline
$r_{\rm A}+r_{\rm b}$ &  0.1468  & 0.0091 & 0.188  & 0.012  & 0.163  & 0.011  & 0.175  & 0.010  &  0.1260           &     \mc{ }      &   \mc{ }      \\
$k$                   &  0.1022  & 0.0024 & 0.1174 & 0.0020 & 0.1172 & 0.0008 & 0.1172 & 0.0007 & 0.127 $\pm$ 0.018 & 0.0990 & 0.0021 & 0.112 & 0.002 \\
$i$ ($^\circ$)        & 86.0     & 1.1    & 83.00  & 0.94   & 84.64  & 0.90   & 83.87  & 0.69   & 89.2  $\pm$ 2.0   &  88.1  & 1.2    &   \mc{ }      \\
$r_{\rm A}$           &  0.1332  & 0.0081 & 0.168  & 0.010  & 0.146  & 0.010  & 0.157  & 0.009  &  0.1118           &     \mc{ }      & 0.183 & 0.018 \\
$r_{\rm b}$           &  0.0136  & 0.0011 & 0.0198 & 0.0014 & 0.0171 & 0.0012 & 0.0182 & 0.0011 &  0.0142           &     \mc{ }      &   \mc{ }      \\
\hline \hline \end{tabular} \end{table*}

\begin{figure} \includegraphics[width=0.48\textwidth,angle=0]{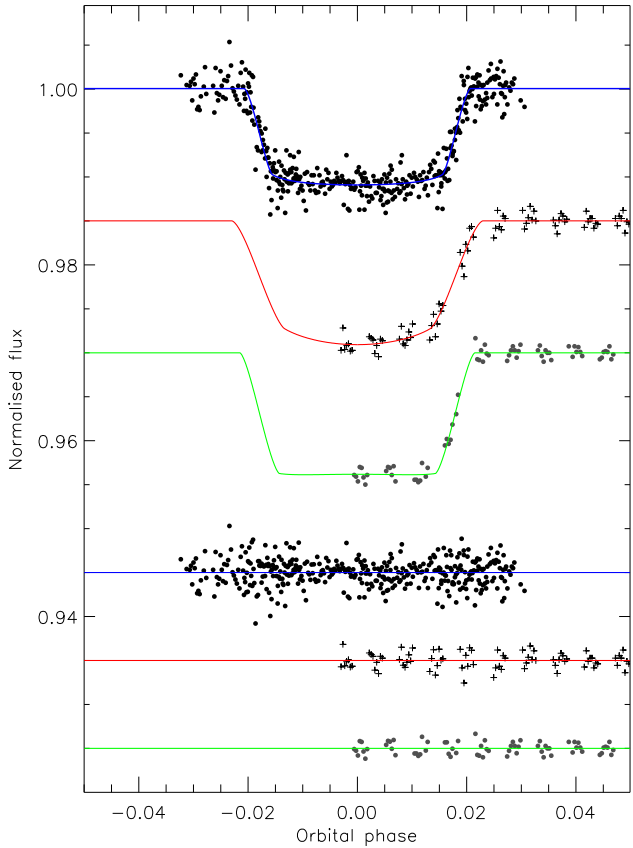}
\caption{\label{fig:lc:ogle10} Phased light curves of the transits of OGLE-TR-10
from \citet{Holman+07apj2} (top) and \citet{Pont+07aa} ($V$ middle, $R$ lower)
compared to the best fits found using {\sc jktebop} and the quadratic LD law
with the linear LDC included as a fitted parameter. The residuals
of the fit are plotted at the bottom of the figure, offset from zero.} \end{figure}

\reff{For the Magellan data the best results were found by fitting for only one LDC (Table\,\ref{tab:lc:ogle10:I}) and the largest parameter uncertainties were obtained by the residual-shift method -- the presence of significant correlated errors is unsurprising given that OGLE-TR-10 is in a crowded stellar field. The best results for the VLT also come from fitting only one LDC, but in these cases the residual-shift method indicates that correlated errors are not important.}

\reff{The solutions of the two VLT light curves are not in great agreement (Tables \ref{tab:lc:ogle10:V} and \ref{tab:lc:ogle10:R}), but the differences are less than 1.7\,$\sigma$ for each parameter. For the final solution I have taken a weighted mean of the VLT $V$ and $R$ light curve solutions, with uncertainties slightly increased to take into account the differences. This solution does not agree well with the Magellan data, as was also found by \citet{Pont+07aa}. The best fits are plotted in Fig.\,\ref{fig:lc:ogle10}, where it can be seen that the two VLT light curves give noticeably different solutions, but both have clearly deeper transits than the Magellan data. The residuals of the best fit confirm that in each case the model provides a good representation of the observations, so any discrepancies are due to the data themselves.}

\reff{The agreement between our final solution (based on the VLT $V$ and $R$ data) and published analyses is not good. The results of \citet{Konacki+05apj} were based on the low-precision OGLE photometry, so it is not surprising that these disagree with later analyses. Those of \citet{Holman+07apj2} are in acceptable agreement with our own solution of the Magellan data, whereas the results from \citet{Pont+07aa} are not in accord with our analysis. Whilst the solution of the VLT data found here should be robust, OGLE-TR-10 would clearly benefit from further photometric study.}

\subsection{OGLE-TR-56}

This object was the second TEP discovered{\reff, and the first} using the extensive photometric database built up by OGLE. Its eclipsing nature was discovered by \citet{Udalski+02aca3} and it was confirmed to be a planet by \citet{Konacki+03natur} from Keck radial velocity observations. Further radial velocities, and re-analyses of the OGLE light curve, have been presented by \citet{Torres+04apj} and \citet{Bouchy+05aa2}.

Aside from the survey-quality photometry obtained by OGLE, the only light curves of OGLE-TR-56 were obtained by \citet{Pont+07aa}. Observations covering one full transit were taken in the $V$ and $R$ passbands using the VLT and FORS1 instrument. The two light curves have a very good photometric precision, particularly considering the apparent magnitude of OGLE-TR-56, but only a small number of datapoints. I have studied these two light curves, graciously provided by F.\ Pont, using $\Porb = 1.211909$\,d and a mass ratio of 0.001 \citep{Pont+07aa}.

\begin{table*} \caption{\label{tab:lcfit:ogle56} Final parameters
of the fit to the \citet{Pont+07aa} light curves of OGLE-TR-56 from
the {\sc jktebop} analysis. These have been compared to the properties
of the system found by \citet{Konacki+03natur}, \citet{Torres+04apj},
\citet{Bouchy+05aa2}, \citet{Bouchy+05aa2} and \citet{Pont+07aa}.
Quantities without quoted uncertainties were not given by those
authors but have been calculated from other parameters which were.}
\begin{tabular}{l r@{\,$\pm$\,}l r@{\,$\pm$\,}l r@{\,$\pm$\,}l c c c c}
\hline \hline
      & \mc{This work} & \mc{This work} & \mc{This work}
        & Konacki et al. & Torres et al. & Bouchy et al. & Pont et al. \\
      & \mc{($V$)} & \mc{($R$)} & \mc{(final)}
        & \citeyearpar{Konacki+03natur} & \citeyearpar{Torres+04apj}
          & \citeyearpar{Bouchy+05aa2} & \citeyearpar{Pont+07aa} \\
\hline
$r_{\rm A}+r_{\rm b}$ & 0.275 & 0.038 & 0.259 & 0.048 & 0.269 & 0.030 & 0.255  &  0.242  &            &             \\
$k$                   & 0.1014& 0.0061& 0.0907& 0.0116& 0.0991& 0.0054& 0.121 &0.124&$0.114\pm0.004$&$0.101\pm0.002$\\
$i$ ($^\circ$)        & 78.8  & 3.0   & 81.8  & 4.1   & 79.8  & 2.4   &$86\pm2$&$81.0\pm2.2$& 81 to 90&             \\
$r_{\rm A}$           & 0.250 & 0.034 & 0.237 & 0.041 & 0.245 & 0.026 & 0.227  &  0.215  &            &$0.26\pm0.01$\\
$r_{\rm b}$           & 0.0253& 0.0041& 0.0215& 0.0062& 0.0241& 0.0034& 0.0276 &  0.0268 &            &             \\
\hline \hline \end{tabular} \end{table*}

\begin{figure} \includegraphics[width=0.48\textwidth,angle=0]{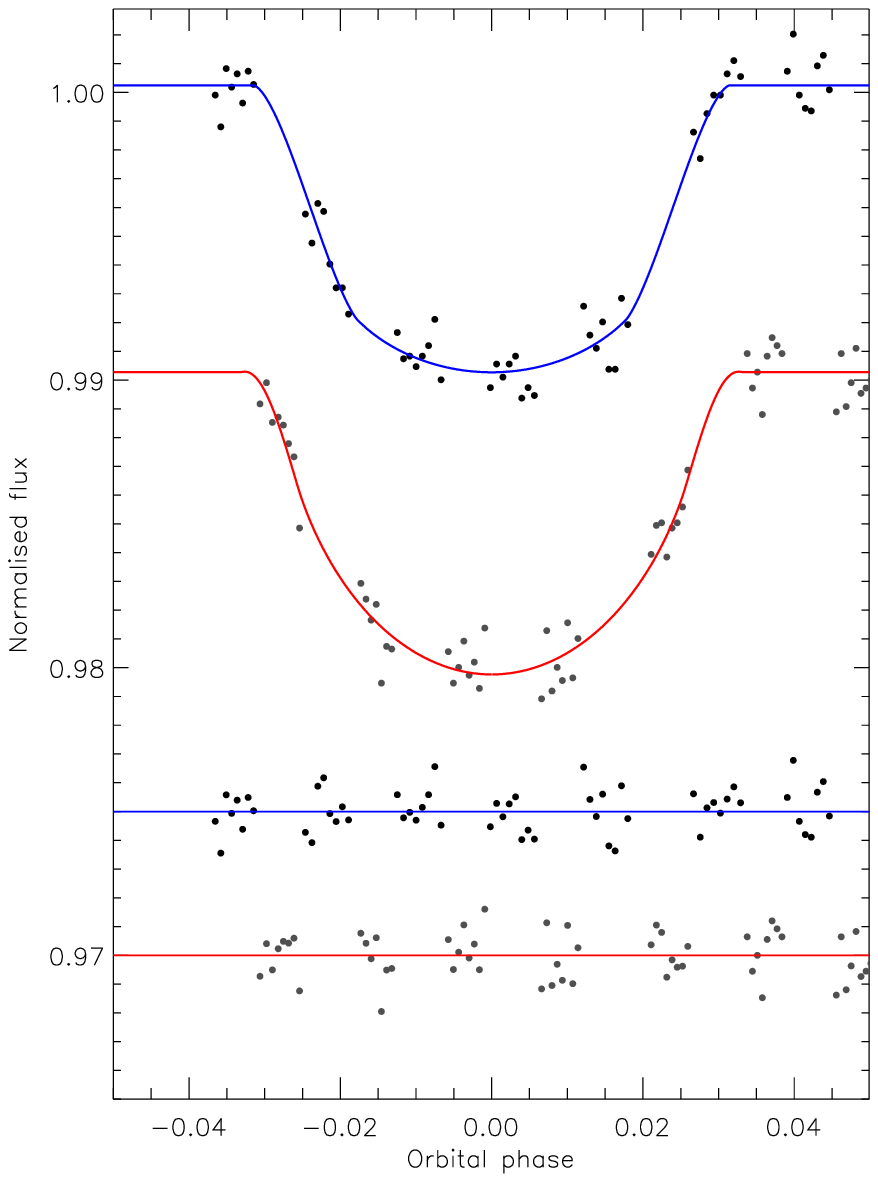}
\caption{\label{fig:lc:ogle56} Phased light curve of the transits of
OGLE-TR-56 from \citet{Pont+07aa} compared to the best fit found using
{\sc jktebop} and the quadratic LD law with the linear LDC
included as a fitted parameter. The upper (blue) curve is for the $V$
band and the lower (red) for the $R$ band data. The residuals of the fit
are plotted at the bottom of the figure, offset from zero.} \end{figure}

For both light curves, solutions with both LDCs free gave poor results (see Table\,\ref{tab:lc:ogle56:v} and Table\,\ref{tab:lc:ogle56:r}). The final results have therefore been taken from the solutions with the linear LDC free and the nonlinear one fixed at theoretically expected values (and perturbed during the Monte Carlo analysis). For both datasets the residual-shift algorithm gave similar results to the Monte Carlo, indicating that correlated errors are not important. For the $R$-band data the solution with the linear LD law is unphysical ($u_{\rm A} > 1.0$) -- the LDC must be fixed when solving this light curve. In addition, the logarithmic and cubic LD law solutions are discrepant so have not been included in the final solution.

The best fits to the two VLT light curves are shown in Fig.\,\ref{fig:lc:ogle56}; the two fits are not visually in good agreement. This can be seen in the final fits to each dataset given in Table\,\ref{tab:lcfit:ogle56}, but the two solutions are actually consistent when the sizes of the errorbars are taken into account. I have therefore used the weighted average of the two for the final solution for OGLE-TR-56 (Table\,\ref{tab:lcfit:ogle56}). A comparison with published results for this TEP shows reasonable agreement when the quality of the OGLE data are taken into account. This system would benefit from more extensive photometry.


\subsection{OGLE-TR-111}

This was also found to be transiting by OGLE \citep{Udalski+02aca2} and has a low density. Its TEP nature was confirmed by \citet{Pont+04aa} and light curve studies have been published by \citet{Winn++07aj} and \citet{Minniti+07apj}. The dataset obtained by \citet{Winn++07aj} is the best and is easily available so was chosen for analysis. It was obtained in the (presumably Cousins) $I$ band and contains 386 observations. I adopted $\Porb = 4.0144479$\,d and a mass ratio of 0.0007 \citep{Winn++07aj} (remember that the value of the mass ratio has extremely little influence on the light curve solutions).

\begin{table*} \caption{\label{tab:lcfit:ogle111} Final parameters
of the fit to the \citet{Winn++07aj} light curve of OGLE-TR-111 from
the {\sc jktebop} analysis. These have been compared to the properties
of OGLE-TR-111 found by \citet{Pont+04aa}, \citet{Winn++07aj} (from
the same data as analysed here) and \citet{Minniti+07apj}.}
\begin{tabular}{l r@{\,$\pm$\,}l r@{\,$\pm$\,}l r@{\,$\pm$\,}l r@{\,$\pm$\,}l}
\hline \hline
& \mc{This work} & \mc{\citet{Pont+04aa}} & \mc{\citet{Winn++07aj}} & \mc{\citet{Minniti+07apj}} \\
\hline
$r_{\rm A}+r_{\rm b}$ &   0.0952  & 0.0044   &   \mc{0.0943}     &    \mc{ }     & \mc{0.09288}                \\
$k$                   &   0.1316  & 0.0023   &   \mc{0.1209}     & 0.132 & 0.002 & \erm{0.1245}{0.0050}{0.0030}\\
$i$ ($^\circ$)        &  88.11    & 0.66     & \mc{86.5 to 90.0} &  88.1 & 0.3   & \mc{87.0 to 90.0}           \\
$r_{\rm A}$           &   0.0842  & 0.0038   &   \mc{0.0841}     &    \mc{ }     & \erm{0.0926}{0.006}{0.011}  \\
$r_{\rm b}$           &   0.01107 & 0.00067  &   \mc{0.01017}    &    \mc{ }     & \mc{0.1028}                 \\
\hline \hline \end{tabular} \end{table*}

\begin{figure} \includegraphics[width=0.48\textwidth,angle=0]{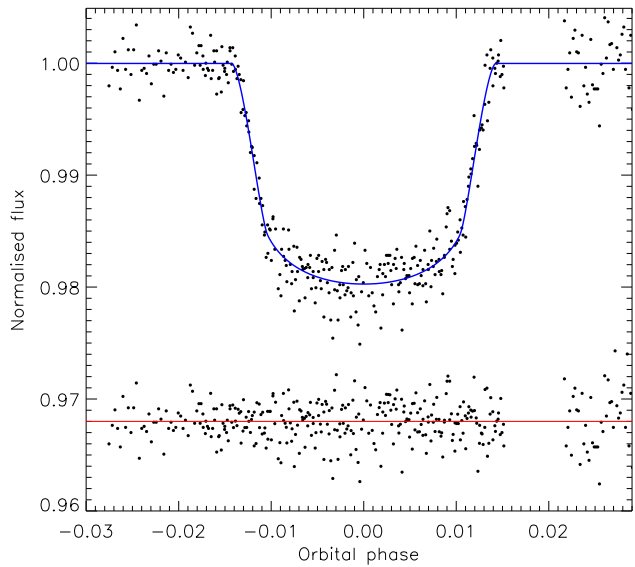}
\caption{\label{fig:lc:ogle111} Phased light curve of the transits of OGLE-TR-111
from \citet{Winn++07aj} compared to the best fit found using {\sc jktebop}
and the quadratic LD law with the linear LDC included as a fitted parameter.
The residuals of the fit are plotted at the bottom of the figure, offset
from zero.} \end{figure}

As for OGLE-TR-10 and OGLE-TR-56 no substantial gain was found in fitting for both LDCs (Table\,\ref{tab:lc:ogle111}), so the nonlinear LDC was fixed for the final solution (and perturbed as usual during the error analysis). \reff{The residual-shift method gave slightly larger uncertainties, so correlated errors are important for these data.} The final results are given in Table\,\ref{tab:lcfit:ogle111} and plotted in Fig.\,\ref{fig:lc:ogle111}, and show good agreement with the study by \citet{Winn++07aj}.


\subsection{OGLE-TR-132}

OGLE-TR-132 was originally discovered as a candidate transiting system by \citet{Udalski+03aca} and confirmed to be a TEP by \citet{Bouchy+04aa}. A light curve from the VLT was presented by \citet{Moutou++04aa}. The data were re-analysed using image deconvolution by \citet{Gillon+07aa3}, who found that the previous light curve was afflicted by systematic errors from the data reduction. An excellent discussion and examples of this problem are given by Gillon et al.

I have solved the VLT light curve from \citet{Gillon+07aa3} (274 observations), which have good precision but do not quite cover the full transit, finishing immediately before the point of fourth contact. Unfortunately this lowers the accuracy of the results, and make correlated noise more important. I adopted a period of 1.689868\,d and a mass ratio of 0.0009 \citep{Gillon+07aa3}.

\begin{table} \caption{\label{tab:lcfit:ogle132} Final parameters
of the fit to the \citet{Gillon+07aa3} light curve of OGLE-TR-132 from
the {\sc jktebop} analysis. These have been compared to the properties
of OGLE-TR-132 found by \citet{Moutou++04aa} and \citet{Gillon+07aa3}.
Quantities without quoted uncertainties were not given by those
authors but have been calculated from other parameters which were.}
\begin{tabular}{l r@{\,$\pm$\,}l r@{\,$\pm$\,}l r@{\,$\pm$\,}l}
\hline \hline
                      & \mc{This work} & \mc{Moutou et al.} & \mc{Gillon et al.} \\
                      & \mc{ } & \mc{\citeyearpar{Moutou++04aa}} & \mc{\citeyearpar{Gillon+07aa3}} \\
\hline
$r_{\rm A}+r_{\rm b}$ &   0.231  & 0.023   &      \mc{ }     &      \mc{ }     \\
$k$                   &   0.0937 & 0.0026  & 0.0812 & 0.0017 & 0.0916 & 0.0014 \\
$i$ ($^\circ$)        &  83.3    & 2.4     &      \mc{ }     &      \mc{ }     \\
$r_{\rm A}$           &   0.211  & 0.020   &      \mc{ }     &      \mc{ }     \\
$r_{\rm b}$           &   0.0198 & 0.0024  &      \mc{ }     &      \mc{ }     \\
\hline \hline \end{tabular} \end{table}

\begin{figure} \includegraphics[width=0.48\textwidth,angle=0]{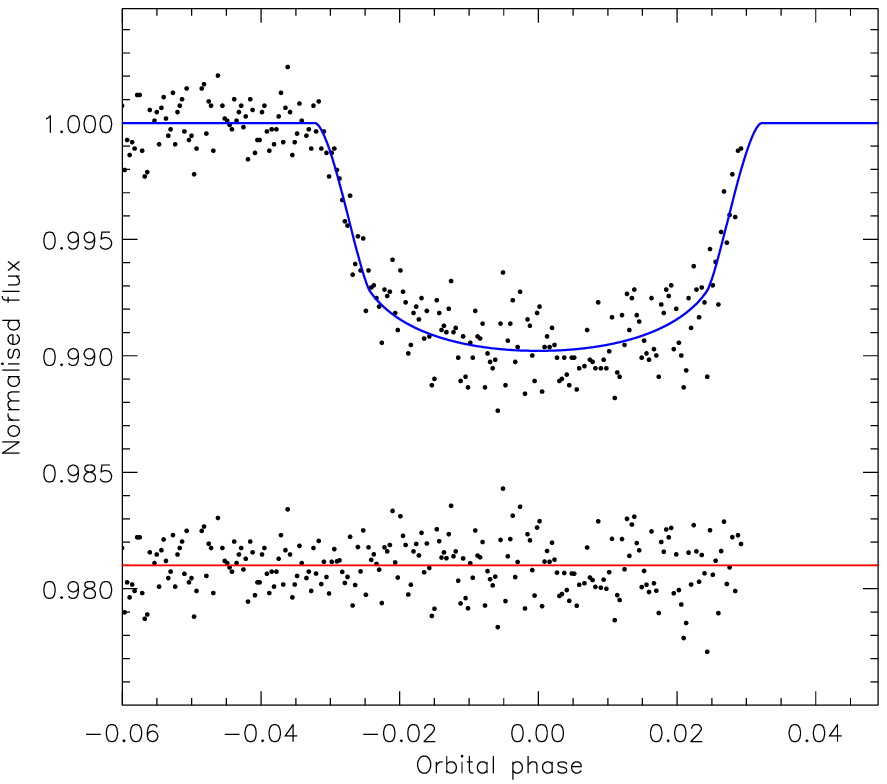}
\caption{\label{fig:lc:ogle132} Phased light curve of the transit of OGLE-TR-132
from \citet{Gillon+07aa3} compared to the best fit found using {\sc jktebop}
and the quadratic LD law with the linear LDC included as a fitted parameter.
The residuals of the fit are plotted at the bottom of the figure, offset
from zero.} \end{figure}

Solutions with two fitted LDCs gave poorly defined coefficients (Table\,\ref{tab:lc:ogle132}) so I have adopted the results of the solutions with the nonlinear LDC fixed. The values of $r_{\rm A}$ and $r_{\rm b}$ are in good agreement between the different solution sets. The orbital inclination is not measured very accurately, and it is important to obtain a new light curve which fully covers the transit. The largest parameter uncertainties are found by the residual-shift method, and this is probably because the light curve does not fully cover the transit. Table\,\ref{tab:lcfit:ogle132} contains the final parameters and the best-fit model is plotted in Fig.\,\ref{fig:lc:ogle132}.


\subsection{GJ 436}

The M\,dwarf system GJ\,436 is a unique object among the known TEPs as both the planet and star are by far the smallest and least massive; the system also has a currently unexplained eccentric orbit. This was known to be a planetary system from spectroscopic radial velocity measurements \citep{Butler+04apj,Maness+07pasp} and transits were first detected by \citet{Gillon+07aa2}. \reff{A good light curve of the transit has been obtained using the {\it Spitzer} Space Telescope \citep{Gillon+07aa1,Deming+07apj}}. {\it Spitzer} has also been used to study the secondary eclipse \citep{Deming+07apj,Demory+07aa}.

\citet{Torres07apj} has measured the parameters of the system from its light and radial velocity curves, theoretical stellar models and its measured absolute $K$-band magnitude and $J-K$ colour index. He found that the star has a radius inflated by approximately 10\% over that predicted by the models, an anomaly which has previously been noticed for low-mass stars in eclipsing binary systems \citep{Ribas03aa,Ribas06apss,lopez07apj}, single low-mass field stars studied using interferometry \citep{Berger+06apj}, and the low-mass secondary components of interacting binaries (\citealt{Littlefair+06sci}).

I have chosen to analyse the {\it Spitzer} light curve presented by \citet{Gillon+07aa1}, which contains 355 datapoints covering one transit at a wavelength of 8\,$\mu$m. It is not possible to constrain the orbital eccentricity and longitude of periastron, $e$ and $\omega$, from one transit so I have adopted the values $e = 0.14 \pm 0.01$ \citep{Demory+07aa} and $\omega = 351^\circ \pm 1.2^\circ$ \citep{Maness+07pasp}. I have also taken $\Porb = 2.64385$\,d from \citet{Gillon+07aa1} and a mass ratio of 0.00016 from \citet{Torres07apj}. An additional complication is that the brightness of the planet is not negligible at 8\,$\mu$m. From \citet[][Figs.\ 1 and 2]{Deming+07apj} I find a surface brightness ratio of approximately 7\%. The surface brightness ratio has been fixed at this value; changes of a factor of two have a smaller effect on the radii than the final error bars. Similarly, changes of $e$ and $\omega$ within their quoted error bars have a negligible effect on the final results.


\begin{table*} \caption{\label{tab:lcfit:gj436} Final parameters
of the fit to the {\it Spitzer} light curve of GJ\,436 from the
{\sc jktebop} analysis, compared to those found by other authors.
Quantities without quoted uncertainties were not given by those
authors but have been calculated from other parameters which were.}
\begin{tabular}{l r@{\,$\pm$\,}l r@{\,$\pm$\,}l c c r@{\,$\pm$\,}l}
\hline \hline
                      & \mc{This work} & \mc{\citet{Gillon+07aa1}} & \citet{Gillon+07aa2}
                        & \citet{Deming+07apj} & \mc{\citet{Torres07apj}} \\
\hline
$r_{\rm A}+r_{\rm b}$ & 0.0791 & 0.0029   & \mc{ }        &                      &        & \mc{ }          \\
$k$                   & 0.08284 & 0.00090 & 0.082 & 0.005 &                      & 0.0836 & 0.0834 & 0.0007 \\
$i$ ($^\circ$)        & 86.43 & 0.18      &  86.5 & 0.2   &\er{85.90}{0.19}{0.18}&        & \mc{ }          \\
$r_{\rm A}$           & 0.0731 & 0.0027   & \mc{ }        &         & 0.0758 $\pm$ 0.0035 & 0.0750 & 0.0033 \\
$r_{\rm b}$           & 0.00605 & 0.00023 & \mc{ }        &                      & 0.00634& \mc{0.00626}    \\
\hline \hline \end{tabular} \end{table*}

\begin{figure} \includegraphics[width=0.48\textwidth,angle=0]{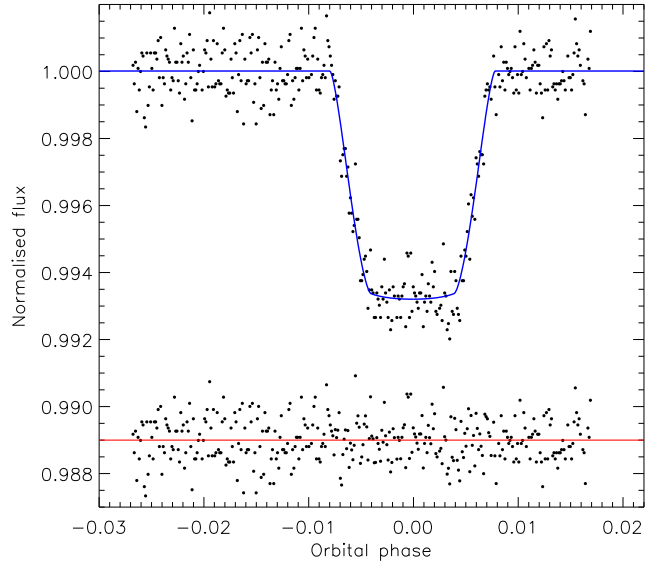}
\caption{\label{fig:lc:gj436} Phased light curve of the transit of GJ\,436
from {\it Spitzer} compared to the best fit found using {\sc jktebop}
and the quadratic LD law with both LDCs included as fitted parameters.
The residuals of the fit are plotted at the bottom of the figure, offset
from zero.} \end{figure}

Attempts to fit for LD did not lead to realistic results (Table\,\ref{tab:lc:gj436}). Whilst the photometry achieves a remarkable precision (0.7\,mmag per point), the observational scatter is still a significant fraction of the very shallow transit depth. The {\it Spitzer} data are taken at 8\,$\mu$m where the effect of LD is minor, so for the final solutions it was acceptable to fix the LDCs to predicted values. The effects of correlated noise are not important. The final results are given in Table\,\ref{tab:lcfit:gj436} and shown in Fig.\,\ref{fig:lc:gj436}. The agreement with literature values is good, but the solution does not lead to particularly accurate radii for either star or planet. Improved photometry is needed, and may be provided in the near future by the Transit Light Curve project \citep[see][]{Winn07xxx}.


\subsection{HD 149026}

HD\,149026\,b is a very anomalous objects, with one of the highest densities of the known TEPs. Its small size suggests that much of its mass lies in a rock/ice core, and HD\,149026 has been the subject of several theoretical investigations concerning its formation and chemical composition \citep[e.g.]{Johnson++07apj, BroegWuchterl07mn, Matsuo+07apj, ZhouLin07apj}.

The TEP nature of HD\,149026 was discovered by \citet{Sato+05apj}, and light curves have also been presented by \citep{Charbonneau+06apj} from the FLWO 1.2\,m telescope
\footnote{Some of the Monte Carlo simulation results presented by \citet[][their Fig.\,2]{Charbonneau+06apj} are concentrated in a sharp peak at $i \approx 90^\circ$ whilst most of the results occupy a much broader distribution at $i \sim 86^\circ$. A probably similar effect was seen whilst writing the Monte Carlo algorithm for {\sc jktebop} and is caused by the value of $i$ becoming almost exactly 90$^\circ$ during the model fitting process. When this happens a numerical derivative will vanish (or an alternative solution will not be taken) if it is calculated symmetrically, because the symmetrical nature of the situation means that a solution with e.g.\ $i = 89.9$ is equivalent to $i = 90.1$. Thus the fit becomes stuck at $i = 90^\circ$ without physical justification. This effect can be avoided by not allowing $i > 90^\circ$. \citet{Charbonneau+06apj} correctly identify their sharp peak at $i \approx 90^\circ$ as a numerical problem and do not use the relevant simulations, so their results will be unaffected.}%
. However, the very shallow transit (depth 0.3\%) means that its light curve parameters remain ill-defined despite the reasonably good quality of the available photometry. \citet{Wolf+07apj} have studied the Rossiter-McLaughlin effect in the system and \citet{Bozorgnia+06pasp} have searched for the presence of lithium and potassium in the planetary atmosphere. \citet{Harrington+07natur} observed a secondary eclipse of HD\,149026 with {\it Spitzer} and found a high brightness temperature at 8\,$\mu$m of $2300 \pm 200$\,K for the planet.

\begin{table*} \caption{\label{tab:lcfit:149026} Final parameters
of the fit to the \citet{Sato+05apj} and \citet{Charbonneau+06apj}
light curves of HD\,149026 from the {\sc jktebop} analysis. These
have been compared to the properties of HD\,149026 found by these
authors. Quantities without quoted uncertainties were not given by
those authors but have been calculated from other parameters which were.}
\begin{tabular}{lccccccc}
\hline \hline
                      & This work                   & This work                   & This work                 & This work                 & \citeauthor{Sato+05apj}  & \citeauthor{Charbonneau+06apj}  & \citeauthor{Wolf+07apj}  \\                  %
                      & (Sato data)                 & (FLWO $g$ data)             & (FLWO $r$ data)           & (final)                   & \citeyearpar{Sato+05apj} & \citeyearpar{Charbonneau+06apj} & \citeyearpar{Wolf+07apj} \\
\hline
$r_{\rm A}+r_{\rm b}$ & \er{0.144}{0.064}{0.014}    & \er{0.158}{0.010}{0.005}    & \er{0.147}{0.011}{0.005}  & \er{0.147}{0.014}{0.008}  & 0.169                    &                                 &                          \\
$k$                   & \er{0.047}{0.005}{0.002}    & \er{0.0533}{0.0023}{0.0014} & \er{0.048}{0.003}{0.003}  & \er{0.048}{0.006}{0.004}  & 0.0514                   &                                 & 0.507 $\pm$ 0.0015       \\
$i$ ($^\circ$)        & 83.0 to 90.0                & 86.0 to 90.0                & 87.0 to 90.0              & 86.0 to 90.0              & 85.3 $\pm$ 1.0           & \er{85.8}{1.6}{1.3}             & 86.1 $\pm$ 1.4           \\
$r_{\rm A}$           & \er{0.138}{0.061}{0.013}    & \er{0.150}{0.009}{0.005}    & \er{0.140}{0.010}{0.004}  & \er{0.140}{0.012}{0.006}  & 0.161                    &                                 &                          \\
$r_{\rm b}$           & \er{0.0065}{0.0039}{0.0008} & \er{0.0080}{0.0007}{0.0004} & \er{0.0068}{0.008}{0.005} & \er{0.0068}{0.011}{0.008} & 0.0082                   & 0.0083                          &                          \\
\hline \hline \end{tabular} \end{table*}

\begin{figure} \includegraphics[width=0.48\textwidth,angle=0]{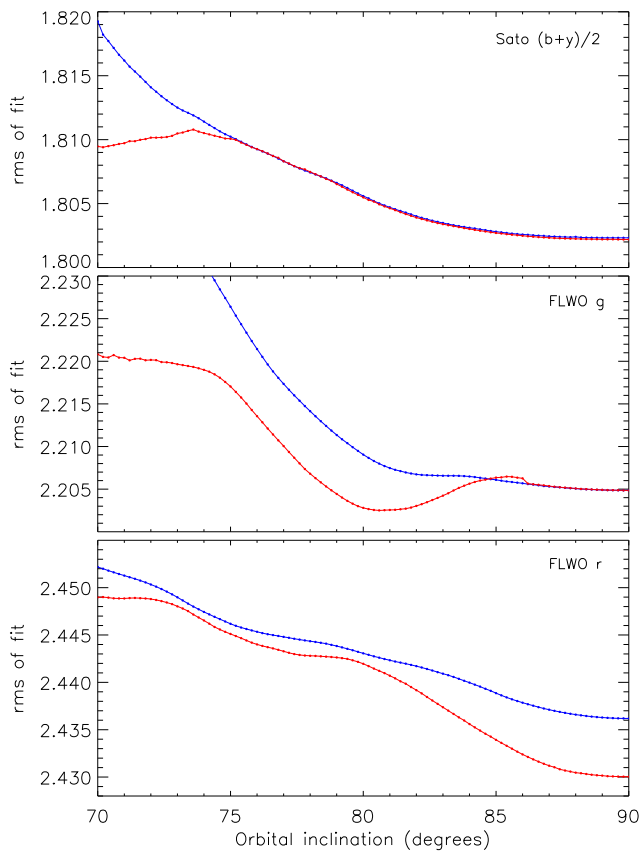}
\caption{\label{fig:rms:149026} Variation of the root-mean-square of
the residuals of the fits to the \citet{Sato+05apj} (top panel) and
the \citet{Charbonneau+06apj} FLWO $g$ (middle) and $r$ (bottom) light
curves as a function of the orbital inclination. In each case only the
quadratic LD law was used and the upper (blue) line is for fits where
both LDCs were fixed and the lower (red) line is for fits
where the linear LDC was fitted for.} \end{figure}

\begin{figure} \includegraphics[width=0.48\textwidth,angle=0]{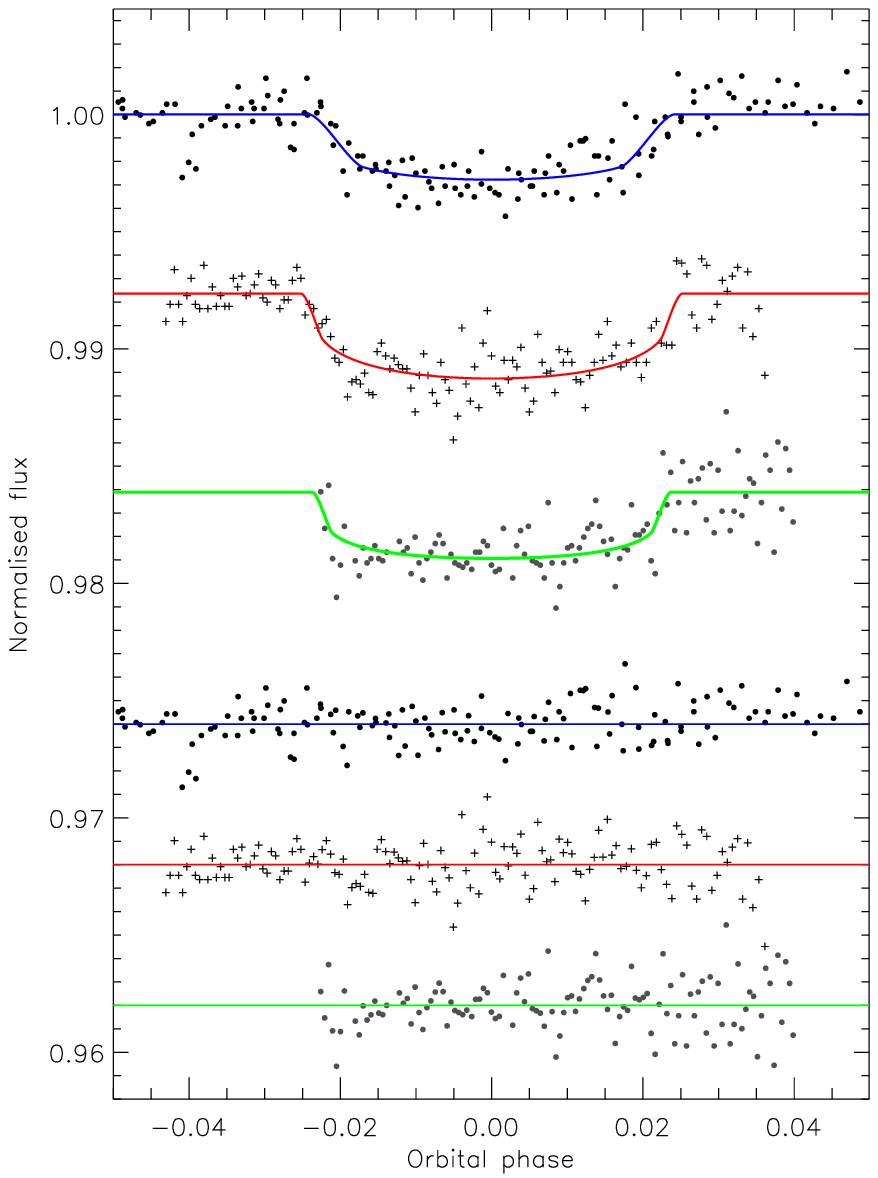}
\caption{\label{fig:lc:149026} Phased light curves of the transits of HD\,149026
from \citet{Sato+05apj} (top) and \citet{Charbonneau+06apj} ($g$ middle, $r$
lower) compared to the best fit found using {\sc jktebop} and the quadratic LD
law. For the \citet{Sato+05apj} data the linear LDC was fitted for whereas for
the \citet{Charbonneau+06apj} LCs all LDCs were fixed at theoretically expected
values. The residuals of the fit are plotted at the bottom of the figure, offset
from zero. All three datasets have been binned by a factor of 5 for display
purposes. The differences between the best-fitting solutions for the three light
curves are clear through the partial eclipse phases, and show that better data
are needed.} \end{figure}

In this work I have solved the ($b$$+$$y$)/2 light curve from \citet{Sato+05apj} and the $g$- and $r$- band data from \citet{Charbonneau+06apj}. \reff{My initial solutions converged to values for the orbital inclination of either $i \approx 81^\circ$ or $i \approx 90^\circ$. To investigate this, additional solutions were calculated with $i$ fixed to values between 90$^\circ$ and 70$^\circ$ in steps of 0.2$^\circ$. The quality of the fit for these additional solutions is shown in Fig.\ref{fig:rms:149026}, and the results for the three light curves are not in good agreement with each other or with the analogous diagram presented by \citet[][their Fig.\,5]{Sato+05apj}.}

\reff{The solutions for the $g$ light curve were very poorly-behaved: standard approaches tended to go to an ill-defined solution with $i \approx 81^\circ$, but the additional solutions did not reproduce the corresponding fit parameters even when $i$ was fixed at values near $81^\circ$. This indicates that the $\chi^2$ surface in parameter space has substantial structure, and these data have not been considered further.}

\reff{The solutions for the $r$ and ($b$$+$$y$)/2 light curves are better-behaved and the resulting parameters are in reasonable agreement (even though the best fits are not; see Fig.\,\ref{fig:lc:149026}). The ($b$$+$$y$)/2 light curve is the best dataset, and it was possible to obtain meaningful values for the linear LDCs when the nonlinear LDCs were fixed. These solutions were therefore adopted as the final results for the ($b$$+$$y$)/2 data. Analyses of the $r$ light curve tended to give unphysical LDCs, so these had to be fixed at theoretically-predicted values for the final solutions of these data. Attempts to fit for both LDCs for either light curve resulted in an impressive variety of physically impossible solutions.}

\reff{The nature of the transit of HD\,149026 means that the inclination remains very poorly constrained. This uncertainty over which part of the stellar disc the planet eclipses causes the fractional radii of both components to be ill-defined, and the upper and lower confidence intervals for individual measurements are quite unbalanced. I have adopted the Charbonneau $r$ outcome as the final results for HD\,149026 (Table\,\ref{tab:lcfit:149026}), but have increased the error estimates to reflect the difference between the $r$ and ($b$$+$$y$)/2 solutions. This choice results in more precise parameter values, compared to the ($b$$+$$y$)/2 solution, but is dependent on theoretically-derived LDCs. However, any systematic error arising from the use of theoretical LDCs should be much smaller than the random errors of the current solution.}

\reff{The final solution presented here is not in good agreement with the $i$, $r_{\rm A}$ and $r_{\rm b}$ values found in previous studies \citep{Sato+05apj,Charbonneau+06apj,Wolf+07apj}. It corresponds to a higher (but still poorly-defined) $i$, which means smaller $r_{\rm A}$ and $r_{\rm b}$ values are needed to give a transit of the observed duration. The substantially smaller $r_{\rm b}$ found here indicates that the planet in the HD\,149026 system has an even {\em larger} density than previously thought.} The only reasonable conclusion that can be drawn is that better light curves are needed before we can attain a precise understanding of the HD\,149026 system.

After the above analysis was performed, another study of HD\,149026 became available. \citet{Winn+07xxx} have obtained new photometry and arrive at a high-$i$ solution which is in good agreement with my own results. The fractional radii of the star and planet are still not well defined, so additional high-precision photometry remains a priority for observers.


\subsection{HD 189733}

HD\,189733 was found to be a TEP by \citet{Bouchy+05aa} and is now one of the best-studied systems due to its brightness and relatively deep transit. It has a faint stellar companion separated by 11.2$^{\prime\prime}$ on the sky \citep{Bakos+06apj}. HD\,189733\,A has a detectable magnetic field \citep{Moutou+07aa} so is an interesting target for studying the activity of planet host stars. A rotational period of $11.953 \pm 0.009$\,d was determined by \citet{HenryWinn08aj} from photometry of its starspot-induced brightness modulation. \citet{HebrardLDE06aa} found eclipses in its {\it Hipparcos} light curve, which has helped to refine \Porb.

Ground-based photometry of HD\,189733 has been presented and studied by \citet{Bakos+06apj2}, \citet{Winn+06apj} and \citet{Winn+07aj}. The primary and secondary minima have also been studied from space using {\it Spitzer} \citep{Deming+06apj, Knutson+07nature, Tinetti+07nature}. In the near future new light curves from HST \citep{Pont+07aa2} and MOST \citep{Croll+07apj2} will become available. Finally, \citet{Baines+07apj} measured the angular diameter and thus linear radius of HD\,189733\,A, which means that it is possible to determine the physical properties of both star and planet using purely empirical methods (Paper\,II).

I have solved the light curves presented by \citet{Bakos+06apj2} (620 datapoints), and the FLWO 1.2\,m $z$-band data (1662 observations) and Wise 1.0\,m $I$-band light curve (344 datapoints) from \citet{Winn+06apj}. A slight eccentricity has not been ruled out \citep{Deming+06apj,Knutson+07nature} but such a small value has a negligible effect on the parameters derived in this work. I have taken $\Porb = 2.2185733$\,d from \citet{Winn+06apj} and a mass ratio of 0.0014 from \citet{Winn+06apj}.

\begin{table*} \caption{\label{tab:lcfit:189733} Final parameters of
the fits to the \citet{Bakos+06apj2} (FLWO $r$) and \citet{Winn+07aj}
(FLWO $z$ and WISE $I$) light curves of HD\,189733 from the {\sc jktebop}
analysis. These have been compared to the properties of HD\,189733 found
by other workers. Quantities without quoted uncertainties were not given
by those authors but have been calculated from other parameters which were.}
\begin{tabular}{l r@{\,$\pm$\,}l r@{\,$\pm$\,}l r@{\,$\pm$\,}l r@{\,$\pm$\,}l@{\hspace{5pt}}r@{\,$\pm$\,}l r@{\,$\pm$\,}l r@{\,$\pm$\,}l}
\hline \hline
\ &  \mc{This work}    &  \mc{This work}    &  \mc{This work}    &   \mc{This work}  & \mc{Bouchy et} & \mc{Bakos et} & \mc{Winn et} \\
\ &\mc{(FLWO $r$ data)}&\mc{(FLWO $z$ data)}&\mc{(WISE $I$ data)}&\mc{(final result)}& \mc{ al.\ \citeyearpar{Bouchy+05aa}} & \mc{al.\ \citeyearpar{Bakos+06apj2}} & \mc{al.\ \citeyearpar{Winn+07aj}} \\
\hline
$r_{\rm A}+r_{\rm b}$ & 0.1184 & 0.0022 & 0.1297 & 0.0023 & 0.1326 & 0.0083 & 0.1287 & 0.0036 & \mc{0.1311}   &    \mc{ }    & \mc{0.1301}     \\
$k$                   & 0.1542 & 0.0019 & 0.1577 & 0.0010 & 0.1499 & 0.0029 & 0.1568 & 0.0024 & 0.172 & 0.003 &    \mc{ }    & \mc{0.1574}     \\
$i$ ($^\circ$)        &  86.50 & 0.59   &  85.73 & 0.20   &  85.38 & 0.68   &  85.78 & 0.25   &  85.3 & 0.1   & 85.79 & 0.24 &  85.76 & 0.29   \\
$r_{\rm A}$           & 0.1026 & 0.0054 & 0.1121 & 0.0019 & 0.1153 & 0.0071 & 0.1113 & 0.0031 & \mc{0.1118}   &    \mc{ }    & 0.1124 & 0.0034 \\
$r_{\rm b}$           & 0.0158 & 0.0012 & 0.0177 & 0.0004 & 0.0173 & 0.0012 & 0.0175 & 0.0005 & \mc{0.0192}   &    \mc{ }    & 0.0177 & 0.0007 \\
\hline \hline \end{tabular} \end{table*}

\begin{figure} \includegraphics[width=0.48\textwidth,angle=0]{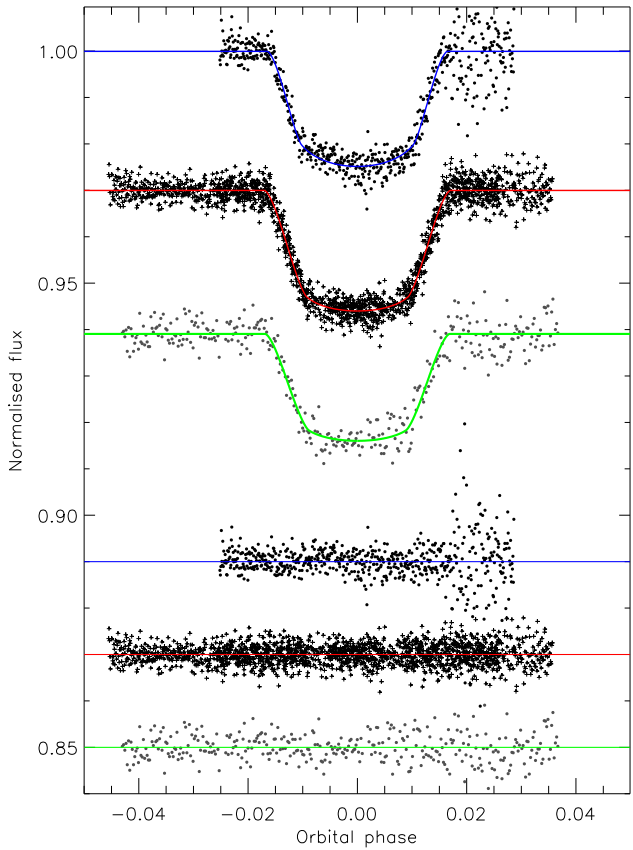}
\caption{\label{fig:lc:189733} Phased light curves of the transits of
HD\,189733 from \citet{Bakos+06apj2} (top) and \citet{Winn+07aj} (middle:
FLWO $z$; bottom: WISE $I$) compared to the best fit found using {\sc jktebop}
and the quadratic LD law with the linear LDCs included as fitted parameters.
The residuals of the fit are plotted at the bottom of the figure, offset from
zero.} \end{figure}

For each of the light curves it was difficult to extract reliable values for both LDCs so I have adopted the solutions with one LDC fixed (Tables \ref{tab:lc:189733:flwor}, \ref{tab:lc:189733:flwoz} and \ref{tab:lc:189733:wiseI}). The Monte Carlo and residual-permutation uncertainties were similar so correlated noise is not important for these datasets. The best-fitting light curves are compared with the data in Fig.\,\ref{fig:lc:189733}.

The solutions of the three light curves are inconsistent (Table\,\ref{tab:lcfit:189733}), due mainly to a mismatch of 6.7\hspace{1pt}$\sigma$ in the ratio of the radii, $k$. This results in discrepancies of 3.1\hspace{1pt}$\sigma$ in $r_{\rm A}$ and 2.1\hspace{1pt}$\sigma$ in $r_{\rm b}$. As $k$ depends essentially on the eclipse depth, this problem is related to the data rather than to the analysis. It is not very surprising that this discrepancy exists, because obtaining light curves of TEPs is an observationally demanding activity. An example of systematic problems with the eclipse depth in OGLE-TR-132 was presented and discussed by \citet{Gillon+07aa3}. For HD\,189733 there are three high-precision light curves available, so the effects of systematic errors can be estimated in detail. HD\,189733 is therefore one of the best-understood of the TEPs. A rather unhappy implication of this discrepancy is that we can only fully understand the intrinsic properties of a TEP if three independent light curves are available. \reff{It is possible that the presence of starspots contributes to the different $k$ values found for the light curves; this could be investigated using the excellent HST light curve which has since been presented by \citet{Pont+07aa2}.}

For the final solution I have calculated the weighted mean for each parameter from the three light curves studied here, and have increased the uncertainties to account for the discrepancies discussed above. Because of this, the resulting solution is very robust. More pieces of the jigsaw can be added to this study in the near future by including the forthcoming HST and MOST datasets. The study of systematic biases in light curve parameters is a vital part of understanding the extent of our knowledge of the properties of TEPs.


\subsection{HD 209458}                                                               \label{sec:tep:209458}

In clear contrast to its position as the final object considered in this work, HD\,209458 was the first known TEP, with discovery light curves published by \citet{Charbonneau+00apj} and \citet{Henry+00apj}. It has since become the most heavily studied, and the observational data available for this system constitutes an embarrassment of riches. Transit light curves have been obtained by HST (twice: \citealt{Brown+01apj} and \citealt{Knutson+07apj}), MOST \citep{Rowe+06apj,Croll+07apj}, {\it Spitzer} \citep{Richardson+06apj} and from terrestrial facilities \citep{Wittenmyer+05apj}. Secondary eclipses have been obtained by {\it Spitzer} \citep{Deming+05natur,Knutson+07xxx} but ground-based attempts have been only partially successful \citep{Snellen05mn,Deming++07mn}. Spectroscopic observations of the Rossiter-McLaughlin effect through transit have added to the understanding of HD\,209458 \citep{Queloz+00aa,Winn+05apj,Wittenmyer+05apj} and no less than three works have obtained precise $\Porb$ measurements using {\it Hipparcos} data \citep{Soderhjelm99ibvs, RobichonArenou00aa, Castellano+00apj}

Given such a large amount of data, I have decided to analyse both sets of HST light curves \citep{Brown+01apj,Knutson+07apj} and the MOST light curve \citep{Rowe+06apj}%
. Studying these datasets allows careful cross-checks of the success of the different solutions and the interagreement of the results, and so indicate whether or not systematic errors (arising particularly from red noise) are important. This is particularly relevant in light of the results for HD\,189733 presented above. The HST data of \citet{Brown+01apj} have been previously solved using {\sc jktebop} \citep{Me++07mn} but are revisited here in order to ensure homogeneity of the results. I have adopted $\Porb = 3.52474859$\,d and a mass ratio of 0.00056 from \citet{Knutson+07apj}.

\begin{figure} \includegraphics[width=0.48\textwidth,angle=0]{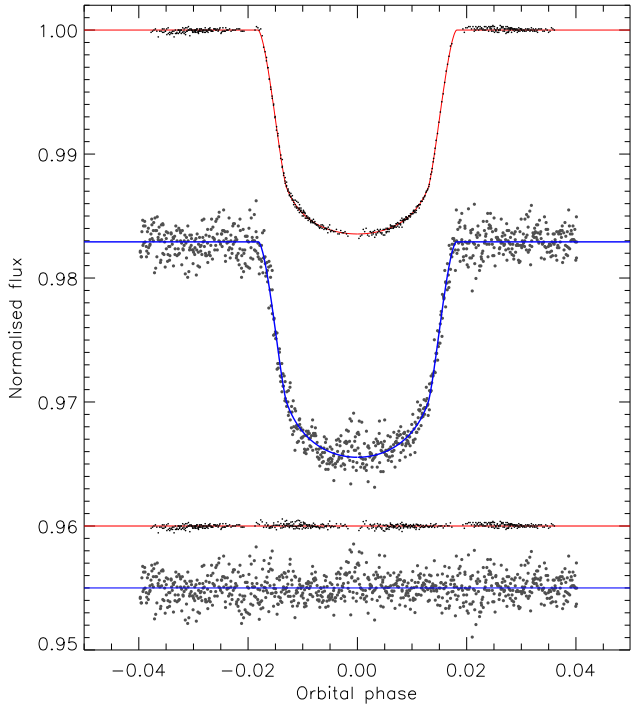}
\caption{\label{fig:lc:209458:hstmost} Phased light curves of the transit
of HD\,209458 from \citet{Brown+01apj} (upper) and \citet{Rowe+06apj} (lower)
compared to the best fits found using {\sc jktebop} and the quadratic LD law.
Each consecutive set of 10 points of the MOST light curve have been combined
to make the plot clearer. The HST light curve is unbinned. The residuals of
the fit are plotted at the bottom of the figure, offset from zero.} \end{figure}

The HST data obtained by \citet{Brown+01apj} (556 datapoints) give good values for both LDCs when the nonlinear LD laws are used (Table\,\ref{tab:lc:209458:brown}), so these solutions have been adopted. The high quality and relatively blue passband of these data allow the linear LD law to be ruled out at a high significance level (\reff{greater than 99.99\%}). There is less to choose between the nonlinear laws, except that the quadratic law gives marginally the \reff{best} fit. The low Poisson noise means that correlated errors dominate here, so I have adopted the uncertainties from the residual-shift algorithm.

When analysing the MOST data \citep{Rowe+06apj} I rejected the datapoints which were distant from transit (Fig.\,\ref{fig:lc:209458:hstmost}) as these add nothing to the solution except increased computational requirements. I also did not consider data from the second of the four transits as only part of this eclipse was observed. This left a total of 8772 datapoints. Solutions with two fitted LDCs were not great so I have adopted those with one fitted LDC. As with the HST data, correlated noise is important so I have adopted the uncertainties from the residual-shift algorithm.

\subsubsection{HST observations of \citet{Knutson+07apj}}

\begin{figure} \includegraphics[width=0.48\textwidth,angle=0]{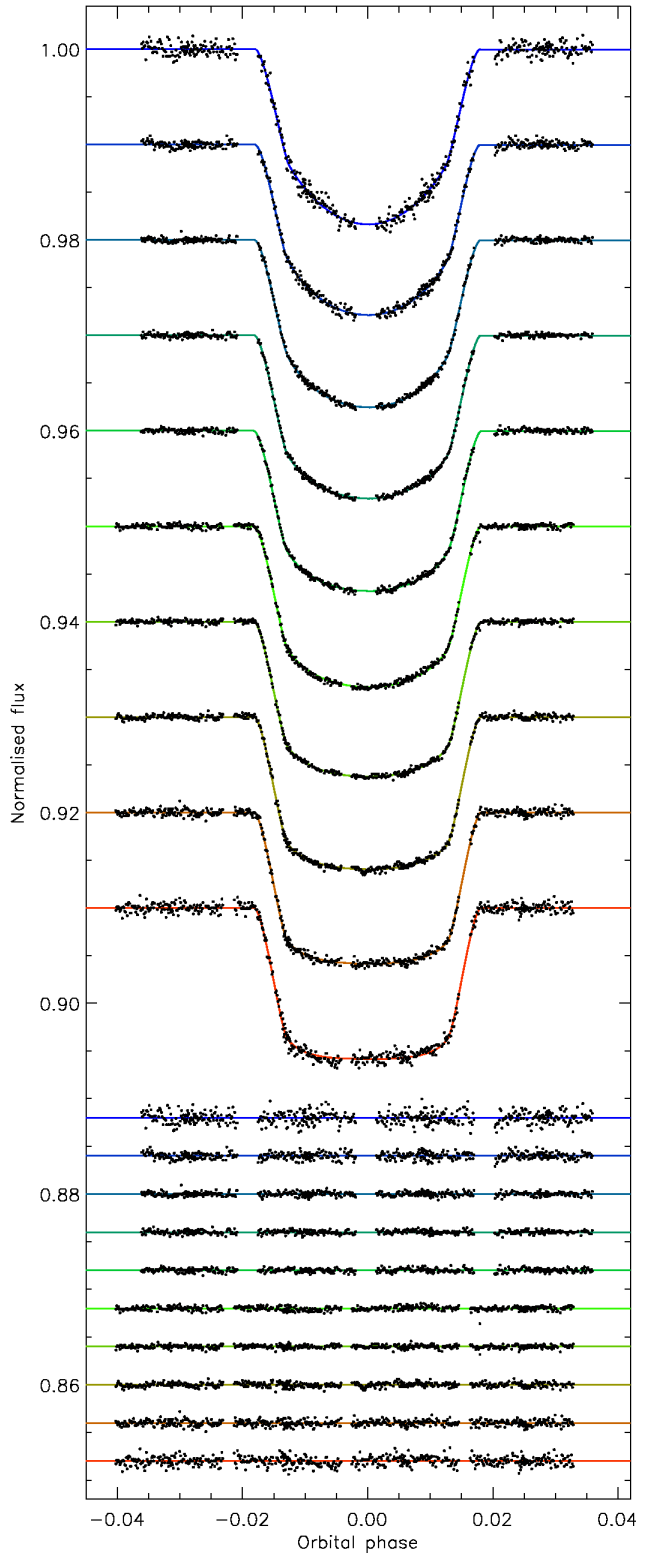}
\caption{\label{fig:lc:209458:knutson} Phased light curves of the transit
of HD\,209458 from \citet{Knutson+07apj} compared to the best fits found
using {\sc jktebop} and the quadratic LD law (both LDCs fitted). The data
are plotted with the bluest passband at the top and the reddest at the
bottom, with offsets to make the plot clear. The residuals of the fits
follow the same sequence and are plotted at the bottom of the figure with
offsets from zero.} \end{figure}

The beautiful HST observations obtained by \citet{Knutson+07apj} allow the properties of the HD\,209458 system to be determined to a very high precision. These data comprise two separate sets of observations obtained with a blue (504 epochs) and a red grating (548 epochs) using HST/STIS. Each set has been split into five different passbands, resulting in ten separate light curves evenly distributed between 320\,nm and 980\,nm. The variation of LD strength with wavelength is clear and can be a strong test of the predictions of stellar model atmospheres.

The ten light curves have been solved separated using linear and quadratic LD (Tables \ref{tab:lc:209458:lin} and \ref{tab:lc:209458:quad}), and the best fits and residuals are plotted in Fig.\,\ref{fig:lc:209458:knutson}. The light curves have a low photon noise so the residual-shift algorithm gives the largest uncertainties. As with the HST observations of \citet{Brown+01apj}, the linear LD law gives a significantly worse fit than the quadratic LD law for the bluer wavelength bands. \reff{In particular, the referee has correctly pointed out that, for the linear-law case, the orbital inclination values are correlated with the wavelength of observation. This is clearly unphysical and arises from the inadequacy of the linear law being more important at shorter wavelengths where LD is stronger. No correlation is seen for the quadratic-law case.}

\begin{figure} \includegraphics[width=0.48\textwidth,angle=0]{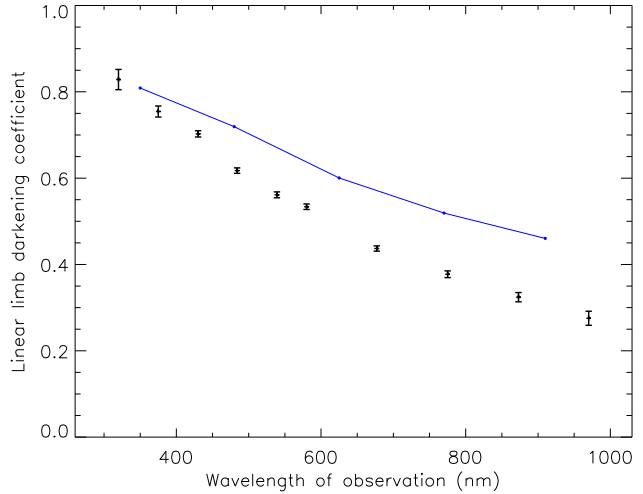}
\caption{\label{fig:lc:209458:lin} Variation of the linear LD law coefficient
with wavelength of observation for the {\sc jktebop} solutions of the ten HST
LCs of HD\,209458 presented by \citet{Knutson+07apj}. The uncertainty on each
point comes from Monte Carlo simulations. The filled circles connected by a
solid line show the predicted LDCs for the SDSS $ugriz$ passbands
from \citet{Claret04aa2}. In each case the central wavelength of the passband
is plotted; detailed conclusions from these figures must take into account
the passband response function as well as central wavelength.} \end{figure}

\reff{The variation with wavelength of the linear LDC is compared in Fig.\,\ref{fig:lc:209458:lin} to predictions for the SDSS $ugriz$ passbands from \citet{Claret04aa2}. The disagreement between observation and theory is clear. However, detailed conclusions should not be drawn from Fig.\,\ref{fig:lc:209458:lin} for two reasons: the linear law is not a good representation of the LD of stellar discs, and this is more important at bluer wavelengths; and the plotted LDCs are not monochromatic but refer to different passbands. I find that the dependence of the light curve solutions on wavelength, noted in the previous paragraph, does not have a significant effect on the conclusions drawn from Fig.\,\ref{fig:lc:209458:lin}.}

\begin{figure} \includegraphics[width=0.48\textwidth,angle=0]{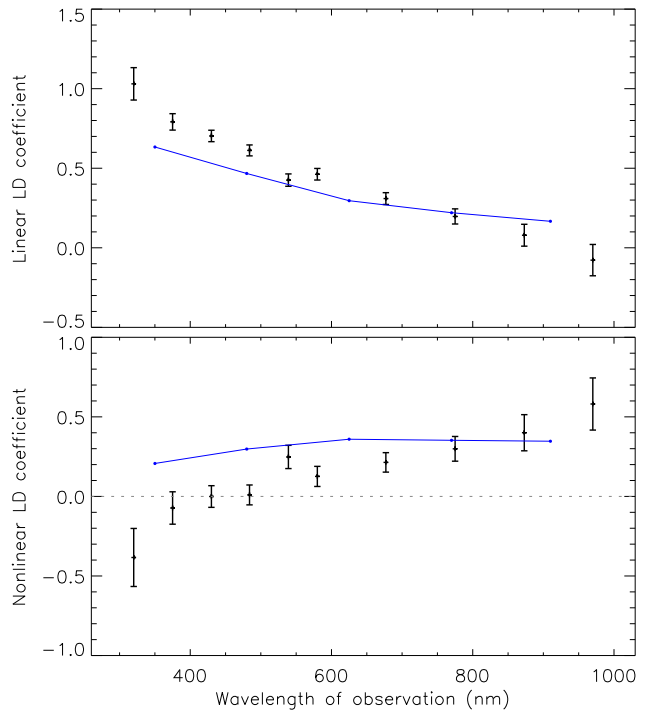}
\caption{\label{fig:lc:209458:quad} Variation with wavelength of the linear
and nonlinear coefficients of the quadratic LD law for the ten HST light
curves of HD\,209458 \citep{Knutson+07apj}. The uncertainty on each point
comes from Monte Carlo simulations. The filled circles connected by a solid
line show the predicted LDCs for the SDSS $ugriz$ passbands from
\citet{Claret04aa2}. The central wavelengths of the passbands are plotted;
detailed conclusions from these figures must take into account the passband
response function as well as central wavelength.} \end{figure}

\reff{In Fig.\,\ref{fig:lc:209458:quad} the observed and predicted LDCs are compared for the quadratic law. The observed LDCs again do not agree with the predicted ones, but investigation of this effect is beyond the scope of the current paper. One interesting point is that the nonlinear LDC for the quadratic law dips below zero at the shortest wavelengths, an effect which is not present in any of the LDC predictions used in this work.}

%

When combining the parameter values obtained for each of the ten Knutson light curves, their scatter was much larger than the derived uncertainties. The scatter is again relatively the largest for $k$ (11.3\hspace{1pt}$\sigma$ for the solutions with linear LD and 5.6\hspace{1pt}$\sigma$ for those with quadratic LD) so for the final result I adopt a weighted mean with uncertainties increased to reflect this greater scatter. Note that this effect is expected because the individual light curves are not independent (they come from the same set of observations). \reff{This does not necessarily imply that there are any problems with the observations or the analysis for the quadratic-law case, but the much worse agreement for linear LD does indicate that this function is a poor representation of reality.}

I have investigated how the accuracy of the radius measurements varies with wavelength, and find (after compensating for the variation in the precision of the data) that the precision at the reddest wavelengths is roughly twice as good as at the bluest. The quality of the LDC determinations does not change much, but their importance to $r_{\rm A}$ and $r_{\rm b}$ decreases with increasing wavelength. This is in agreement with the general belief that redder passbands give more precise results because the effect of LD is less important.

\subsubsection{Final results for HD\,209458}

\begin{table*} \caption{\label{tab:lcfit:209458} Final parameters of the
fit to the \citet{Brown+01apj}, \citet{Knutson+07apj} and \citet{Rowe+06apj}
light curves of HD\,209458 from the {\sc jktebop} analysis presented here.
These have been compared to the properties of HD\,209458 found in the
literature. The uncertainties quoted for the final results from this
work have been increased to yield $\chi_{\rm red}^{\ 2} = 1.0$.
Quantities without quoted uncertainties were not given directly
but have been calculated from other parameters which were.}
\begin{tabular}{l r@{\,$\pm$\,}l r@{\,$\pm$\,}l r@{\,$\pm$\,}l r@{\,$\pm$\,}l r@{\,$\pm$\,}l r@{\,$\pm$\,}l}
\hline \hline
\                     & \mc{This work} & \mc{This work} & \mc{This work} & \mc{This work}
                        & \mc{ } & \mc{Southworth et} \\
\                     & \mc{HST data (Brown)} & \mc{MOST data} & \mc{HST data (Knutson)} & \mc{(final)}
                        & \mc{ } & \mc{al.\ \citeyearpar{Me++07mn}} \\
\hline
$r_{\rm A}+r_{\rm b}$ & 0.12779 & 0.00078 & 0.12637 & 0.00367 & 0.12775 & 0.00046 & 0.12805 & 0.00031 & \mc{ } & 0.12785 & 0.00050 \\
$k$                   & 0.12077 & 0.00045 & 0.12352 & 0.00132 & 0.12175 & 0.00030 & 0.12225 & 0.00065 & \mc{ } & 0.12074 & 0.00037 \\
$i$ ($^\circ$)        & 86.681  & 0.082   & 86.790  & 0.407   & 86.590  & 0.046   & 86.589  & 0.076   & \mc{ } & 86.677  & 0.060   \\
$r_{\rm A}$           & 0.11402 & 0.00069 & 0.11248 & 0.00314 & 0.11384 & 0.00041 & 0.11414 & 0.00024 & \mc{ } & 0.11405 & 0.00042 \\
$r_{\rm b}$           & 0.01377 & 0.00011 & 0.01389 & 0.00052 & 0.01389 & 0.00006 & 0.01392 & 0.00010 & \mc{ } & 0.01377 & 0.00008 \\
\hline
\                     & \mc{Brown et al.} & \mc{Mandel \& Agol} & \mc{Richardson et al.}
                        & \mc{Gim\'enez et al.} & \mc{Knutson et al.} & \mc{Rowe et al.} \\
\                     & \mc{\citeyearpar{Brown+01apj}} & \mc{\citeyearpar{MandelAgol02apj}}
                        & \mc{\citeyearpar{Richardson+06apj}} & \mc{\citeyearpar{Gimenez06aa}}
                          & \mc{\citeyearpar{Knutson+07apj}} & \mc{\citeyearpar{Rowe+07xxx}} \\
\hline
$r_{\rm A}+r_{\rm b}$ &    \mc{ }    &      \mc{ }       &    \mc{ }    &    \mc{ }       &    \mc{ }      &  \mc{ }    \\
$k$                   &    \mc{ }    & 0.12070 & 0.00027 &    \mc{ }    & 0.1208 & 0.0002 &    \mc{ }      &  \mc{ }    \\
$i$ ($^\circ$)        & 86.68 & 0.14 &      \mc{ }       & 87.97 & 0.85 & 86.71  & 0.04   & 86.929 & 0.010 & 86.9 & 0.2 \\
$r_{\rm A}$           &    \mc{ }    & 0.11391 & 0.00042 &    \mc{ }    & 0.1136 & 0.0003 &    \mc{ }      &  \mc{ }    \\
$r_{\rm b}$           &    \mc{ }    &      \mc{ }       &    \mc{ }    & 0.0137 & 0.0001 &    \mc{ }      &  \mc{ }    \\
\hline \hline \end{tabular} \end{table*}


The results of the three light curve analyses for HD\,209458 are given in Table\,\ref{tab:lcfit:209458} and are not in good agreement. There is again a difference between the results which is larger than expected given the derived errors. This is worst for the ratio of the radii (3.3\hspace{1pt}$\sigma$) but surprisingly does not affect $r_{\rm A}$ or $r_{\rm b}$ (0.5\hspace{1pt}$\sigma$). The fact that the discrepancy remains the worst for $k$ is disquieting because this should be very well determined from the observed eclipse depths. Our understanding of HD\,209458 is good, like HD\,189733, because sufficient independent data exist for us to make a useful estimate of the importance of systematic errors.

For the final result I have again adopted a weighted mean of the parameters from the three separate light curves but increased the uncertainties to account for the discrepancy between the measurements. The final parameters found here are in reasonable agreement with most literature values. The orbital inclinations found in some published studies are not in good agreement within the errors; this can be attributed to those studies fixing LDCs at theoretical values and so ending up with biased best-fit parameters and underestimated errorbars.


\section{Surface gravities of each system}                                            \label{sec:logg}

\citet{Me+04mn3} showed that it is possible to measure the surface gravity of the secondary component of an eclipsing system from only the results of a light curve analysis and the measured velocity amplitude of the primary component%
\footnote{This concept was also independently discovered by \citet{Winn+07aj} and \citet{Beatty+07apj} -- note that this does not necessarily require the eclipses to be total as claimed by \citet{Beatty+07apj}.}%
. A simplified formula was given by \citet{Me++07mn}:
\begin{equation} \label{eq:g}
g_{\rm b} = \frac{ 2 \pi }{ \Porb } \,\frac{ (1-e^2)^\frac{1}{2} K_{\rm A} }{ r_{\rm b}^{\ 2} \sin i }
\end{equation}
where $K_{\rm A}$ is the stellar velocity amplitude, and applied to the fourteen TEPs then known. The importance of this is that whilst $g_{\rm b}$ can be found directly for a transiting planet, measuring its other physical properties (mass, radius, density) requires additional constraints such as are usually obtained from stellar theory, so are not empirical. \citet{Me++07mn} found a significant correlation between $g_{\rm b}$ and \Porb, although it is not yet clear whether this is a real effect or arises from selection biases. A plausible explanation is that short-period planets are close to their parent star and much of their atmosphere has been evaporated off by UV and X-ray irradiation. High-energy irradiation is believed to cause close-in extrasolar planets to lose their atmospheres \citep{Lammer+03apj}.

\begin{figure*} \includegraphics[width=\textwidth,angle=0]{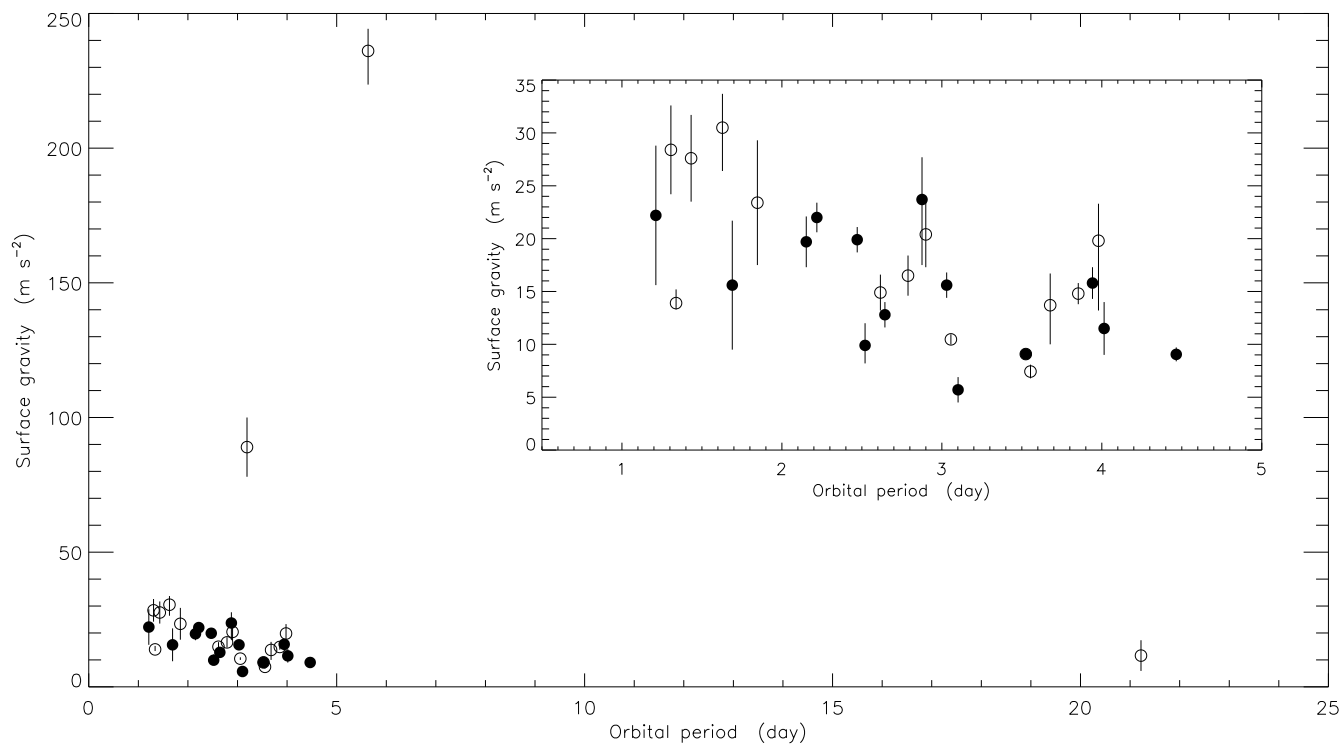} \\
\caption{\label{fig:abspar:g2} Comparison between the surface gravities and
orbital periods of the known transiting exoplanets. Filled circles denote
the values obtained in this work and and open circles the systems for which
values have been taken from the literature. The error bars for HD\,209458
($\Porb = 3.52$\,d) are smaller than the plotted symbol. \reff{A close-up
of the main population of known TEPs is shown in the inset panel.} }
\end{figure*}

The $g_{\rm b}$ values for the TEPs studied in this work have been determined from the light curve analyses and published measurements of $K_{\rm A}$ (Table\,\ref{tab:abspar:k1}). These have been augmented by literature values for the other known TEPs, and are given in Table\,\ref{tab:abspar:g2} and plotted in Fig.\,\ref{fig:abspar:g2}. The correlation is still clearly visible but is quite weak. The error bars for some of the planets could be decreased by performing light curve analyses similar to those presented in this work. The fractional radii from the light curve analyses are sensitive to the amount of stellar irradiation intercepted by the transiting planets, but plots of $r_{\rm A}$ and $r_{\rm b}$ show no correlation with \Porb.

\begin{table} \caption{\label{tab:abspar:k1} Stellar
velocity amplitudes for the systems studied here.}
\begin{tabular}{l r@{\,$\pm$\,}l l}
\hline \hline
System & \mc{$K_{\rm A}$ (\ms)} & Reference \\
\hline
TrES-1      & 115.2  &  6.2  & \citet{Alonso+04apj}     \\
TrES-2      & 181.3  &  2.6  & \citet{Odonovan+06apj}   \\
XO-1        & 116.0  &  9.0  & \citet{Mccullough+06apj} \\
WASP-1      & 114    & 13    & \citet{Cameron+07mn}     \\
WASP-2      & 155    &  7    & \citet{Cameron+07mn}     \\
HAT-P-1     &  60.3  &  2.1  & \citet{Bakos+07apj}      \\
OGLE-TR-10  &  81    & 17    & \citet{Bouchy+05aa2}     \\
OGLE-TR-10  &  80    & 25    & \citet{Konacki+05apj}    \\[-1pt]
OGLE-TR-10  &  80.3  & 14.1  & (adopted in this work)   \\[-1pt]
OGLE-TR-56  & 212    & 22    & \citet{Bouchy+05aa2}     \\
OGLE-TR-111 &  78    & 14    & \citet{Pont+04aa}        \\
OGLE-TR-132 & 141    & 42    & \citet{Bouchy+04aa}      \\
GJ 436      &  18.34 &  0.52 & \citet{Maness+07pasp}    \\
HD 149026   &  43.3  &  1.2  & \citet{Sato+05apj}       \\
HD 189733   & 205    &  6    & \citet{Bouchy+05aa}      \\
HD 209458   &  85.1  &  1.0  & \citet{Naef+04aa}        \\
\hline \hline \end{tabular} \end{table}

\begin{table} \caption{\label{tab:abspar:g2} Measured planetary surface
gravities for the systems studied here (upper part of table) and for
the other known TEPs (lower part of table). References are entirely
parenthesised if the surface gravity has been calculated from the
planetary properties ($M_{\rm b}$ and $R_{\rm b}$) given in that work.}
\begin{tabular}{l r@{\,$\pm$\,}l l}
\hline \hline
System & \mc{$g_{\rm b}$ (\ms)} & Reference \\
\hline
TrES-1      &    15.6  & 1.2        & This work \\
TrES-2      &    19.9  & 1.2        & This work \\
XO-1        &    15.8  & 1.5        & This work \\
WASP-1      & \erm{9.9}{2.1}{1.7}   & This work \\
WASP-2      &    19.7  & 2.4        & This work \\
HAT-P-1     &     9.05 & 0.66       & This work \\
OGLE-TR-10  &     5.7  & 1.2        & This work \\
OGLE-TR-56  &    22.2  & 6.6        & This work \\
OGLE-TR-111 &    11.5  & 2.5        & This work \\
OGLE-TR-132 &    15.6  & 6.1        & This work \\
GJ 436      &    12.8  & 1.2        & This work \\
HD 149026   & \erm{23.7}{4.0}{6.2}  & This work \\
HD 189733   &    22.0  & 1.4        & This work \\
HD 209458   &    9.08  & 0.17       & This work \\
\hline
TrES-3      &   28.4   & 4.2        & \citep{Odonovan+06apj}      \\
TrES-4      &    7.43  & 0.65       & \citet{Odonovan+07apj}      \\
XO-2        &   14.9   & 1.7        & \citep{Burke+07apj}         \\
XO-3        &   89     & 11         & \citep{Johnskrull+07xxx}    \\
WASP-3      &   23.4   & 5.9        & \citet{Pollacco+07xxx}      \\
WASP-4      & \erm{13.9}{1.3}{0.6}  & \citet{Wilson+07xxx}        \\
WASP-5      & \erm{30.5}{3.2}{4.1}  & \citet{Anderson+07xxx}      \\
OGLE-TR-113 &   27.6   & 4.1        & \citep{Gillon+06aa}         \\
OGLE-TR-182 & \erm{19.8}{3.5}{6.6}  & \citep{Pont+07xxx2}         \\
OGLE-TR-211 & \erm{13.7}{3.0}{3.7}  & \citep{Udalski+07xxx}       \\
HAT-P-2     & \erm{236.1}{8.2}{12.5}& \citet{Loeillet+07xxx}      \\
HAT-P-3     &   20.4   & 3.1        & \citet{Torres+07apj}        \\
HAT-P-4     &   10.47  & 0.48       & \citet{Kovacs+07apj}        \\
HAT-P-5     &   16.5   & 1.9        & \citet{Bakos+07apj2}        \\
HAT-P-6     &   14.8   & 1.0        & \citet{Noyes+08apj}         \\
HD 17156    &   11.6   & 5.7        & \citep{Barbieri+07aa}       \\
\hline \hline \end{tabular} \end{table}

                                                                                   \label{sec:conclusion}
\section{Summary and conclusions}

At this time over thirty transiting extrasolar planetary systems are known, but their properties have been obtained using a variety of different methods. I have presented detailed light curve analyses of fourteen TEPs for which good light curves are available, paying particular attention to the importance of limb darkening (LD) and the assessment of random and systematic errors. In a second paper the results of these analyses will be used to determine the physical properties of each system, in an homogeneous way, using the predictions of theoretical stellar models and concentrating on obtaining the most reliable values and uncertainties.

The parameters found in my light curve analyses are generally in good agreement with published studies. However, the uncertainties found in this study are often significantly larger (by factors of up to 3). This stems from including several sources of error (Monte Carlo simulations for random errors, a residual permutation algorithm to assess systematic errors, and the effect of using different LD laws), some of which are not incorporated into previous works.

Although my uncertainty estimates are generally larger than those from other studies, solutions of different light curves for individual systems still do {\em not} agree to within these errors. This problem is clear for the two well-studied systems HD\,189733 and HD\,209458, both of which have three different high-quality light curves available. The worst discrepancy occurs for the ratio of the radii (6.7\hspace{1pt}$\sigma$ for HD\,189733 and 3.7\hspace{1pt}$\sigma$ for HD\,209458), which is primarily dependent on the observed transit depth. This indicates that systematic errors remain in the light curves which cannot be detected through analysis of only one dataset; definitive results for an individual TEP can only be obtained when at least {\em three} independent light curves are obtained and analysed.

I was not able to reproduce published solutions of the light curves for HD\,149026, which contains a high-density planet. The three light curves of this system give quite different results and it is clear that further observations are needed. My solution has a higher orbital inclination (corresponding to a central transit) and leads to smaller radii for both the star and planet, which suggests that HD\,149026\,b is even more dense than currently thought.

An improved light curve is also needed for OGLE-TR-132 because the current best dataset does not cover the full transit. Further observations would also be useful for TrES-2, the only system where light curve solutions with one or two fitted LD coefficients (LDCs) do not agree well. Additional investigations of OGLE-TR-10 would also be useful, because published studies of this object give widely varying results.

The correlation between orbital period and planetary surface gravity (first noticed by \citealt{Me++07mn}) remains visible when the new TEPs are included, although there are a few outliers with much higher gravities. Several of the surface gravity measurements are quite imprecise, with the main contribution to their uncertainty arising from the stellar velocity amplitude rather than the light curve analyses. New spectroscopic radial velocity studies for XO-1, WASP-1, WASP-2 and the OGLE transiting planets are needed to improve our understanding of the physical properties of these systems.

\subsection{Treatment of limb darkening}

One potentially thorny problem in the study of the light curves of TEPs is how much effect the use of different LD laws has on the light curve solutions. The linear LD law is clearly inadequate when studying the high-quality HST light curves of HD\,209458 and can be ruled out at the 99.99\% level. This is in agreement with results from the study of eclipsing binary stars \citep{Me++07aa} and with the predictions of stellar model atmospheres \citep{Vanhamme93aj,Claret00aa}. However, the linear law is adequate for most of the datasets studied in this work (particularly those from longer wavelengths).

Solutions were also done using the two-coefficient quadratic, square-root, logarithmic and cubic LD laws. It is usually not possible to measure both LDCs for one star because they are strongly correlated, but this means that one LDC can be fixed at a reasonable value without making a significant difference to the light curve solution. When both LDCs are included as fitted parameters, the cubic law gives the best-determined LDCs and the square-root law suffers from the biggest correlations. The quadratic and cubic laws generally perform well, but the usefulness of the cubic law is restricted as there are no theoretical predictions available. In general, though, there is little to choose between solutions using the four different LD laws: the solutions of \reff{all but the highest-quality} transit light curves are {\em not} adversely affected by the choice of LD law. This is good news for studies of transiting planets (but bad news for studies of stellar limb darkening).

\reff{Additional solutions with all LDCs fixed were calculated for TrES-1, TrES-2 and XO-1. In general these give results in (mostly) acceptable agreement with other solutions, but with substantially smaller errorbars. It is therefore clear that fixing coefficients at theoretical values not only allows biases to creep in to the results, but also makes the uncertainties too optimistic.}

Despite the previous point, a comparison between measured LDCs and those predicted using stellar model atmospheres is a promising way of improving our knowledge of stellar atmospheres. It is known that current stellar model atmospheres do not reproduce the LD of stars very well (A.\ Claret, private communication). Whilst a detailed study is beyond the scope of the present work, a limited comparison was carried out between the LDCs measured from light curves of HD\,209458 in ten passbands spread through the optical wavelength range. Predicted LDCs were unable to match the measured coefficients for either the linear or quadratic laws. A similar study by \citet{Tingley++06aa} also found a disagreement between the observed and predicted colour dependence of limb-darkened transit light curves.

\subsection{Looking ahead}

Much of our understanding of extrasolar planets rests on the transiting systems, and this work confirms that these can be studied to high accuracy. New light curves of HD\,149026, TrES-2, OGLE-TR-10, OGLE-TR-56, OGLE-TR-132 and GJ\,436 would help to improve our understanding further, as would additional radial velocity studies of XO-1, WASP-1, WASP-2 and the OGLE transiting planets. In Paper\,II (in preparation) the light curve analysis results from the current paper will be used to determine homogeneous sets of physical properties for the fourteen TEPs studied here, with the aim of improving the prospects for statistical studies of these and other systems.


\section{Acknowledgements}

I am grateful to Josh Winn, Matthew Holman, Bun'ei Sato and David Charbonneau for making their data easily available, and to Fr\'ederic Pont for sending me light curves of OGLE-TR-10 and OGLE-TR-56. I would like to thank the anonymous referee for a timely and authoritative report, and Peter Wheatley, Antonio Claret, Tom Marsh, Boris G\"ansicke, Pierre Maxted and Andrew Levan for useful discussions. I acknowledge financial support from PPARC in the form of a postdoctoral research assistant position. The following internet-based resources were used: the NASA Astrophysics Data System; the SIMBAD database operated at CDS, Strasbourg, France; and the ar$\chi$iv scientific paper preprint service operated by Cornell University. Guest User, Canadian Astronomy Data Centre, which is operated by the Herzberg Institute of Astrophysics, National Research Council of Canada. This paper uses data from the MOST satellite, a Canadian Space Agency mission, jointly operated by Dynacon Inc., the University of Toronto Institute for Aerospace Studies and the University of British Columbia, with the assistance of the University of Vienna.


\bibliographystyle{mn_new}

\begin{thebibliography}{159}
\expandafter\ifx\csname natexlab\endcsname\relax\def\natexlab#1{#1}\fi

\bibitem[{{Alonso} et~al.(2004)}]{Alonso+04apj}
{Alonso}, R., et~al., 2004, ApJ, 613, L153

\bibitem[{{Anderson} et~al.(2008)}]{Anderson+07xxx}
{Anderson}, D.~R., et~al., 2008, MNRAS submitted, {\tt arXiv0801.1685}

\bibitem[{{Baines} et~al.(2007){Baines}, {van Belle}, {ten Brummelaar},
  {McAlister}, {Swain}, {Turner}, {Sturmann}, \& {Sturmann}}]{Baines+07apj}
{Baines}, E.~K., {van Belle}, G.~T., {ten Brummelaar}, T.~A., {McAlister},
  H.~A., {Swain}, M., {Turner}, N.~H., {Sturmann}, L., {Sturmann}, J., 2007,
  ApJ, 661, L195

\bibitem[{{Bakos} et~al.(2004){Bakos}, {Noyes}, {Kov{\'a}cs}, {Stanek},
  {Sasselov}, \& {Domsa}}]{Bakos+04pasp}
{Bakos}, G., {Noyes}, R.~W., {Kov{\'a}cs}, G., {Stanek}, K.~Z., {Sasselov},
  D.~D., {Domsa}, I., 2004, PASP, 116, 266

\bibitem[{{Bakos} et~al.(2002){Bakos}, {L{\'a}z{\'a}r}, {Papp}, {S{\'a}ri}, \&
  {Green}}]{Bakos+02pasp}
{Bakos}, G.~{\'A}., {L{\'a}z{\'a}r}, J., {Papp}, I., {S{\'a}ri}, P., {Green},
  E.~M., 2002, PASP, 114, 974

\bibitem[{{Bakos} et~al.(2006{\natexlab{a}}){Bakos}, {P{\'a}l}, {Latham},
  {Noyes}, \& {Stefanik}}]{Bakos+06apj}
{Bakos}, G.~{\'A}., {P{\'a}l}, A., {Latham}, D.~W., {Noyes}, R.~W., {Stefanik},
  R.~P., 2006{\natexlab{a}}, ApJ, 641, L57

\bibitem[{{Bakos} et~al.(2006{\natexlab{b}})}]{Bakos+06apj2}
{Bakos}, G.~{\'A}., et~al., 2006{\natexlab{b}}, ApJ, 650, 1160

\bibitem[{{Bakos} et~al.(2007{\natexlab{a}})}]{Bakos+07apj}
{Bakos}, G.~{\'A}., et~al., 2007{\natexlab{a}}, ApJ, 656, 552

\bibitem[{{Bakos} et~al.(2007{\natexlab{b}})}]{Bakos+07apj2}
{Bakos}, G.~{\'A}., et~al., 2007{\natexlab{b}}, ApJ, 671, L173

\bibitem[{{Barban} et~al.(2003){Barban}, {Goupil}, {Van't Veer-Menneret},
  {Garrido}, {Kupka}, \& {Heiter}}]{Barban+03aa}
{Barban}, C., {Goupil}, M.~J., {Van't Veer-Menneret}, C., {Garrido}, R.,
  {Kupka}, F., {Heiter}, U., 2003, A\&A, 405, 1095

\bibitem[{{Barbieri} et~al.(2007)}]{Barbieri+07aa}
{Barbieri}, M., et~al., 2007, A\&A, 476, L13

\bibitem[{{Beatty} et~al.(2007)}]{Beatty+07apj}
{Beatty}, T.~G., et~al., 2007, ApJ, 663, 573

\bibitem[{{Berger} et~al.(2006)}]{Berger+06apj}
{Berger}, D.~H., et~al., 2006, ApJ, 644, 475

\bibitem[{{Bouchy} et~al.(2004){Bouchy}, {Pont}, {Santos}, {Melo}, {Mayor},
  {Queloz}, \& {Udry}}]{Bouchy+04aa}
{Bouchy}, F., {Pont}, F., {Santos}, N.~C., {Melo}, C., {Mayor}, M., {Queloz},
  D., {Udry}, S., 2004, A\&A, 421, L13

\bibitem[{{Bouchy} et~al.(2005{\natexlab{a}}){Bouchy}, {Pont}, {Melo},
  {Santos}, {Mayor}, {Queloz}, \& {Udry}}]{Bouchy+05aa2}
{Bouchy}, F., {Pont}, F., {Melo}, C., {Santos}, N.~C., {Mayor}, M., {Queloz},
  D., {Udry}, S., 2005{\natexlab{a}}, A\&A, 431, 1105

\bibitem[{{Bouchy} et~al.(2005{\natexlab{b}})}]{Bouchy+05aa}
{Bouchy}, F., et~al., 2005{\natexlab{b}}, A\&A, 444, L15

\bibitem[{{Bozorgnia} et~al.(2006){Bozorgnia}, {Fortney}, {McCarthy},
  {Fischer}, \& {Marcy}}]{Bozorgnia+06pasp}
{Bozorgnia}, N., {Fortney}, J.~J., {McCarthy}, C., {Fischer}, D.~A., {Marcy},
  G.~W., 2006, PASP, 118, 1249

\bibitem[{{Broeg} \& {Wuchterl}(2007)}]{BroegWuchterl07mn}
{Broeg}, C., {Wuchterl}, G., 2007, MNRAS, 376, L62

\bibitem[{{Brown} et~al.(2001){Brown}, {Charbonneau}, {Gilliland}, {Noyes}, \&
  {Burrows}}]{Brown+01apj}
{Brown}, T.~M., {Charbonneau}, D., {Gilliland}, R.~L., {Noyes}, R.~W.,
  {Burrows}, A., 2001, ApJ, 552, 699

\bibitem[{{Bruntt} et~al.(2006){Bruntt}, {Southworth}, {Torres}, {Penny},
  {Clausen}, \& {Buzasi}}]{Bruntt+06aa}
{Bruntt}, H., {Southworth}, J., {Torres}, G., {Penny}, A.~J., {Clausen}, J.~V.,
  {Buzasi}, D.~L., 2006, A\&A, 456, 651

\bibitem[{{Burke} et~al.(2007)}]{Burke+07apj}
{Burke}, C.~J., et~al., 2007, ApJ, 671, 2115

\bibitem[{{Burrows} et~al.(2007){Burrows}, {Hubeny}, {Budaj}, \&
  {Hubbard}}]{Burrows+07apj}
{Burrows}, A., {Hubeny}, I., {Budaj}, J., {Hubbard}, W.~B., 2007, ApJ, 661, 502

\bibitem[{{Butler} et~al.(1996){Butler}, {Marcy}, {Williams}, {McCarthy},
  {Dosanjh}, \& {Vogt}}]{Butler+96pasp}
{Butler}, R.~P., {Marcy}, G.~W., {Williams}, E., {McCarthy}, C., {Dosanjh}, P.,
  {Vogt}, S.~S., 1996, PASP, 108, 500

\bibitem[{{Butler} et~al.(2004){Butler}, {Vogt}, {Marcy}, {Fischer}, {Wright},
  {Henry}, {Laughlin}, \& {Lissauer}}]{Butler+04apj}
{Butler}, R.~P., {Vogt}, S.~S., {Marcy}, G.~W., {Fischer}, D.~A., {Wright},
  J.~T., {Henry}, G.~W., {Laughlin}, G., {Lissauer}, J.~J., 2004, ApJ, 617, 580

\bibitem[{{Castellano} et~al.(2000){Castellano}, {Jenkins}, {Trilling},
  {Doyle}, \& {Koch}}]{Castellano+00apj}
{Castellano}, T., {Jenkins}, J., {Trilling}, D.~E., {Doyle}, L., {Koch}, D.,
  2000, ApJ, 532, L51

\bibitem[{{Charbonneau} et~al.(2000){Charbonneau}, {Brown}, {Latham}, \&
  {Mayor}}]{Charbonneau+00apj}
{Charbonneau}, D., {Brown}, T.~M., {Latham}, D.~W., {Mayor}, M., 2000, ApJ,
  529, L45

\bibitem[{{Charbonneau} et~al.(2007){Charbonneau}, {Winn}, {Everett}, {Latham},
  {Holman}, {Esquerdo}, \& {O'Donovan}}]{Charbonneau+07apj}
{Charbonneau}, D., {Winn}, J.~N., {Everett}, M.~E., {Latham}, D.~W., {Holman},
  M.~J., {Esquerdo}, G.~A., {O'Donovan}, F.~T., 2007, ApJ, 658, 1322

\bibitem[{{Charbonneau} et~al.(2006)}]{Charbonneau+06apj}
{Charbonneau}, D., et~al., 2006, ApJ, 636, 445

\bibitem[{{Christian} et~al.(2006)}]{Christian+06mn}
{Christian}, D.~J., et~al., 2006, MNRAS, 372, 1117

\bibitem[{{Claret}(2000)}]{Claret00aa}
{Claret}, A., 2000, A\&A, 363, 1081

\bibitem[{{Claret}(2004)}]{Claret04aa2}
{Claret}, A., 2004, A\&A, 428, 1001

\bibitem[{{Claret} \& {Hauschildt}(2003)}]{ClaretHauschildt03aa}
{Claret}, A., {Hauschildt}, P.~H., 2003, A\&A, 412, 241

\bibitem[{{Claret} et~al.(1995){Claret}, {D{\'{\i}}az-Cordov{\'e}s}, \&
  {Gim{\'e}nez}}]{Claret++95aas}
{Claret}, A., {D{\'{\i}}az-Cordov{\'e}s}, J., {Gim{\'e}nez}, A., 1995, A\&AS,
  114, 247

\bibitem[{{Collier Cameron} et~al.(2007)}]{Cameron+07mn}
{Collier Cameron}, A., et~al., 2007, MNRAS, 375, 951

\bibitem[{{Croll} et~al.(2007{\natexlab{a}})}]{Croll+07apj}
{Croll}, B., et~al., 2007{\natexlab{a}}, ApJ, 658, 1328

\bibitem[{{Croll} et~al.(2007{\natexlab{b}})}]{Croll+07apj2}
{Croll}, B., et~al., 2007{\natexlab{b}}, ApJ, 671, 2129

\bibitem[{{Deming} et~al.(2005){Deming}, {Seager}, {Richardson}, \&
  {Harrington}}]{Deming+05natur}
{Deming}, D., {Seager}, S., {Richardson}, L.~J., {Harrington}, J., 2005,
  Nature, 434, 740

\bibitem[{{Deming} et~al.(2006){Deming}, {Harrington}, {Seager}, \&
  {Richardson}}]{Deming+06apj}
{Deming}, D., {Harrington}, J., {Seager}, S., {Richardson}, L.~J., 2006, ApJ,
  644, 560

\bibitem[{{Deming} et~al.(2007{\natexlab{a}}){Deming}, {Harrington},
  {Laughlin}, {Seager}, {Navarro}, {Bowman}, \& {Horning}}]{Deming+07apj}
{Deming}, D., {Harrington}, J., {Laughlin}, G., {Seager}, S., {Navarro}, S.~B.,
  {Bowman}, W.~C., {Horning}, K., 2007{\natexlab{a}}, ApJ, 667, L199

\bibitem[{{Deming} et~al.(2007{\natexlab{b}}){Deming}, {Richardson}, \&
  {Harrington}}]{Deming++07mn}
{Deming}, D., {Richardson}, L.~J., {Harrington}, J., 2007{\natexlab{b}}, MNRAS,
  378, 148

\bibitem[{{Demory} et~al.(2007)}]{Demory+07aa}
{Demory}, B.-O., et~al., 2007, A\&A, 475, 1125

\bibitem[{{D{\'{\i}}az-Cordov{\'e}s} \& {Gim{\'e}nez}(1992)}]{DiazGimenez92aa}
{D{\'{\i}}az-Cordov{\'e}s}, J., {Gim{\'e}nez}, A., 1992, A\&A, 259, 227

\bibitem[{{D{\'{\i}}az-Cordov{\'e}s} et~al.(1995){D{\'{\i}}az-Cordov{\'e}s},
  {Claret}, \& {Gim{\'e}nez}}]{Diaz++95aas}
{D{\'{\i}}az-Cordov{\'e}s}, J., {Claret}, A., {Gim{\'e}nez}, A., 1995, A\&AS,
  110, 329

\bibitem[{{Etzel}(1975)}]{Etzel75}
{Etzel}, P.~B., 1975, Masters Thesis, San Diego State University

\bibitem[{{Etzel}(1981)}]{Etzel81conf}
{Etzel}, P.~B., 1981, in {Carling}, E.~B., {Kopal}, Z., eds., Photometric and
  Spectroscopic Binary Systems, NATO ASI Ser. C., 69, Dordrecht, p. 111

\bibitem[{{Gillon} et~al.(2006){Gillon}, {Pont}, {Moutou}, {Bouchy}, {Courbin},
  {Sohy}, \& {Magain}}]{Gillon+06aa}
{Gillon}, M., {Pont}, F., {Moutou}, C., {Bouchy}, F., {Courbin}, F., {Sohy},
  S., {Magain}, P., 2006, A\&A, 459, 249

\bibitem[{{Gillon} et~al.(2007{\natexlab{a}})}]{Gillon+07aa1}
{Gillon}, M., et~al., 2007{\natexlab{a}}, A\&A, 471, L51

\bibitem[{{Gillon} et~al.(2007{\natexlab{b}})}]{Gillon+07aa2}
{Gillon}, M., et~al., 2007{\natexlab{b}}, A\&A, 472, L13

\bibitem[{{Gillon} et~al.(2007{\natexlab{c}})}]{Gillon+07aa3}
{Gillon}, M., et~al., 2007{\natexlab{c}}, A\&A, 466, 743

\bibitem[{{Gim{\'e}nez}(2006)}]{Gimenez06aa}
{Gim{\'e}nez}, A., 2006, A\&A, 450, 1231

\bibitem[{{Hadrava}(1990)}]{Hadrava90coska}
{Hadrava}, P., 1990, Contributions of the Astronomical Observatory Skalnate
  Pleso, 20, 23

\bibitem[{{Harrington} et~al.(2007){Harrington}, {Luszcz}, {Seager}, {Deming},
  \& {Richardson}}]{Harrington+07natur}
{Harrington}, J., {Luszcz}, S., {Seager}, S., {Deming}, D., {Richardson},
  L.~J., 2007, Nature, 447, 691

\bibitem[{{H{\'e}brard} \& {Lecavelier Des Etangs}(2006)}]{HebrardLDE06aa}
{H{\'e}brard}, G., {Lecavelier Des Etangs}, A., 2006, A\&A, 445, 341

\bibitem[{{Henry} \& {Winn}(2008)}]{HenryWinn08aj}
{Henry}, G.~W., {Winn}, J.~N., 2008, AJ, 135, 68

\bibitem[{{Henry} et~al.(2000){Henry}, {Marcy}, {Butler}, \&
  {Vogt}}]{Henry+00apj}
{Henry}, G.~W., {Marcy}, G.~W., {Butler}, R.~P., {Vogt}, S.~S., 2000, ApJ, 529,
  L41

\bibitem[{{Hill}(1979)}]{Hill79pdao}
{Hill}, G., 1979, Publications of the Dominion Astrophysical Observatory
  Victoria, 15, 297

\bibitem[{{Holman} et~al.(2006)}]{Holman+06apj}
{Holman}, M.~J., et~al., 2006, ApJ, 652, 1715

\bibitem[{{Holman} et~al.(2007{\natexlab{a}})}]{Holman+07apj2}
{Holman}, M.~J., et~al., 2007{\natexlab{a}}, ApJ, 655, 1103

\bibitem[{{Holman} et~al.(2007{\natexlab{b}})}]{Holman+07apj}
{Holman}, M.~J., et~al., 2007{\natexlab{b}}, ApJ, 664, 1185

\bibitem[{{Jenkins} et~al.(2002){Jenkins}, {Caldwell}, \&
  {Borucki}}]{Jenkins++02apj}
{Jenkins}, J.~M., {Caldwell}, D.~A., {Borucki}, W.~J., 2002, ApJ, 564, 495

\bibitem[{{Johns-Krull} et~al.(2007)}]{Johnskrull+07xxx}
{Johns-Krull}, C.~M., et~al., 2007, ApJ in press, {\tt arXiv:0712.4283}

\bibitem[{{Johnson} et~al.(2006){Johnson}, {Goodman}, \&
  {Menou}}]{Johnson++07apj}
{Johnson}, E.~T., {Goodman}, J., {Menou}, K., 2006, ApJ, 647, 1413

\bibitem[{{Klinglesmith} \& {Sobieski}(1970)}]{KlinglesmithSobieski70aj}
{Klinglesmith}, D.~A., {Sobieski}, S., 1970, AJ, 75, 175

\bibitem[{{Knutson} et~al.(2007{\natexlab{a}}){Knutson}, {Charbonneau},
  {Allen}, {Burrows}, \& {Megeath}}]{Knutson+07xxx}
{Knutson}, H.~A., {Charbonneau}, D., {Allen}, L.~E., {Burrows}, A., {Megeath},
  S.~T., 2007{\natexlab{a}}, ApJ, in press, {\tt arXiv:0709.3984}

\bibitem[{{Knutson} et~al.(2007{\natexlab{b}}){Knutson}, {Charbonneau},
  {Deming}, \& {Richardson}}]{Knutson+07pasp}
{Knutson}, H.~A., {Charbonneau}, D., {Deming}, D., {Richardson}, L.~J.,
  2007{\natexlab{b}}, PASP, 119, 616

\bibitem[{{Knutson} et~al.(2007{\natexlab{c}}){Knutson}, {Charbonneau},
  {Noyes}, {Brown}, \& {Gilliland}}]{Knutson+07apj}
{Knutson}, H.~A., {Charbonneau}, D., {Noyes}, R.~W., {Brown}, T.~M.,
  {Gilliland}, R.~L., 2007{\natexlab{c}}, ApJ, 655, 564

\bibitem[{{Knutson} et~al.(2007{\natexlab{d}})}]{Knutson+07nature}
{Knutson}, H.~A., et~al., 2007{\natexlab{d}}, Nature, 447, 183

\bibitem[{{Konacki} et~al.(2003{\natexlab{a}}){Konacki}, {Torres}, {Jha}, \&
  {Sasselov}}]{Konacki+03natur}
{Konacki}, M., {Torres}, G., {Jha}, S., {Sasselov}, D.~D., 2003{\natexlab{a}},
  Nature, 421, 507

\bibitem[{{Konacki} et~al.(2003{\natexlab{b}}){Konacki}, {Torres}, {Sasselov},
  \& {Jha}}]{Konacki+03apj}
{Konacki}, M., {Torres}, G., {Sasselov}, D.~D., {Jha}, S., 2003{\natexlab{b}},
  ApJ, 597, 1076

\bibitem[{{Konacki} et~al.(2005){Konacki}, {Torres}, {Sasselov}, \&
  {Jha}}]{Konacki+05apj}
{Konacki}, M., {Torres}, G., {Sasselov}, D.~D., {Jha}, S., 2005, ApJ, 624, 372

\bibitem[{{Kopal}(1950)}]{Kopal51}
{Kopal}, Z., 1950, Harvard College Observatory Circular, 454, 1

\bibitem[{{Kopal}(1959)}]{Kopal59book}
{Kopal}, Z., 1959, {Close binary systems}, The International Astrophysics
  Series, London: Chapman \& Hall, 1959

\bibitem[{{Kov{\'a}cs} et~al.(2007)}]{Kovacs+07apj}
{Kov{\'a}cs}, G., et~al., 2007, ApJ, 670, L41

\bibitem[{{Kruszewski} \& {Semeniuk}(2003)}]{KruszewskiSemeniuk03aca}
{Kruszewski}, A., {Semeniuk}, I., 2003, Acta Astronomica, 53, 241

\bibitem[{{Lammer} et~al.(2003){Lammer}, {Selsis}, {Ribas}, {Guinan}, {Bauer},
  \& {Weiss}}]{Lammer+03apj}
{Lammer}, H., {Selsis}, F., {Ribas}, I., {Guinan}, E.~F., {Bauer}, S.~J.,
  {Weiss}, W.~W., 2003, ApJ, 598, L121

\bibitem[{{Laughlin} et~al.(2005){Laughlin}, {Marcy}, {Vogt}, {Fischer}, \&
  {Butler}}]{Laughlin+05apj}
{Laughlin}, G., {Marcy}, G.~W., {Vogt}, S.~S., {Fischer}, D.~A., {Butler},
  R.~P., 2005, ApJ, 629, L121

\bibitem[{{Littlefair} et~al.(2006){Littlefair}, {Dhillon}, {Marsh},
  {G{\"a}nsicke}, {Southworth}, \& {Watson}}]{Littlefair+06sci}
{Littlefair}, S.~P., {Dhillon}, V.~S., {Marsh}, T.~R., {G{\"a}nsicke}, B.~T.,
  {Southworth}, J., {Watson}, C.~A., 2006, Science, 314, 1578

\bibitem[{{Loeillet} et~al.(2007)}]{Loeillet+07xxx}
{Loeillet}, B., et~al., 2007, A\&A, submitted, {\tt arXiv:0707.0679}

\bibitem[{{L{\'o}pez-Morales}(2007)}]{lopez07apj}
{L{\'o}pez-Morales}, M., 2007, ApJ, 660, 732

\bibitem[{{Maceroni} \& {Rucinski}(1997)}]{MaceroniRucinski97pasp}
{Maceroni}, C., {Rucinski}, S.~M., 1997, PASP, 109, 782

\bibitem[{{Mandel} \& {Agol}(2002)}]{MandelAgol02apj}
{Mandel}, K., {Agol}, E., 2002, ApJ, 580, L171

\bibitem[{{Maness} et~al.(2007){Maness}, {Marcy}, {Ford}, {Hauschildt},
  {Shreve}, {Basri}, {Butler}, \& {Vogt}}]{Maness+07pasp}
{Maness}, H.~L., {Marcy}, G.~W., {Ford}, E.~B., {Hauschildt}, P.~H., {Shreve},
  A.~T., {Basri}, G.~B., {Butler}, R.~P., {Vogt}, S.~S., 2007, PASP, 119, 90

\bibitem[{{Matsuo} et~al.(2007){Matsuo}, {Shibai}, {Ootsubo}, \&
  {Tamura}}]{Matsuo+07apj}
{Matsuo}, T., {Shibai}, H., {Ootsubo}, T., {Tamura}, M., 2007, ApJ, 662, 1282

\bibitem[{{Mayor} \& {Queloz}(1995)}]{MayorQueloz95natur}
{Mayor}, M., {Queloz}, D., 1995, Nature, 378, 355

\bibitem[{{McCullough} et~al.(2006)}]{Mccullough+06apj}
{McCullough}, P.~R., et~al., 2006, ApJ, 648, 1228

\bibitem[{{Minniti} et~al.(2007)}]{Minniti+07apj}
{Minniti}, D., et~al., 2007, ApJ, 660, 858

\bibitem[{{Moutou} et~al.(2004){Moutou}, {Pont}, {Bouchy}, \&
  {Mayor}}]{Moutou++04aa}
{Moutou}, C., {Pont}, F., {Bouchy}, F., {Mayor}, M., 2004, A\&A, 424, L31

\bibitem[{{Moutou} et~al.(2007)}]{Moutou+07aa}
{Moutou}, C., et~al., 2007, A\&A, 473, 651

\bibitem[{{Naef} et~al.(2004){Naef}, {Mayor}, {Beuzit}, {Perrier}, {Queloz},
  {Sivan}, \& {Udry}}]{Naef+04aa}
{Naef}, D., {Mayor}, M., {Beuzit}, J.~L., {Perrier}, C., {Queloz}, D., {Sivan},
  J.~P., {Udry}, S., 2004, A\&A, 414, 351

\bibitem[{{Nelson} \& {Davis}(1972)}]{NelsonDavis72apj}
{Nelson}, B., {Davis}, W.~D., 1972, ApJ, 174, 617

\bibitem[{{Noyes} et~al.(2008)}]{Noyes+08apj}
{Noyes}, R.~W., et~al., 2008, ApJ, 673, L79

\bibitem[{{O'Donovan} et~al.(2006)}]{Odonovan+06apj}
{O'Donovan}, F.~T., et~al., 2006, ApJ, 651, L61

\bibitem[{{O'Donovan} et~al.(2007)}]{Odonovan+07apj}
{O'Donovan}, F.~T., et~al., 2007, ApJ, 663, L37

\bibitem[{{Pollacco} et~al.(2007)}]{Pollacco+07xxx}
{Pollacco}, D., et~al., 2007, MNRAS, submitted, {\tt arXiv:0711.0126}

\bibitem[{{Pont} et~al.(2004){Pont}, {Bouchy}, {Queloz}, {Santos}, {Melo},
  {Mayor}, \& {Udry}}]{Pont+04aa}
{Pont}, F., {Bouchy}, F., {Queloz}, D., {Santos}, N.~C., {Melo}, C., {Mayor},
  M., {Udry}, S., 2004, A\&A, 426, L15

\bibitem[{{Pont} et~al.(2005){Pont}, {Bouchy}, {Melo}, {Santos}, {Mayor},
  {Queloz}, \& {Udry}}]{Pont+05aa}
{Pont}, F., {Bouchy}, F., {Melo}, C., {Santos}, N.~C., {Mayor}, M., {Queloz},
  D., {Udry}, S., 2005, A\&A, 438, 1123

\bibitem[{{Pont} et~al.(2006){Pont}, {Zucker}, \& {Queloz}}]{Pont++06mn}
{Pont}, F., {Zucker}, S., {Queloz}, D., 2006, MNRAS, 373, 231

\bibitem[{{Pont} et~al.(2007{\natexlab{a}})}]{Pont+07xxx2}
{Pont}, F., et~al., 2007{\natexlab{a}}, A\&A, submitted, {\tt arXiv:0710.5278}

\bibitem[{{Pont} et~al.(2007{\natexlab{b}})}]{Pont+07aa2}
{Pont}, F., et~al., 2007{\natexlab{b}}, A\&A, 476, 1347

\bibitem[{{Pont} et~al.(2007{\natexlab{c}})}]{Pont+07aa}
{Pont}, F., et~al., 2007{\natexlab{c}}, A\&A, 465, 1069

\bibitem[{{Popper}(1984)}]{Popper84aj}
{Popper}, D.~M., 1984, AJ, 89, 132

\bibitem[{{Popper} \& {Etzel}(1981)}]{PopperEtzel81aj}
{Popper}, D.~M., {Etzel}, P.~B., 1981, AJ, 86, 102

\bibitem[{{Press} et~al.(1992){Press}, {Teukolsky}, {Vetterling}, \&
  {Flannery}}]{Press+92book}
{Press}, W.~H., {Teukolsky}, S.~A., {Vetterling}, W.~T., {Flannery}, B.~P.,
  1992, {Numerical recipes in FORTRAN 77. The art of scientific computing},
  Cambridge: University Press, 2nd ed.

\bibitem[{{Queloz} et~al.(2000){Queloz}, {Eggenberger}, {Mayor}, {Perrier},
  {Beuzit}, {Naef}, {Sivan}, \& {Udry}}]{Queloz+00aa}
{Queloz}, D., {Eggenberger}, A., {Mayor}, M., {Perrier}, C., {Beuzit}, J.~L.,
  {Naef}, D., {Sivan}, J.~P., {Udry}, S., 2000, A\&A, 359, L13

\bibitem[{{Ribas}(2003)}]{Ribas03aa}
{Ribas}, I., 2003, A\&A, 398, 239

\bibitem[{{Ribas}(2006)}]{Ribas06apss}
{Ribas}, I., 2006, Ap\&SS, 304, 89

\bibitem[{{Richardson} et~al.(2006){Richardson}, {Harrington}, {Seager}, \&
  {Deming}}]{Richardson+06apj}
{Richardson}, L.~J., {Harrington}, J., {Seager}, S., {Deming}, D., 2006, ApJ,
  649, 1043

\bibitem[{{Robichon} \& {Arenou}(2000)}]{RobichonArenou00aa}
{Robichon}, N., {Arenou}, F., 2000, A\&A, 355, 295

\bibitem[{{Rowe} et~al.(2006)}]{Rowe+06apj}
{Rowe}, J.~F., et~al., 2006, ApJ, 646, 1241

\bibitem[{{Rowe} et~al.(2007)}]{Rowe+07xxx}
{Rowe}, J.~F., et~al., 2007, ApJ submitted, {\tt arXiv:0711.4111}

\bibitem[{{Russell}(1912{\natexlab{a}})}]{Russell12apj}
{Russell}, H.~N., 1912{\natexlab{a}}, ApJ, 35, 315

\bibitem[{{Russell}(1912{\natexlab{b}})}]{Russell12apj2}
{Russell}, H.~N., 1912{\natexlab{b}}, ApJ, 36, 54

\bibitem[{{Russell} \& {Merrill}(1952)}]{RussellMerrill52book}
{Russell}, H.~N., {Merrill}, J.~E., 1952, {The determination of the elements of
  eclipsing binaries}, Contributions from the Princeton University Observatory,
  Princeton: Observatory, 1952

\bibitem[{{Sato} et~al.(2005)}]{Sato+05apj}
{Sato}, B., et~al., 2005, ApJ, 633, 465

\bibitem[{{Shporer} et~al.(2007){Shporer}, {Tamuz}, {Zucker}, \&
  {Mazeh}}]{Shporer+07mn}
{Shporer}, A., {Tamuz}, O., {Zucker}, S., {Mazeh}, T., 2007, MNRAS, 376, 1296

\bibitem[{{Snellen}(2005)}]{Snellen05mn}
{Snellen}, I.~A.~G., 2005, MNRAS, 363, 211

\bibitem[{{S{\"o}derhjelm}(1999)}]{Soderhjelm99ibvs}
{S{\"o}derhjelm}, S., 1999, IBVS, 4816, 1

\bibitem[{{Southworth} \& {Clausen}(2007)}]{MeClausen07aa}
{Southworth}, J., {Clausen}, J.~V., 2007, A\&A, 461, 1077

\bibitem[{{Southworth} et~al.(2004{\natexlab{a}}){Southworth}, {Maxted}, \&
  {Smalley}}]{Me++04mn}
{Southworth}, J., {Maxted}, P.~F.~L., {Smalley}, B., 2004{\natexlab{a}}, MNRAS,
  349, 547

\bibitem[{{Southworth} et~al.(2004{\natexlab{b}}){Southworth}, {Maxted}, \&
  {Smalley}}]{Me++04mn2}
{Southworth}, J., {Maxted}, P.~F.~L., {Smalley}, B., 2004{\natexlab{b}}, MNRAS,
  351, 1277

\bibitem[{{Southworth} et~al.(2004{\natexlab{c}}){Southworth}, {Zucker},
  {Maxted}, \& {Smalley}}]{Me+04mn3}
{Southworth}, J., {Zucker}, S., {Maxted}, P.~F.~L., {Smalley}, B.,
  2004{\natexlab{c}}, MNRAS, 355, 986

\bibitem[{{Southworth} et~al.(2005){Southworth}, {Smalley}, {Maxted}, {Claret},
  \& {Etzel}}]{Me+05mn}
{Southworth}, J., {Smalley}, B., {Maxted}, P.~F.~L., {Claret}, A., {Etzel},
  P.~B., 2005, MNRAS, 363, 529

\bibitem[{{Southworth} et~al.(2007{\natexlab{a}}){Southworth}, {Bruntt}, \&
  {Buzasi}}]{Me++07aa}
{Southworth}, J., {Bruntt}, H., {Buzasi}, D.~L., 2007{\natexlab{a}}, A\&A, 467,
  1215

\bibitem[{{Southworth} et~al.(2007{\natexlab{b}}){Southworth}, {Wheatley}, \&
  {Sams}}]{Me++07mn}
{Southworth}, J., {Wheatley}, P.~J., {Sams}, G., 2007{\natexlab{b}}, MNRAS,
  379, L11

\bibitem[{{Sozzetti} et~al.(2007){Sozzetti}, {Torres}, {Charbonneau}, {Latham},
  {Holman}, {Winn}, {Laird}, \& {O'Donovan}}]{Sozzetti+07apj}
{Sozzetti}, A., {Torres}, G., {Charbonneau}, D., {Latham}, D.~W., {Holman},
  M.~J., {Winn}, J.~N., {Laird}, J.~B., {O'Donovan}, F.~T., 2007, ApJ, 664,
  1190

\bibitem[{{Steffen} \& {Agol}(2005)}]{SteffenAgol05mn}
{Steffen}, J.~H., {Agol}, E., 2005, MNRAS, 364, L96

\bibitem[{{Street} et~al.(2007)}]{Street+07mn}
{Street}, R.~A., et~al., 2007, MNRAS, 379, 816

\bibitem[{{Tinetti} et~al.(2007)}]{Tinetti+07nature}
{Tinetti}, G., et~al., 2007, Nature, 448, 169

\bibitem[{{Tingley} et~al.(2006){Tingley}, {Thurl}, \&
  {Sackett}}]{Tingley++06aa}
{Tingley}, B., {Thurl}, C., {Sackett}, P., 2006, A\&A, 445, L27

\bibitem[{{Torres}(2007)}]{Torres07apj}
{Torres}, G., 2007, ApJ, 671, L65

\bibitem[{{Torres} et~al.(2004){Torres}, {Konacki}, {Sasselov}, \&
  {Jha}}]{Torres+04apj}
{Torres}, G., {Konacki}, M., {Sasselov}, D.~D., {Jha}, S., 2004, ApJ, 609, 1071

\bibitem[{{Torres} et~al.(2007)}]{Torres+07apj}
{Torres}, G., et~al., 2007, ApJ, 666, L121

\bibitem[{{Udalski} et~al.(2002{\natexlab{a}}){Udalski}, {Szewczyk}, {Zebrun},
  {Pietrzynski}, {Szymanski}, {Kubiak}, {Soszynski}, \&
  {Wyrzykowski}}]{Udalski+02aca2}
{Udalski}, A., {Szewczyk}, O., {Zebrun}, K., {Pietrzynski}, G., {Szymanski},
  M., {Kubiak}, M., {Soszynski}, I., {Wyrzykowski}, L., 2002{\natexlab{a}},
  Acta Astronomica, 52, 317

\bibitem[{{Udalski} et~al.(2002{\natexlab{b}}){Udalski}, {Zebrun}, {Szymanski},
  {Kubiak}, {Soszynski}, {Szewczyk}, {Wyrzykowski}, \&
  {Pietrzynski}}]{Udalski+02aca3}
{Udalski}, A., {Zebrun}, K., {Szymanski}, M., {Kubiak}, M., {Soszynski}, I.,
  {Szewczyk}, O., {Wyrzykowski}, L., {Pietrzynski}, G., 2002{\natexlab{b}},
  Acta Astronomica, 52, 115

\bibitem[{{Udalski} et~al.(2003){Udalski}, {Pietrzynski}, {Szymanski},
  {Kubiak}, {Zebrun}, {Soszynski}, {Szewczyk}, \&
  {Wyrzykowski}}]{Udalski+03aca}
{Udalski}, A., {Pietrzynski}, G., {Szymanski}, M., {Kubiak}, M., {Zebrun}, K.,
  {Soszynski}, I., {Szewczyk}, O., {Wyrzykowski}, L., 2003, Acta Astronomica,
  53, 133

\bibitem[{{Udalski} et~al.(2002{\natexlab{c}})}]{Udalski+02aca}
{Udalski}, A., et~al., 2002{\natexlab{c}}, Acta Astronomica, 52, 1

\bibitem[{{Udalski} et~al.(2007)}]{Udalski+07xxx}
{Udalski}, A., et~al., 2007, A\&A submitted, {\tt arXiv:0711.3978}

\bibitem[{{Udry} \& {Santos}(2007)}]{UdrySantos07araa}
{Udry}, S., {Santos}, N.~C., 2007, ARA\&A, 45, 397

\bibitem[{{Van Hamme}(1993)}]{Vanhamme93aj}
{Van Hamme}, W., 1993, AJ, 106, 2096

\bibitem[{{van't Veer}(1960)}]{Vantveer60book}
{van't Veer}, F., 1960, {L'assombrissement centre-bord des etoiles.}, Doctoral
  thesis, University of Utrecht

\bibitem[{{Weldrake} et~al.(2007){Weldrake}, {Bayliss}, {Sackett}, {Tingley},
  {Gillon}, \& {Setiawan}}]{Weldrake+07xxx}
{Weldrake}, D.~T.~F., {Bayliss}, D.~D.~R., {Sackett}, P.~D., {Tingley}, B.~W.,
  {Gillon}, M., {Setiawan}, J., 2007, ApJL submitted, {\tt arXiv:0711.1746}

\bibitem[{{Wilson} et~al.(2006)}]{Wilson+06pasp}
{Wilson}, D.~M., et~al., 2006, PASP, 118, 1245

\bibitem[{{Wilson} et~al.(2008)}]{Wilson+07xxx}
{Wilson}, D.~M., et~al., 2008, ApJL in press, {\tt arXiv:0801.1509}

\bibitem[{{Wilson}(1979)}]{Wilson79apj}
{Wilson}, R.~E., 1979, ApJ, 234, 1054

\bibitem[{{Wilson}(1993)}]{Wilson93aspc}
{Wilson}, R.~E., 1993, in {Leung}, K.-C., {Nha}, I.-S., eds., ASP Conf. Ser.
  38: New Frontiers in Binary Star Research, p.~91

\bibitem[{{Wilson} \& {Devinney}(1971)}]{WilsonDevinney71apj}
{Wilson}, R.~E., {Devinney}, E.~J., 1971, ApJ, 166, 605

\bibitem[{{Winn}(2007)}]{Winn07xxx}
{Winn}, J.~N., 2007, in {Fischer}, D., {Rasio}, F., {Thorsett}, S.,
  {Wolszczan}, A., eds., Extreme Solar Systems, ASP Conference Series, in
  press, {\tt arXiv:0710.1098}

\bibitem[{{Winn} et~al.(2007{\natexlab{a}}){Winn}, {Henry}, {Torres}, \&
  {Holman}}]{Winn+07xxx}
{Winn}, J.~N., {Henry}, G.~W., {Torres}, G., {Holman}, M.~J.,
  2007{\natexlab{a}}, ApJ, in press, {\tt arXiv:0711.1888}

\bibitem[{{Winn} et~al.(2007{\natexlab{b}}){Winn}, {Holman}, \&
  {Fuentes}}]{Winn++07aj}
{Winn}, J.~N., {Holman}, M.~J., {Fuentes}, C.~I., 2007{\natexlab{b}}, AJ, 133,
  11

\bibitem[{{Winn} et~al.(2007{\natexlab{c}}){Winn}, {Holman}, \&
  {Roussanova}}]{Winn++07apj}
{Winn}, J.~N., {Holman}, M.~J., {Roussanova}, A., 2007{\natexlab{c}}, ApJ, 657,
  1098

\bibitem[{{Winn} et~al.(2005)}]{Winn+05apj}
{Winn}, J.~N., et~al., 2005, ApJ, 631, 1215

\bibitem[{{Winn} et~al.(2006)}]{Winn+06apj}
{Winn}, J.~N., et~al., 2006, ApJ, 653, L69

\bibitem[{{Winn} et~al.(2007{\natexlab{d}})}]{Winn+07aj}
{Winn}, J.~N., et~al., 2007{\natexlab{d}}, AJ, 133, 1828

\bibitem[{{Winn} et~al.(2007{\natexlab{e}})}]{Winn+07aj2}
{Winn}, J.~N., et~al., 2007{\natexlab{e}}, AJ, 134, 1707

\bibitem[{{Wittenmyer} et~al.(2005)}]{Wittenmyer+05apj}
{Wittenmyer}, R.~A., et~al., 2005, ApJ, 632, 1157

\bibitem[{{Wolf} et~al.(2007){Wolf}, {Laughlin}, {Henry}, {Fischer}, {Marcy},
  {Butler}, \& {Vogt}}]{Wolf+07apj}
{Wolf}, A.~S., {Laughlin}, G., {Henry}, G.~W., {Fischer}, D.~A., {Marcy}, G.,
  {Butler}, P., {Vogt}, S., 2007, ApJ, 667, 549

\bibitem[{{Wood}(1971)}]{Wood71aj}
{Wood}, D.~B., 1971, AJ, 76, 701

\bibitem[{{Wood}(1973)}]{Wood73pasp}
{Wood}, D.~B., 1973, PASP, 85, 253

\bibitem[{{Zhou} \& {Lin}(2007)}]{ZhouLin07apj}
{Zhou}, J.-L., {Lin}, D.~N.~C., 2007, ApJ, 666, 447

\end{thebibliography}

\label{lastpage}


\appendix

\section{Results of the light curve analyses}

The tables in this section contain the detailed results of the modelling of each light curve. \reff{Note that whilst all the results are best fits to the relevant data, some parameters are unphysical (for example the limb darkening coefficients imply that the limb of the star produces a negative amount of light). In these cases the unphysical results have {\em not} been used but are retained in the table for completeness.}

\begin{table*} \caption{\label{tab:lc:tres1} Parameters of the {\sc jktebop}
best fits of the TrES-1 light curve from \citet{Winn++07apj} ($\Nobs =
1149$) for different treatment of LD. The parameter symbols are as given
in Section\,\ref{sec:analysis:lc}. The transit midpoint $T_0$ is expressed
as HJD $-$ 2\,453\,000. \reff{The perturbations of the LD coefficients are
discussed in Section\,\ref{sec:analysis:ld}.}}
\begin{tabular}{l r@{\,$\pm$\,}l r@{\,$\pm$\,}l r@{\,$\pm$\,}l r@{\,$\pm$\,}l r@{\,$\pm$\,}l}
\hline \hline
\                   &\mc{Linear LD law}&\mc{Quadratic LD law}&\mc{Square-root LD law}&\mc{Logarithmic LD law}&\mc{Cubic LD law}   \\
\hline
\multicolumn{11}{l}{All LD coefficients fixed} \\
\hline
$r_{\rm A}+r_{\rm b}$ & \erm{0.1082}{0.0018}{0.0004}    & \erm{0.1078}{0.0015}{0.0004}    & \erm{0.1078}{0.0016}{0.0005}    & \erm{0.1077}{0.0016}{0.0004}    & \mc{ } \\[3pt]
$k$                   & \erm{0.13444}{0.00059}{0.00038} & 0.13568 & 0.00039               & 0.13533 & 0.00040               & 0.13566 & 0.00041               & \mc{ } \\[3pt]
$i$ (deg.)            & \erm{89.85}{0.15}{0.92}         & \erm{89.96}{0.04}{0.90}         & \erm{89.91}{0.09}{0.91}         & \erm{89.99}{0.01}{0.97}         & \mc{ } \\
$u_{\rm A}$           & \mc{0.53 fixed}                 & \mc{0.28 fixed}                 & \mc{0.14 fixed}                 & \mc{0.61 fixed}                       & \mc{ } \\
$v_{\rm A}$           & \mc{ }                          & \mc{0.29 fixed}                 & \mc{0.52 fixed}                 & \mc{0.26 fixed}                       & \mc{ } \\
$T_0$                 & 898.87336 & 0.00009             & 898.87336 & 0.00008             & 898.87336 & 0.00008             & 898.87336 & 0.00008             & \mc{ } \\
\hline \hline \end{tabular} \end{table*}

\begin{table*} \caption{\label{tab:lc:tres2} Parameters of the {\sc jktebop}
best fits of the TrES-2 light curve ($z$ band, $\Nobs = 1033$) from
\citet{Holman+07apj} for different treatments of LD. The parameter symbols
are as given in Section\,\ref{sec:analysis:lc}. The transit midpoint $T_0$
is expressed as HJD $-$ 2\,453\,000. For each part of the table the upper
parameters are fitted quantities and the lower parameters are derived quantities.}
\begin{tabular}{l r@{\,$\pm$\,}l r@{\,$\pm$\,}l r@{\,$\pm$\,}l r@{\,$\pm$\,}l r@{\,$\pm$\,}l}
\hline \hline
\                   &\mc{Linear LD law}&\mc{Quadratic LD law}&\mc{Square-root LD law}&\mc{Logarithmic LD law}&\mc{Cubic LD law}   \\
\hline
\multicolumn{11}{l}{All LD coefficients fixed} \\
\hline
$r_{\rm A}+r_{\rm b}$ &   0.1468  & 0.0020  &   0.1475  & 0.0020  &   0.1479  & 0.0020  &   0.1478  & 0.0021  &  \mc{ }  \\
$k$                   &   0.12681 & 0.00086 &   0.12537 & 0.00086 &   0.12537 & 0.00087 &   0.12532 & 0.00089 &  \mc{ }  \\
$i$ (deg.)            &  83.61    & 0.12    &  83.58    & 0.12    &  83.54    & 0.12    &  83.55    & 0.12    &  \mc{ }  \\
$u_{\rm A}$           &   \mc{0.49 fixed}   &   \mc{0.22 fixed}   &   \mc{0.06 fixed}   &    \mc{0.57 fixed}  &  \mc{ }  \\
$v_{\rm A}$           &        \mc{ }       &   \mc{0.31 fixed}   &   \mc{0.56 fixed}   &    \mc{0.27 fixed}  &  \mc{ }  \\
$T_0$                 & 989.75282 & 0.00017 & 989.75298 & 0.00018 & 989.75279 & 0.00017 & 989.75279 & 0.00018 &  \mc{ }  \\
\hline \hline \end{tabular} \end{table*}

\begin{table*} \caption{\label{tab:lc:xo1} Parameters of the {\sc jktebop} best
fits of the XO-1 light curve from the FLWO 1.2m telescope \citep{Holman+06apj}
($z$ band, $\Nobs = 821$) for different treatment of LD. The parameter symbols
are as given in Section\,\ref{sec:analysis:lc}. The transit midpoint $T_0$ is
expressed as HJD $-$ 2\,453\,000. For each part of the table the upper parameters
are fitted quantities and the lower parameters are derived quantities.}
\begin{tabular}{l r@{\,$\pm$\,}l r@{\,$\pm$\,}l r@{\,$\pm$\,}l r@{\,$\pm$\,}l r@{\,$\pm$\,}l}
\hline \hline
\                    &\mc{Linear LD law}&\mc{Quadratic LD law}&\mc{Square-root LD law}&\mc{Logarithmic LD law}&\mc{Cubic LD law} \\
\hline
\multicolumn{11}{l}{All LD coefficients fixed} \\
\hline
$r_{\rm A}+r_{\rm b}$ &   0.1003  & 0.0019  &   0.1004  & 0.0017  &   0.1012  & 0.0019  &   0.1010  & 0.0018  & \mc{ } \\
$k$                   &   0.1317  & 0.0007  &   0.1327  & 0.0005  &   0.1331  & 0.0006  &   0.1330  & 0.0006  & \mc{ } \\
$i$ (deg.)            &  88.94    & 0.57    &  88.87    & 0.42    &  88.66    & 0.37    &  88.71    & 0.38    & \mc{ } \\
$u_{\rm A}$           &  \mc{0.49 fixed}    &   \mc{0.23 fixed}   &  \mc{0.07 fixed}    &  \mc{0.57 fixed}    & \mc{ } \\
$v_{\rm A}$           &      \mc{    }      &   \mc{0.31 fixed}   &  \mc{0.55 fixed}    &  \mc{0.27 fixed}    & \mc{ } \\
$T_0$                 & 875.92236 & 0.00010 & 875.92236 & 0.00011 & 875.92235 & 0.00011 & 875.92236 & 0.00011 & \mc{ } \\
\hline \hline \end{tabular} \end{table*}

\begin{table*} \caption{\label{tab:lc:wasp1:shporer} Parameters of the
{\sc jktebop} best fits of the WASP-1 light curve from \citet{Shporer+07mn}
($I$ band, $\Nobs = 583$) for different treatment of LD. The parameter symbols
are as given in Section\,\ref{sec:analysis:lc}. The transit midpoint $T_0$
is expressed as HJD $-$ 2\,454\,000. For each part of the table the upper
parameters are fitted quantities and the lower parameters are derived quantities.}
\begin{tabular}{l r@{\,$\pm$\,}l r@{\,$\pm$\,}l r@{\,$\pm$\,}l r@{\,$\pm$\,}l r@{\,$\pm$\,}l}
\hline \hline
\                   &\mc{Linear LD law}&\mc{Quadratic LD law}&\mc{Square-root LD law}&\mc{Logarithmic LD law}&\mc{Cubic LD law}   \\
\hline
\multicolumn{11}{l}{Fitting for one LD coefficient} \\
\hline
$r_{\rm A}+r_{\rm b}$ &   0.2036  & 0.0104  &   0.1986  & 0.0102  &   0.2032  & 0.0106  &   0.1978  & 0.0096  &   0.2027  & 0.0114  \\
$k$                 &   0.1042  & 0.0011  &   0.1031  & 0.0012  &   0.1038  & 0.0012  &   0.1025  & 0.0013  &   0.1037  & 0.0012  \\
$i$ (deg.)          &  86.0     & 1.6     &  87.0     & 2.4     &  86.1     & 2.2     &  87.3     & 2.3     &  86.2     & 2.3     \\
$u_{\rm A}$           &   0.215   & 0.070   &   0.043   & 0.082   &$-$0.114   & 0.079   &   0.644   & 0.088   &   0.158   & 0.074   \\
$v_{\rm A}$           &        \mc{ }       & \mc{0.35 perturbed} & \mc{0.55 perturbed} & \mc{0.60 perturbed} & \mc{0.20 perturbed} \\
$T_0$               &  13.3127  & 0.0004  &  13.3127  & 0.0004  &  13.3127  & 0.0004  &  13.3127  & 0.0004  &  13.3127  & 0.0004  \\
\hline \hline \end{tabular} \end{table*}

\begin{table*} \caption{\label{tab:lc:wasp1:charb} Parameters
of the {\sc jktebop} best fits of the WASP-1 light curve from
\citet{Charbonneau+07apj} ($z$ band, $\Nobs = 657$) for
different treatment of LD. The parameter symbols are as given
in Section\,\ref{sec:analysis:lc}. The transit midpoint $T_0$
is expressed as HJD $-$ 2\,450\,000. For each part of the
table the upper parameters are fitted quantities and the
lower parameters are derived quantities.}
\begin{tabular}{l r@{\,$\pm$\,}l r@{\,$\pm$\,}l r@{\,$\pm$\,}l r@{\,$\pm$\,}l r@{\,$\pm$\,}l}
\hline \hline
\                   &\mc{Linear LD law}&\mc{Quadratic LD law}&\mc{Square-root LD law}&\mc{Logarithmic LD law}&\mc{Cubic LD law}   \\
\hline
\multicolumn{11}{l}{Fitting for one LD coefficient} \\
\hline
$r_{\rm A}+r_{\rm b}$ &   0.1941  & 0.0065  &   0.1918  & 0.0052  &   0.1922  & 0.0057  &   0.1912  & 0.0047  &   0.1919  & 0.0052  \\
$k$                 &   0.10479 & 0.00097 &   0.10405 & 0.00093 &   0.10429 & 0.00093 &   0.10411 & 0.00092 &   0.10402 & 0.00089 \\
$i$ (deg.)          &  88.0     & 1.9     &  89.3     & 1.6     &  88.8     & 1.7     &  89.6     & 1.6     &  89.2     & 1.6     \\
$u_{\rm A}$           &   0.211   & 0.54    &   0.055   & 0.060   &$-$0.141   & 0.058   &   0.408   & 0.059   &   0.143   & 0.061   \\
$v_{\rm A}$           &        \mc{ }       & \mc{0.32 perturbed} & \mc{0.60 perturbed} & \mc{0.28 perturbed} & \mc{0.28 perturbed} \\
$T_0$               &   5.7520  & 0.0003  &   5.7520  & 0.0003  &   5.7520  & 0.0003  &   5.7520  & 0.0003  &   5.7520  & 0.0003  \\
\hline \hline \end{tabular} \end{table*}

\begin{table*} \caption{\label{tab:lc:wasp2} Parameters
of the {\sc jktebop} best fits of the WASP-2 light curve
from \citet{Charbonneau+07apj} ($z$ band, $\Nobs = 426$)
for different treatment of LD. The parameter symbols are
as given in Section\,\ref{sec:analysis:lc}. The transit
midpoint $T_0$ is expressed as HJD $-$ 2\,454\,000. For
each part of the table the upper parameters are fitted
quantities and the lower parameters are derived quantities.}
\begin{tabular}{l r@{\,$\pm$\,}l r@{\,$\pm$\,}l r@{\,$\pm$\,}l r@{\,$\pm$\,}l r@{\,$\pm$\,}l}
\hline \hline
\                   &\mc{Linear LD law}&\mc{Quadratic LD law}&\mc{Square-root LD law}&\mc{Logarithmic LD law}&\mc{Cubic LD law}   \\
\hline
\multicolumn{11}{l}{Fitting for one LD coefficient} \\
\hline
$r_{\rm A}+r_{\rm b}$ &   0.1403  & 0.0058  &   0.1409  & 0.0059  &   0.1408  & 0.0060  &   0.1409  & 0.0060  &   0.1409  & 0.0060  \\
$k$                 &   0.1322  & 0.0014  &   0.1310  & 0.0015  &   0.1315  & 0.0015  &   0.1313  & 0.0015  &   0.1313  & 0.0016  \\
$i$ (deg.)          &  84.86    & 0.45    &  84.83    & 0.46    &  84.84    & 0.47    &  84.83    & 0.48    &  84.83    & 0.47    \\
$u_{\rm A}$           &   0.476   & 0.147   &   0.240   & 0.171   &   0.128   & 0.163   &   0.594   & 0.169   &   0.369   & 0.167   \\
$v_{\rm A}$           &        \mc{ }       & \mc{0.30 perturbed} & \mc{0.53 perturbed} & \mc{0.25 perturbed} & \mc{0.20 perturbed} \\
$T_0$               &   8.73204 & 0.00021 &   8.73205 & 0.00022 &   8.73205 & 0.00022 &   8.73205 & 0.00022 &   8.73205 & 0.00021 \\
\hline \hline \end{tabular} \end{table*}

\begin{table*} \caption{\label{tab:lc:hat1a} Parameters of the
{\sc jktebop} best fits of the FLWO light curve of HAT-P-1 from
\citet{Winn+07aj2} for different treatments of LD. The parameter
symbols are as given in Section\,\ref{sec:analysis:lc}. The
transit midpoint $T_0$ is expressed as HJD $-$ 2\,453\,000. For
each part of the table the upper parameters are fitted quantities
and the lower parameters are derived quantities.}
\begin{tabular}{l r@{\,$\pm$\,}l r@{\,$\pm$\,}l r@{\,$\pm$\,}l r@{\,$\pm$\,}l r@{\,$\pm$\,}l}
\hline \hline
\                     &\mc{Linear LD law}&\mc{Quadratic LD law}&\mc{Square-root LD law}&\mc{Logarithmic LD law}&\mc{Cubic LD law}   \\
\hline
\multicolumn{11}{l}{Fitting for one LD coefficient} \\
\hline
$r_{\rm A}+r_{\rm b}$ &   0.1025  & 0.0029  &   0.1026  & 0.0031  &   0.1026  & 0.0030  &   0.1026  & 0.0031  &   0.1026  & 0.0029  \\
$k$                   &   0.11328 & 0.00082 &   0.11217 & 0.00092 &   0.1126  & 0.00092 &   0.1124  & 0.00087 &   0.1128  & 0.00091 \\
$i$ (deg.)            &  86.36    & 0.22    &  86.38    & 0.26    &  86.37    & 0.25    &  86.37    & 0.25    &  86.37    & 0.23    \\
$u_{\rm A}$           &   0.485   & 0.063   &   0.273   & 0.076   &   0.117   & 0.070   &   0.629   & 0.075   &   0.439   & 0.067   \\
$v_{\rm A}$           &        \mc{ }       & \mc{0.30 perturbed} & \mc{0.58 perturbed} & \mc{0.28 perturbed} & \mc{0.10 perturbed} \\
$T_0$                 & 997.79205 & 0.00030 & 997.79205 & 0.00030 & 997.79205 & 0.00026 & 997.79202 & 0.00028 & 997.79206 & 0.00030 \\
\hline \hline \end{tabular} \end{table*}

\begin{table*} \caption{\label{tab:lc:hat1b} Parameters of the
{\sc jktebop} best fits of the Lick light curve of HAT-P-1 from
\citet{Winn+07aj2} for different treatments of LD. The parameter
symbols are as given in Section\,\ref{sec:analysis:lc}. The
transit midpoint $T_0$ is expressed as HJD $-$ 2\,453\,000. For
each part of the table the upper parameters are fitted quantities
and the lower parameters are derived quantities.}
\begin{tabular}{l r@{\,$\pm$\,}l r@{\,$\pm$\,}l r@{\,$\pm$\,}l r@{\,$\pm$\,}l r@{\,$\pm$\,}l}
\hline \hline
\                     &\mc{Linear LD law}&\mc{Quadratic LD law}&\mc{Square-root LD law}&\mc{Logarithmic LD law}&\mc{Cubic LD law}   \\
\hline
\multicolumn{11}{l}{Fitting for one LD coefficient} \\
\hline
$r_{\rm A}+r_{\rm b}$ &   0.1065  & 0.0028  &   0.1055  & 0.0027  &   0.1056  & 0.0030  &   0.1056  & 0.0029  &   0.1056  & 0.0030  \\
$k$                   &   0.11108 & 0.00062 &   0.10992 & 0.00059 &   0.11031 & 0.00066 &   0.11012 & 0.00068 &   0.11054 & 0.00066 \\
$i$ (deg.)            &  86.01    & 0.22    &  86.11    & 0.22    &  86.10    & 0.24    &  86.10    & 0.22    &  86.09    & 0.23    \\
$u_{\rm A}$           &   0.339   & 0.076   &   0.124   & 0.081   &$-$0.022   & 0.081   &   0.486   & 0.083   &   0.304   & 0.079   \\
$v_{\rm A}$           &        \mc{ }       & \mc{0.30 perturbed} & \mc{0.58 perturbed} & \mc{0.28 perturbed} & \mc{0.10 perturbed} \\
$T_0$                 & 997.79357 & 0.00028 & 997.79355 & 0.00026 & 997.79355 & 0.00026 & 997.79357 & 0.00028 & 997.79356 & 0.00026 \\
\hline \hline \end{tabular} \end{table*}

\begin{table*} \caption{\label{tab:lc:ogle10:I} Parameters of the {\sc jktebop}
best fits of the OGLE-TR-10 light curve from \citet{Holman+07apj2} ($I$ band,
$\Nobs = 388$) for different treatment of LD. The parameter symbols are as
given in Section\,\ref{sec:analysis:lc}. The transit midpoint $T_0$ is
expressed as HJD $-$ 2\,453\,00. For each part of the table the upper
parameters are fitted quantities and the lower parameters are derived quantities.}
\begin{tabular}{l r@{\,$\pm$\,}l r@{\,$\pm$\,}l r@{\,$\pm$\,}l r@{\,$\pm$\,}l r@{\,$\pm$\,}l}
\hline \hline
\                   &\mc{Linear LD law}&\mc{Quadratic LD law}&\mc{Square-root LD law}&\mc{Logarithmic LD law}&\mc{Cubic LD law}   \\
\hline
\multicolumn{11}{l}{Fitting for one LD coefficient} \\
\hline
$r_{\rm A}+r_{\rm b}$ &   0.1469  & 0.0058  &   0.1468  & 0.0064  &   0.1469  & 0.0060  &   0.1468  & 0.0052  &   0.1469  & 0.0061  \\
$k$                 &   0.1026  & 0.0010  &   0.1021  & 0.0011  &   0.1023  & 0.0011  &   0.1021  & 0.0011  &   0.1022  & 0.0011  \\
$i$ (deg.)          &  85.98    & 0.59    &  86.05    & 0.68    &  86.02    & 0.61    &  86.04    & 0.65    &  86.02    & 0.62    \\
$u_{\rm A}$           &   0.223   & 0.063   &   0.011   & 0.075   &$-$0.130   & 0.070   &   0.396   & 0.071   &   0.143   & 0.074   \\
$v_{\rm A}$           &        \mc{ }       & \mc{0.32 perturbed} & \mc{0.60 perturbed} & \mc{0.30 perturbed} & \mc{0.20 perturbed} \\
$P$ (days)          & 3.101257 & 0.000005 & 3.101258 & 0.000005 & 3.101257 & 0.000005 & 3.101257 & 0.000005 & 3.101258 & 0.000005 \\
\hline \hline \end{tabular} \end{table*}

\begin{table*} \caption{\label{tab:lc:ogle10:V} Parameters of the {\sc jktebop}
best fits of the VLT $V$-band light curve of OGLE-TR-10 ($\Nobs = 78$) from
\citet{Pont+07aa} for different treatment of LD. The parameter symbols are as
given in Section\,\ref{sec:analysis:lc}. The transit midpoint $T_0$ is expressed
as HJD $-$ 2\,453\,000. For each part of the table the upper parameters are
fitted quantities and the lower parameters are derived quantities.}
\begin{tabular}{l r@{\,$\pm$\,}l r@{\,$\pm$\,}l r@{\,$\pm$\,}l r@{\,$\pm$\,}l r@{\,$\pm$\,}l}
\hline \hline
\                     &\mc{Linear LD law}&\mc{Quadratic LD law}&\mc{Square-root LD law}&\mc{Logarithmic LD law}&\mc{Cubic LD law}   \\
\hline
\multicolumn{11}{l}{Fitting for one LD coefficient} \\
\hline
$r_{\rm A}+r_{\rm b}$ &   0.1881  & 0.0080  &   0.1885  & 0.0098  &   0.1880  & 0.0088  &   0.1882  & 0.0089  &   0.1873  & 0.0092  \\
$k$                   &   0.1183  & 0.0014  &   0.1174  & 0.0017  &   0.1175  & 0.0015  &   0.1176  & 0.0015  &   0.1172  & 0.0016  \\
$i$ (deg.)            &  82.96    & 0.65    &  82.95    & 0.79    &  83.00    & 0.74    &  82.98    & 0.74    &  83.07    & 0.76    \\
$u_{\rm A}$           &   0.48    & 0.14    &   0.29    & 0.19    &   0.13    & 0.16    &   0.58    & 0.17    &   0.40    & 0.17    \\
$v_{\rm A}$           &        \mc{ }       & \mc{0.25 perturbed} & \mc{0.55 perturbed} & \mc{0.20 perturbed} & \mc{0.20 perturbed} \\
$T_0$                 & 890.67095 & 0.00047 & 890.67095 & 0.00049 & 890.67095 & 0.00047 & 890.67095 & 0.00045 & 890.67096 & 0.00047 \\
\hline \hline \end{tabular} \end{table*}

\begin{table*} \caption{\label{tab:lc:ogle10:R} Parameters of the {\sc jktebop}
best fits of the VLT $R$-band light curve of OGLE-TR-10 ($\Nobs = 67$) from
\citet{Pont+07aa} for different treatment of LD. The parameter symbols are
as given in Section\,\ref{sec:analysis:lc}. The transit midpoint $T_0$ is
expressed as HJD $-$ 2\,453\,000. For each part of the table the upper
parameters are fitted quantities and the lower parameters are derived quantities.}
\begin{tabular}{l r@{\,$\pm$\,}l r@{\,$\pm$\,}l r@{\,$\pm$\,}l r@{\,$\pm$\,}l r@{\,$\pm$\,}l}
\hline \hline
\                     &\mc{Linear LD law}&\mc{Quadratic LD law}&\mc{Square-root LD law}&\mc{Logarithmic LD law}&\mc{Cubic LD law}   \\
\hline
\multicolumn{11}{l}{Fitting for one LD coefficient} \\
\hline
$r_{\rm A}+r_{\rm b}$ &   0.1636  & 0.0093  &   0.1634  & 0.0095  &   0.1634  & 0.0096  &   0.1633  & 0.0093  &   0.1633  & 0.0092  \\
$k$                   &   0.11764 & 0.00067 &   0.11720 & 0.00068 &   0.11722 & 0.00069 &   0.11710 & 0.00069 &   0.1174  & 0.00067 \\
$i$ (deg.)            &  84.61    & 0.74    &  84.64    & 0.76    &  84.64    & 0.75    &  84.65    & 0.74    &  84.64    & 0.72    \\
$u_{\rm A}$           &   0.00    & 0.11    &$-$0.19    & 0.13    &$-$0.35    & 0.12    &   0.13    & 0.12    &$-$0.05    & 0.11    \\
$v_{\rm A}$           &        \mc{ }       & \mc{0.25 perturbed} & \mc{0.55 perturbed} & \mc{0.25 perturbed} & \mc{0.10 perturbed} \\
$T_0$                 & 890.67095 & 0.00044 & 890.67095 & 0.00047 & 890.67095 & 0.00043 & 890.67095 & 0.00043 & 890.67095 & 0.00044 \\
\hline \hline \end{tabular} \end{table*}

\begin{table*} \caption{\label{tab:lc:ogle56:v} Parameters of the {\sc jktebop}
best fits of the VLT $V$-band light curve of OGLE-TR-56 (68 observations) from
\citet{Pont+07aa} for different treatment of LD. The parameter symbols are as
given in Section\,\ref{sec:analysis:lc}. The transit midpoint $T_0$ is expressed
as HJD $-$ 2\,453\,000. For each part of the table the upper parameters are
fitted quantities and the lower parameters are derived quantities.}
\begin{tabular}{l r@{\,$\pm$\,}l r@{\,$\pm$\,}l r@{\,$\pm$\,}l r@{\,$\pm$\,}l r@{\,$\pm$\,}l}
\hline \hline
\                     &\mc{Linear LD law}&\mc{Quadratic LD law}&\mc{Square-root LD law}&\mc{Logarithmic LD law}&\mc{Cubic LD law}   \\
\hline
\multicolumn{11}{l}{Fitting for one LD coefficient} \\
\hline
$r_{\rm A}+r_{\rm b}$ &   0.273   & 0.033   &   0.275   & 0.035   &   0.275   & 0.033   &   0.275   & 0.031   &   0.274   & 0.033   \\
$k$                   &   0.1022  & 0.0053  &   0.1012  & 0.0056  &   0.1015  & 0.0053  &   0.1013  & 0.0047  &   0.1017  & 0.0052  \\
$i$ (deg.)            &  78.8     & 2.5     &  78.7     & 2.8     &  78.7     & 2.7     &  78.7     & 2.5     &  78.8     & 2.6     \\
$u_{\rm A}$           &   0.69    & 0.28    &   0.49    & 0.33    &   0.32    & 0.32    &   0.79    & 0.35    &   0.63    & 0.30    \\
$v_{\rm A}$           &        \mc{ }       & \mc{0.25 perturbed} & \mc{0.55 perturbed} & \mc{0.23 perturbed} & \mc{0.10 perturbed} \\
$T_0$                 & 936.59816 & 0.00067 & 936.59814 & 0.00066 & 936.59813 & 0.00067 & 936.59814 & 0.00064 & 936.59814 & 0.00064 \\
\hline \hline \end{tabular} \end{table*}

\begin{table*} \caption{\label{tab:lc:ogle56:r} Parameters of the {\sc jktebop}
best fits of the VLT $R$-band light curve of OGLE-TR-56 (67 observations) from
\citet{Pont+07aa} for different treatment of LD. The parameter symbols are as
given in Section\,\ref{sec:analysis:lc}. The transit midpoint $T_0$ is expressed
as HJD $-$ 2\,453\,000. For each part of the table the upper parameters are
fitted quantities and the lower parameters are derived quantities.}
\begin{tabular}{l r@{\,$\pm$\,}l r@{\,$\pm$\,}l r@{\,$\pm$\,}l r@{\,$\pm$\,}l r@{\,$\pm$\,}l}
\hline \hline
\                     &\mc{Linear LD law}&\mc{Quadratic LD law}&\mc{Square-root LD law}&\mc{Logarithmic LD law}&\mc{Cubic LD law}   \\
\hline
\multicolumn{11}{l}{Fitting for one LD coefficient} \\
\hline
$r_{\rm A}+r_{\rm b}$ &   0.236   & 0.043   &   0.265   & 0.039   &   0.252   & 0.041   &   0.347   & 0.018   &   0. 333  & 0.042   \\
$k$                   &   0.0890  & 0.0104  &   0.0912  & 0.0090  &   0.0900  & 0.0099  &   0.0918  & 0.0113  &   0.0999  & 0.0081  \\
$i$ (deg.)            &  83.3     & 4.3     &  81.4     & 3.3     &  82.3     & 3.5     &  73.4     & 1.5     &  77.3     & 2.5     \\
$u_{\rm A}$           &   1.14    & 0.31    &   1.06    & 0.30    &   1.06    & 0.29    &$-$0.84    & 3.43    &   1.36    & 0.11    \\
$v_{\rm A}$           &        \mc{ }       & \mc{0.28 perturbed} & \mc{0.60 perturbed} & \mc{0.25 perturbed} & \mc{0.10  perturbed} \\
$T_0$                 & 936.59811 & 0.00049 & 936.59811 & 0.00049 & 936.59814 & 0.00049 & 936.59852 & 0.00046 & 936.59843 & 0.00046 \\
\hline \hline \end{tabular} \end{table*}

\begin{table*} \caption{\label{tab:lc:ogle111} Parameters of the {\sc jktebop}
best fits of the OGLE-TR-111 light curve from \citet{Winn++07aj} ($I$ band,
$\Nobs = 386$) for different treatment of LD. The parameter symbols are as given
in Section\,\ref{sec:analysis:lc}. The transit midpoint $T_0$ is expressed as
HJD $-$ 2\,453\,000. For each part of the table the upper parameters are fitted
quantities and the lower parameters are derived quantities.}
\begin{tabular}{l r@{\,$\pm$\,}l r@{\,$\pm$\,}l r@{\,$\pm$\,}l r@{\,$\pm$\,}l r@{\,$\pm$\,}l}
\hline \hline
\                   &\mc{Linear LD law}&\mc{Quadratic LD law}&\mc{Square-root LD law}&\mc{Logarithmic LD law}&\mc{Cubic LD law}   \\
\hline
\multicolumn{11}{l}{Fitting for one LD coefficient} \\
\hline
$r_{\rm A}+r_{\rm b}$ &   0.0954  & 0.0036  &   0.0950  & 0.0040  &   0.0954  & 0.0037  &   0.0953  & 0.0036  &   0.0953  & 0.0039  \\
$k$                 &   0.1324  & 0.0017  &   0.1313  & 0.0021  &   0.1318  & 0.0018  &   0.1316  & 0.0018  &   0.1314  & 0.0019  \\
$i$ (deg.)          &  88.03    & 0.47    &  88.15    & 0.61    &  88.08    & 0.51    &  88.11    & 0.51    &  88.12    & 0.56    \\
$u_{\rm A}$           &   0.454   & 0.051   &   0.320   & 0.064   &   0.189   & 0.057   &   0.604   & 0.061   &   0.405   & 0.060   \\
$v_{\rm A}$           &        \mc{ }       & \mc{0.26 perturbed} & \mc{0.45 perturbed} & \mc{0.22 perturbed} & \mc{0.20 perturbed} \\
$T_0$               & 799.75165 & 0.00019 & 799.75165 & 0.00021 & 799.75165 & 0.00020 & 799.75165 & 0.00020 & 799.75165 & 0.00020 \\
\hline \hline \end{tabular} \end{table*}

\begin{table*} \caption{\label{tab:lc:ogle132} Parameters of
the {\sc jktebop} best fits of the OGLE-TR-132 light curve from
\citet{Gillon+07aa3} ($R_{\rm special}$ filter, $\Nobs = 274$)
for different treatment of LD. The parameter symbols are as
given in Section\,\ref{sec:analysis:lc}. The transit midpoint
$T_0$ is expressed as HJD $-$ 2\,453\,000. For each part of the
table the upper parameters are fitted quantities and the lower
parameters are derived quantities.}
\begin{tabular}{l r@{\,$\pm$\,}l r@{\,$\pm$\,}l r@{\,$\pm$\,}l r@{\,$\pm$\,}l r@{\,$\pm$\,}l}
\hline \hline
\                   &\mc{Linear LD law}&\mc{Quadratic LD law}&\mc{Square-root LD law}&\mc{Logarithmic LD law}&\mc{Cubic LD law}   \\
\hline
\multicolumn{11}{l}{Fitting for one LD coefficient} \\
\hline
$r_{\rm A}+r_{\rm b}$ &   0.228   & 0.017   &   0.231   & 0.020   &   0.231   & 0.018   &   0.231   & 0.019   &   0.231   & 0.020   \\
$k$                   &   0.0941  & 0.0021  &   0.0937  & 0.0022  &   0.0938  & 0.0020  &   0.0937  & 0.0022  &   0.0937  & 0.0023  \\
$i$ (deg.)            &  83.5     & 1.7     &  83.2     & 2.0     &  83.3     & 1.8     &  83.3     & 2.0     &  83.3     & 2.0     \\
$u_{\rm A}$           &   0.534   & 0.071   &   0.341   & 0.099   &   0.166   & 0.088   &   0.684   & 0.086   &   0.464   & 0.088   \\
$v_{\rm A}$           &        \mc{ }       & \mc{0.30 perturbed} & \mc{0.60 perturbed} & \mc{0.25 perturbed} & \mc{0.20 perturbed} \\
$T_0$                 & 142.59118 & 0.00033 & 142.59121 & 0.00035 & 142.59120 & 0.00033 & 142.59121 & 0.00034 & 142.59121 & 0.00034 \\
\hline \hline \end{tabular} \end{table*}

\begin{table*} \caption{\label{tab:lc:gj436} Parameters of the
{\sc jktebop} best fits of the {\it Spitzer} light curve of
GJ\,436 for different LD laws. The parameter symbols are as given
in Section\,\ref{sec:analysis:lc}. The transit midpoint $T_0$
is expressed as HJD $-$ 2454000. For each part of the table the
upper parameters are fitted quantities and the lower parameters
are derived quantities.}
\begin{tabular}{l r@{\,$\pm$\,}l r@{\,$\pm$\,}l r@{\,$\pm$\,}l r@{\,$\pm$\,}l r@{\,$\pm$\,}l}
\hline \hline
\                   &\mc{Linear LD law}&\mc{Quadratic LD law}&\mc{Square-root LD law}&\mc{Logarithmic LD law}&\mc{Cubic LD law}   \\
\hline
$r_{\rm A}+r_{\rm b}$ &   0.0794  & 0.0024  &   0.0790  & 0.0024  &   0.0794  & 0.0023  &   0.0790  & 0.0023  &   0.0791  & 0.0025  \\
$k$                   &   0.08313 & 0.00072 &   0.08283 & 0.00075 &   0.08259 & 0.00078 &   0.08300 & 0.00077 &   0.08291 & 0.00074 \\
$i$ (deg.)            &  86.41    & 0.15    &  86.41    & 0.15    &  86.41    & 0.15    &  86.44    & 0.14    &  86.44    & 0.15    \\
$u_{\rm A}$           & \mc{0.10 perturbed}&\mc{$-$0.05 perturbed}&\mc{$-$0.11 perturbed}&\mc{0.18 perturbed} & \mc{0.05 perturbed} \\
$v_{\rm A}$           &        \mc{ }       & \mc{0.20 perturbed} & \mc{0.25 perturbed} & \mc{0.14 perturbed} & \mc{0.10 perturbed} \\
$T_0$                 & 280.78212 & 0.00011 & 280.78211 & 0.00011 & 280.78212 & 0.00011 & 280.78212 & 0.00010 & 280.78212 & 0.00010 \\
\hline
$r_{\rm A}$           &   0.07332 & 0.0022  &   0.0730  & 0.0022  &   0.0734  & 0.0021  &   0.0730  & 0.0021  &   0.0730  & 0.0023  \\
$r_{\rm b}$           &   0.006095& 0.00019 &   0.00604 & 0.00020 &   0.00606 & 0.00018 &   0.00606 & 0.00018 &   0.00606 & 0.00019 \\
$\sigma$ ($m$mag)     &     \mc{0.7043}     &     \mc{0.7051}     &     \mc{0.7041}     &     \mc{0.7049}     &     \mc{0.7048}     \\
$\chi^2_{\rm \ red}$  &     \mc{1.3718}     &     \mc{1.3748}     &     \mc{1.3712}     &     \mc{1.3743}     &     \mc{1.3738}     \\
\hline \hline \end{tabular} \end{table*}

\clearpage

\begin{table*} \caption{\label{tab:lc:149026} Parameters of the {\sc jktebop}
best fits of the HD\,149026 light curve from \citet{Sato+05apj} and
\citet{Charbonneau+06apj} for different treatment of LD. The parameter symbols
are as given in Section\,\ref{sec:analysis:lc}. The transit midpoint $T_0$ is
expressed as HJD $-$ 2\,453\,000. For each part of the table the upper parameters
are fitted quantities and the lower parameters are derived quantities.}
\begin{tabular}{l c c c c c}
\hline \hline
\                   & Linear LD law & Quadratic LD law & Square-root LD law & Logarithmic LD law \\
\hline
\multicolumn{5}{l}{APT ($b$+$y$/2) light curve from \citet{Sato+05apj}} \\
\hline
$r_{\rm A}+r_{\rm b}$ & \er{0.143}{0.062}{0.019}       & \er{0.145}{0.065}{0.017}       & \er{0.144}{0.060}{0.017}       & \er{0.158}{0.065}{0.015}       \\[3pt]
$k$                   & \er{0.0464}{0.0033}{0.0028}    & \er{0.0463}{0.0033}{0.0029}    & \er{0.0482}{0.0031}{0.0035}    & \er{0.0480}{0.0033}{0.0030}    \\[3pt]
$i$ (deg.)            & 82.0 to 90.0                   & 83.0 to 90.0                   & 83.0 to 90.0                   & 83.0 to 90.0                   \\[3pt]
$u_{\rm A}$           & \er{0.735}{0.35}{0.26}         & \er{0.510}{0.392}{0.302}       & \er{0.331}{0.358}{0.271}       & \er{0.780}{0.298}{0.305}       \\[3pt]
$v_{\rm A}$           &                                & 0.25 perturbed                 & 0.55 perturbed                 & 0.20 perturbed                 \\[3pt]
$P$ (days)            & \er{2.87555}{0.00037}{0.00034} & \er{2.87560}{0.00031}{0.00036} & \er{2.87558}{0.00034}{0.00036} & \er{2.87286}{0.00031}{0.00034} \\
\hline \hline \end{tabular} \end{table*}

\begin{table*} \caption{\label{tab:lc:189733:flwor} Parameters of the
{\sc jktebop} best fits of the FLWO $r$-band light curve ($\Nobs = 642$)
of HD\,189733 from \citet{Bakos+06apj2} for different treatments of LD.
The parameter symbols are as given in Section\,\ref{sec:analysis:lc}. The
transit midpoint $T_0$ is expressed as HJD $-$ 2\,453\,000. For each part
of the table the upper parameters are fitted quantities and the lower
parameters are derived quantities. Note that the $\sigma$ is affected by
scatter due to high airmass towards the end of the observations.}
\begin{tabular}{l r@{\,$\pm$\,}l r@{\,$\pm$\,}l r@{\,$\pm$\,}l r@{\,$\pm$\,}l r@{\,$\pm$\,}l}
\hline \hline
\                     &\mc{Linear LD law}&\mc{Quadratic LD law}&\mc{Square-root LD law}&\mc{Logarithmic LD law}&\mc{Cubic LD law}   \\
\hline
\multicolumn{11}{l}{Fitting for one LD coefficient} \\
\hline
$r_{\rm A}+r_{\rm b}$ &   0.1190  & 0.0068  &   0.1194  & 0.0066  &   0.1193  & 0.0069  &   0.1194  & 0.0069  &   0.1192  & 0.0073  \\
$k$                   &   0.1509  & 0.0034  &   0.1500  & 0.0036  &   0.1504  & 0.0037  &   0.1502  & 0.0033  &   0.1497  & 0.0039  \\
$i$ (deg.)            &  86.56    & 0.61    &  86.56    & 0.64    &  86.55    & 0.65    &  86.55    & 0.64    &  86.59    & 0.69    \\
$u_{\rm A}$           &   0.541   & 0.123   &   0.413   & 0.148   &   0.309   & 0.133   &   0.651   & 0.139   &   0.481   & 0.148   \\
$v_{\rm A}$           &        \mc{ }       & \mc{0.21 perturbed} & \mc{0.38 perturbed} & \mc{0.18 perturbed} & \mc{0.20 perturbed} \\
$T_0$                 & 642.70584 & 0.00025 & 642.70584 & 0.00026 & 642.70584 & 0.00025 & 642.70584 & 0.00024 & 642.70584 & 0.00025 \\
\hline \hline \end{tabular} \end{table*}

\begin{table*} \caption{\label{tab:lc:189733:flwoz} Parameters of the
{\sc jktebop} best fits of the FLWO light curve of HD\,189733 ($z$ band,
$\Nobs = 1662$) from \citet{Winn+07aj} for different treatments of LD.
The parameter symbols are as given in Section\,\ref{sec:analysis:lc}.
For each part of the table the upper parameters are fitted quantities
and the lower parameters are derived quantities.}
\begin{tabular}{l r@{\,$\pm$\,}l r@{\,$\pm$\,}l r@{\,$\pm$\,}l r@{\,$\pm$\,}l r@{\,$\pm$\,}l}
\hline \hline
\                   &\mc{Linear LD law}&\mc{Quadratic LD law}&\mc{Square-root LD law}&\mc{Logarithmic LD law}&\mc{Cubic LD law}   \\
\hline
\multicolumn{11}{l}{Fitting for one LD coefficient} \\
\hline
$r_{\rm A}+r_{\rm b}$ &   0.1295  & 0.0019  &   0.1298  & 0.0021  &   0.1297  & 0.0019  &   0.1298  & 0.0020  &   0.1297  & 0.0020  \\
$k$                   &   0.1585  & 0.0009  &   0.1572  & 0.0009  &   0.1579  & 0.0009  &   0.1576  & 0.0009  &   0.1580  & 0.0009  \\
$i$ (deg.)            &  85.73    & 0.16    &  85.73    & 0.16    &  85.73    & 0.16    &  85.74    & 0.17    &  85.73    & 0.17    \\
$u_{\rm A}$           &   0.440   & 0.058   &   0.222   & 0.062   &   0.124   & 0.062   &   0.571   & 0.066   &   0.397   & 0.058   \\
$v_{\rm A}$           &        \mc{ }       & \mc{0.31 perturbed} & \mc{0.50 perturbed} & \mc{0.24 perturbed} & \mc{0.10 perturbed} \\
\hline
$r_{\rm A}$           &   0.1118  & 0.0016  &   0.1122  & 0.0016  &   0.1120  & 0.0016  &   0.1121  & 0.0017  &   0.1120  & 0.0016  \\
\hline \hline \end{tabular} \end{table*}

\begin{table*} \caption{\label{tab:lc:189733:wiseI} Parameters of the
{\sc jktebop} best fits of the WISE Observatory HD\,189733 light curve
($I$ band, $\Nobs = 345$) from \citet{Winn+07aj} for different LD laws.
The parameter symbols are as given in Section\,\ref{sec:analysis:lc}.
The transit midpoint $T_0$ is expressed as HJD $-$ 2\,450\,000. For
each part of the table the upper parameters are fitted quantities and
the lower parameters are derived quantities.}
\begin{tabular}{l r@{\,$\pm$\,}l r@{\,$\pm$\,}l r@{\,$\pm$\,}l r@{\,$\pm$\,}l r@{\,$\pm$\,}l}
\hline \hline
\                     &\mc{Linear LD law}&\mc{Quadratic LD law}&\mc{Square-root LD law}&\mc{Logarithmic LD law}&\mc{Cubic LD law}   \\
\hline
\multicolumn{11}{l}{Fitting for one LD coefficient} \\
\hline
$r_{\rm A}+r_{\rm b}$ &   0.1320  & 0.0067  &   0.1327  & 0.0072  &   0.1327  & 0.0073  &   0.1327  & 0.0071  &   0.1325  & 0.0074  \\
$k$                   &   0.1506  & 0.0025  &   0.1496  & 0.0026  &   0.1500  & 0.0024  &   0.1498  & 0.0025  &   01501   & 0.0026  \\
$i$ (deg.)            &  85.41    & 0.56    &  85.37    & 0.61    &  85.38    & 0.58    &  85.38    & 0.60    &  85.38    & 0.60    \\
$u_{\rm A}$           &   0.42   & 0.22   &   0.22   & 0.27   &   0.11   & 0.25   &   0.52   & 0.27   &   0.36   & 0.25   \\
$v_{\rm A}$           &        \mc{ }       & \mc{0.25 perturbed} & \mc{0.46 perturbed} & \mc{0.22 perturbed} & \mc{0.10 perturbed} \\
$T_0$                 & 984.36572 & 0.00029 & 984.36582 & 0.00028 & 984.36582 & 0.00027 & 984.36582 & 0.00028 & 984.36582 & 0.00029 \\
\hline
$r_{\rm A}$           &   0.1147  & 0.0056  &   0.1154  & 0.0060  &   0.1153  & 0.0062  &   0.1154  & 0.0060  &   0.1152  & 0.0064  \\
$r_{\rm b}$           &   0.0173  & 0.0010  &   0.0173  & 0.0011  &   0.0173  & 0.0011  &   0.0173  & 0.0011  &   0.0173  & 0.0011  \\
$\sigma$ ($m$mag)     &     \mc{3.2212}     &     \mc{3.2204}     &     \mc{3.2207}     &     \mc{3.2205}     &     \mc{3.2207}     \\
$\chi^2_{\rm \ red}$  &     \mc{1.0042}     &     \mc{1.0038}     &     \mc{1.0039}     &     \mc{1.0038}     &     \mc{1.0040}     \\
\hline \hline \end{tabular} \end{table*}

\begin{table*} \caption{\label{tab:lc:209458:brown} Parameters
of the {\sc jktebop} best fits of the HD\,209458 light curve
from \citet{Brown+01apj} ($\Nobs = 556$) for different
treatments of LD. The parameter symbols are as given in
Section\,\ref{sec:analysis:lc}. The transit midpoint $T_0$ is
expressed as HJD $-$ 2\,451\,000. For each part of the table
the upper parameters are fitted quantities and the lower
parameters are derived quantities.}
\begin{tabular}{l r@{\,$\pm$\,}l r@{\,$\pm$\,}l r@{\,$\pm$\,}l r@{\,$\pm$\,}l r@{\,$\pm$\,}l}
\hline \hline
\                     &\mc{Linear LD law}&\mc{Quadratic LD law}&\mc{Square-root LD law}&\mc{Logarithmic LD law}&\mc{Cubic LD law}   \\
\hline
\multicolumn{11}{l}{Fitting for one LD coefficient} \\
\hline
$r_{\rm A}+r_{\rm b}$ &   0.12889 & 0.00041 &   0.12768 & 0.00048 &   0.12840 & 0.00044 &   0.12811 & 0.00048 &   0.12846 & 0.00043 \\
$k$                   &   0.12260 & 0.00011 &   0.12095 & 0.00020 &   0.12172 & 0.00013 &   0.12137 & 0.00018 &   0.12201 & 0.00021 \\
$i$ (deg.)            &  86.472   & 0.038   &  86.668   & 0.052   &  86.567   & 0.044   &  86.612   & 0.049   &  86.543   & 0.046   \\
$u_{\rm A}$           &   0.494   & 0.004   &   0.297   & 0.020   &   0.152   & 0.021   &   0.674   & 0.023   &   0.463   & 0.011   \\
$v_{\rm A}$           &        \mc{ }       & \mc{0.34 perturbed} & \mc{0.57 perturbed} & \mc{0.28 perturbed} & \mc{0.10 perturbed} \\
$T_0$                 & 659.936716& 0.000021& 659.936712& 0.000022& 659.936714& 0.000021& 659.936713& 0.000020& 659.963715& 0.000022\\
\hline \hline \end{tabular} \end{table*}

\begin{table*} \caption{\label{tab:lc:209458:most} Parameters of the {\sc jktebop}
best fits of the MOST light curve of HD\,209458 from \citet{Rowe+06apj}
(unfiltered GOAT?, $\Nobs = 8772$ as analysed here) for different treatments
of LD. The parameter symbols are as given in Section\,\ref{sec:analysis:lc}.
For each part of the table the upper parameters are fitted quantities and
the lower parameters are derived quantities.}
\begin{tabular}{l r@{\,$\pm$\,}l r@{\,$\pm$\,}l r@{\,$\pm$\,}l r@{\,$\pm$\,}l r@{\,$\pm$\,}l}
\hline \hline
\                     &\mc{Linear LD law}&\mc{Quadratic LD law}&\mc{Square-root LD law}&\mc{Logarithmic LD law}&\mc{Cubic LD law}   \\
\hline
\multicolumn{11}{l}{Fitting for one LD coefficient} \\
\hline
$r_{\rm A}+r_{\rm b}$ &   0.1269  & 0.0021  &   0.1259  & 0.0022  &   0.1266  & 0.0022  &   0.1264  & 0.0022  &   0.1266  & 0.0022  \\
$k$                   &   0.12447 & 0.00071 &   0.1231  & 0.00079 &   0.1235  & 0.00080 &   0.1235  & 0.00078 &   0.1238  & 0.00076 \\
$i$ (deg.)            &  86.69    & 0.22    &  86.85    & 0.24    &  86.77    & 0.24    &  86.79    & 0.24    &  86.75    & 0.23    \\
$u_{\rm A}$           &   0.552   & 0.023   &   0.415   & 0.031   &   0.196   & 0.032   &   0.691   & 0.034   &   0.524   & 0.026   \\
$v_{\rm A}$           &        \mc{ }       & \mc{0.25 perturbed} & \mc{0.60 perturbed} & \mc{0.21 perturbed} & \mc{0.10 perturbed} \\
\hline
$r_{\rm A}$           &   0.1128  & 0.0018  &   0.1121  & 0.0019  &   0.1127  & 0.0019  &   0.1125  & 0.0019  &   0.1126  & 0.0019  \\
\hline \hline \end{tabular} \end{table*}

\begin{table*} \caption{\label{tab:lc:209458:lin} Parameters of the
{\sc jktebop} best fits of the HST light curves of HD\,209458 from
\citet{Knutson+07apj} ($\Nobs = 504$ for the five bluer passbands
and $\Nobs = 548$ for the five redder passbands) for the linear LD
law. The parameter symbols are as given in Section\,\ref{sec:analysis:lc}.
For each part of the table the upper parameters are fitted quantities
and the lower parameters are derived quantities.}
\begin{tabular}{l r@{\,$\pm$\,}l r@{\,$\pm$\,}l r@{\,$\pm$\,}l r@{\,$\pm$\,}l r@{\,$\pm$\,}l}
\hline \hline
\                     &    \mc{320\,nm}     &    \mc{375\,nm}     &    \mc{430\,nm}     &    \mc{484\,nm}     &    \mc{539\,nm}     \\
\hline
$r_{\rm A}+r_{\rm b}$ &   0.1270  & 0.0024  &   0.1247  & 0.0024  &   0.1271  & 0.0007  &   0.1272  & 0.0006  &   0.1286  & 0.0006  \\
$k$                   &   0.12230 & 0.00128 &   0.12160 & 0.0013  &   0.12248 & 0.0003  &   0.12238 & 0.0002  &   0.12272 & 0.0002  \\
$i$ (deg.)            &  86.71    & 0.26    &  86.93    & 0.15    &  86.65    & 0.07    &  86.63    & 0.06    &  86.50    & 0.06    \\
$u_{\rm A}$           &   0.828   & 0.023   &   0.754   & 0.013   &   0.703   & 0.007   &   0.618   & 0.006   &   0.561   & 0.0007  \\
\hline
$r_{\rm A}$           &   0.1131  & 0.0021  &   0.1112  & 0.0011  &   0.1133  & 0.0006  &   0.1134  & 0.0005  &   0.1146  & 0.0005  \\
$r_{\rm b}$           &   0.01384 & 0.00039 &   0.01352 & 0.00020 &   0.01387 & 0.00010 &   0.01387 & 0.00008 &   0.01406 & 0.00009 \\
$\sigma$ ($m$mag)     &     \mc{0.6568}     &     \mc{0.3686}     &     \mc{0.2238}     &     \mc{0.1989}     &     \mc{0.2067}     \\
\hline \hline \end{tabular} \end{table*}

\begin{table*} \caption{\label{tab:lc:209458:quad} Parameters of the
{\sc jktebop} best fits of the HST light curves of HD\,209458 from
\citet{Knutson+07apj} ($\Nobs = 504$ for the five bluer passbands
and $\Nobs = 548$ for the five redder passbands) for the quadratic LD
law. The parameter symbols are as given in Section\,\ref{sec:analysis:lc}.
For each part of the table the upper parameters are fitted quantities
and the lower parameters are derived quantities.}
\begin{tabular}{l r@{\,$\pm$\,}l r@{\,$\pm$\,}l r@{\,$\pm$\,}l r@{\,$\pm$\,}l r@{\,$\pm$\,}l}
\hline \hline
\                     &    \mc{320\,nm}     &    \mc{375\,nm}     &    \mc{430\,nm}     &    \mc{484\,nm}     &    \mc{539\,nm}     \\
\hline
$r_{\rm A}+r_{\rm b}$ &   0.1271  & 0.0022  &   0.1248  & 0.0012  &   0.1271  & 0.0007  &   0.1272  & 0.0006  &   0.1277  & 0.0008  \\
$k$                   &   0.1249  & 0.0018  &   0.1221  & 0.0009  &   0.12249 & 0.0006  &   0.1223  & 0.0005  &   0.1213  & 0.0005  \\
$i$ (deg.)            &  86.57    & 0.25    &  86.90    & 0.15    &  86.65    & 0.09    &  86.64    & 0.07    &  86.66    & 0.09    \\
$u_{\rm A}$           &   1.030   & 0.102   &   0.791   & 0.052   &   0.703   & 0.036   &   0.612   & 0.034   &   0.426   & 0.039   \\
$v_{\rm A}$           &$-$0.384   & 0.182   &$-$0.073   & 0.102   &$-$0.001   & 0.068   &   0.009   & 0.062   &   0.248   & 0.092   \\
\hline
$r_{\rm A}$           &   0.1129  & 0.0018  &   0.1112  & 0.0010  &   0.1133  & 0.0006  &   0.1133  & 0.0005  &   0.1139  & 0.0007  \\
$r_{\rm b}$           &   0.01411 & 0.0004  &   0.01357 & 0.00021 &   0.01387 & 0.00013 &   0.01386 & 0.00010 &   0.01382 & 0.00013 \\
\hline \hline \end{tabular} \end{table*}

\end{document}